\documentclass[twocolumn,nofootinbib,prd,aps,superscriptaddress,tightenlines]
{revtex4}

\usepackage{amssymb}
\usepackage{bm}
\usepackage{braket}
\usepackage{slashed}
\usepackage{color}
\usepackage{graphicx}

{\count255=\time\divide\count255 by 60 \xdef\hourmin{\number\count255}
  \multiply\count255 by-60\advance\count255 by\time
  \xdef\hourmin{\hourmin:\ifnum\count255<10 0\fi\the\count255}}

\def\lqcd{ \Lambda_{\text{QCD}}}
\newcommand{\nn}{\nonumber \\ }

\newcommand\lQ{\mathsf{L_Q}}
\newcommand\lM{\mathsf{L_M}}
\newcommand\LL{\mathsf{L}}
\newcommand\lQM{\mathsf{L_{Q/M}}}
\newcommand\lm{\mathsf{L_m}}

\newcommand\lp{\mathsf{L_p}}

\newcommand\conf{\mathcal{C}}
\def\mul{\mu_l}
\newcommand\lMl{\log \frac{M^2}{\mul^2} }
\def\proj{\mathsf{P}}

\def\sceth{ $\text{SCET}_{\text{EW}}$}
\def\scetl{ $\text{SCET}_{\gamma}$}

\def\softm{\mathfrak{S}}

\def\waver{\mathfrak{R}}

\def\amp{{\mathfrak{M}}}
\def\bamp{\boldsymbol{\mathfrak{M}}}
\def\aem{\alpha_{\text{em}}}

\def\msbar{$\overline{\hbox{MS}}$}

\def\abs#1{\left| #1 \right|}
\def\vev#1{\left\langle #1 \right\rangle}

\def\bn{ \bar n}

\def\rd{{\rm d}}

\begin{document}

\title{Factorization Structure of Gauge Theory Amplitudes and Application to\\  Hard 
Scattering Processes at the LHC}

\author{Jui-yu Chiu}
\affiliation{Department of Physics, University of California at San Diego,
  La Jolla, CA 92093}

\author{Andreas Fuhrer}
\affiliation{Department of Physics, University of California at San Diego,
  La Jolla, CA 92093}

\author{Randall Kelley}
\affiliation{Department of Physics, University of California at San Diego,
  La Jolla, CA 92093}

\author{Aneesh V.~Manohar}
\affiliation{Department of Physics, University of California at San Diego,
  La Jolla, CA 92093}

\begin{abstract}
Previous work on electroweak radiative corrections to high energy scattering using soft-collinear effective theory (SCET) has  been extended to include external transverse and longitudinal gauge bosons and Higgs bosons. This allows one to compute radiative corrections to \emph{all} parton-level hard scattering amplitudes in the standard model to NLL order, including QCD and electroweak radiative corrections, mass effects, and Higgs exchange corrections, if the high-scale matching, which is suppressed by two orders in the log counting, and contains no large logs, is known. The factorization structure of the effective theory places strong constraints on the form of gauge theory amplitudes at high energy for massless and massive gauge theories, which are discussed in detail in the paper. The radiative corrections can be written as the sum of process-independent one-particle collinear functions, and a universal soft function. We give plots for the radiative corrections to $q \bar q \to W_T W_T$, $Z_TZ_T$, $W_LW_L$, and $Z_L H$, and $gg \to W_T W_T$ to illustrate our results. The purely electroweak corrections are large, ranging from 12\% at 500~GeV to 37\% at 2~TeV for transverse $W$ pair production, and increasing rapidly with energy. The estimated theoretical uncertainty to the partonic (hard) cross-section in most cases is below one percent, smaller than uncertainties in the parton distribution functions (PDFs). We  discuss the relation between SCET and other factorization methods, and derive the Magnea-Sterman equations for the Sudakov form factor using SCET, for massless and massive gauge theories, and for light and heavy external particles. 
\end{abstract}

\date{\today\quad\hourmin}

\maketitle
%\tableofcontents

\section{Introduction}\label{sec:intro}

Soft collinear effective theory (SCET) \cite{BFL,SCET1,SCET2,BPS} is a field theory which describes the interactions of energetic particles which produce final states with small invariant mass. SCET was developed for QCD, but has recently been extended to spontaneously broken gauge theories with massive gauge bosons~\cite{CGKM1,CGKM2,CKM}.  This allows one to compute electroweak corrections to Standard Model processes at high energies, and to sum electroweak Sudakov logarithms~\cite{CGKM2,CKM}.  One can compute QCD and electroweak corrections to parton-level scattering amplitudes with 1\% precision. To achieve this accuracy, one needs to compute two-loop QCD and one-loop electroweak radiative corrections, and sum the  large logarithms using renormalization group methods in SCET. An extensive analysis with plots of numerical results for $2\to2$ parton processes such as dijet production, Drell-Yan, and top-quark production was given in Ref.~\cite{CKM}. The results included fermion mass effects (e.g. due to the top quark), $\gamma-Z$ mixing (which leads to $M_W \not=M_Z$) as well as Higgs exchange corrections. The results are extended in this paper to include Higgs and gauge bosons. The purely electroweak corrections are substantial --- 20\% for $W$ pair production and 10\% for $Z H$ production at 1~TeV rising to 50\% and 25\% respectively, at 4~TeV, and are much larger than two-loop QCD corrections.

The Sudakov double-logarithms arise as a result of soft and collinear singularities in radiation, and are a feature of exclusive amplitudes such as form-factors. They are absent in totally inclusive processes such as the total $e^+ e^-$ hadronic cross-section. There is some cancellation of the QCD Sudakov corrections in inclusive processes such as jet rates, because the infrared singularities cancel between real and virtual graphs. Jet cross-sections still have a residual double-logarithmic dependence on infrared parameters in the jet definition such as $y_{\text{cut}}$, and can have large radiative corrections.  Unlike QCD, the electroweak Sudakov corrections do not cancel, even for a totally inclusive cross-section, because the initial state particles are not electroweak singlets~\cite{ccc,ciafaloni1,ciafaloni2}.

Radiative corrections to high energy processes have been obtained previously using fixed order computations by many groups~\cite{ccc,ciafaloni1,ciafaloni2,fadin,kps,fkps,jkps,jkps4,beccaria,dp1,dp2,hori,beenakker,dmp,pozzorini,js,melles1,melles2,melles3,Denner:2006jr,kuhnW,Denner:2008yn}. The computations use infrared evolution equations~\cite{fadin}, based on an analysis of the infrared structure of the perturbation theory amplitude and a factorization theorem for the Sudakov form factor~\cite{pqcd}.  We have checked our EFT results against these computations. 

The computation of electroweak radiative corrections is greatly simplified by the use of effective field theory methods. The computation splits into a high-scale matching condition, and the EFT computation. The matching can be computed neglecting all low energy scales such as fermion masses, gauge boson masses, and spontaneous symmetry breaking. Secondly, the effective theory results depend only on the color structure of the amplitude, and are independent of the momentum structure. This is because the large momenta (i.e. the kinematic variables $s,t,u$) are labels in the effective theory, and not dynamical variables. Thus an amplitude $A(s,t,u)$ in the effective theory can be treated as a number, with a different value depending on the kinematics. In contrast, in a computation without the use of effective theory methods, $A$ is a vertex in a Feynman graph, and the arguments of $A$ get integrated over in loop graphs. In gauge boson production, there are 10 possible kinematic structures~\cite{sack}, and they all evolve the same way in the effective theory.

In this paper, we explain in more detail the factorization structure of the high energy amplitudes which follows from the effective theory~\cite{HardScattering,CGKM1,CGKM2,CKM}, and extend our results to include external gauge bosons and Higgs scalars, so that we have the results for \emph{all} standard model particles. The computations for longitudinally polarized gauge  bosons use the Goldstone boson equivalence theorem~\cite{cornwall,vayonakis,leequiggthacker,chanowitz,gounaris,bagger,yaoyuan,bohmbook}. The factorization results hold both for the anomalous dimension, as well as for the low-energy matrix elements in the effective theory, i.e.\ for both the infinite and finite parts of the amplitude.

The factorization structure is rederived using the $\Delta$-regulator introduced in Ref.~\cite{Delta}, because this gives the answer in a form which can be most easily compared with previous results using other methods~\cite{ms,aybat1,aybat2,dms}. We show that the hard scattering amplitude can be written as a sum of process-independent one-particle collinear functions which depend on external particle energies, and a universal soft function which depends only on the external particle directions. The amplitude for a hard scattering process with $r$ external particles is given by combining the collinear functions for each external particle with the universal soft function. The collinear function for all standard model particles, and the universal soft function can be computed using the results given in this paper. The explicit expressions for the standard model are rather lengthy, because of the different possible fermion quantum numbers, and because of custodial $SU(2)$ violation which leads to $\gamma-Z$ mixing, $M_W \not = M_Z$, and to different radiative corrections for the $W$ and $Z$. For this reason, in this paper, we give the results for a broken $SU(2)$ gauge theory, with $2n_F$ massless fermion doublets. The standard model results can be obtained straightforwardly from the $SU(2)$ results, and will be given explicitly in a companion paper~\cite{p2}. All numerical plots shown here are in the full standard model, including Higgs corrections due to the top-quark Yukawa coupling.

The high energy behavior of scattering amplitudes for massive gauge theories is the same as that for massless gauge theories, since symmetry breaking is a soft  effect and does not influence the high energy behavior of the amplitude. Thus our results are related to recent work on the infrared structure of gauge theory amplitudes~\cite{armoni,alday1,alday2,dms,gardi,becherneubert,alday3}, which we discuss in Sec.~\ref{sec:casimir}. Of particular interest is whether the SCET amplitudes obey two properties, Casimir scaling and the sum-on-pairs form. 

We compute the Sudakov form factor in a massive gauge theory, following the analysis of Magnea and Sterman~\cite{ms}. This clarifies the connection between SCET results and the analysis using factorization methods. While SCET and factorization methods~\cite{css} use the same terminology of soft, collinear and hard amplitudes, the different contributions are not identical in the two approaches, even though their sum is the same. In our computation, the infrared sector of the gauge theory is under complete perturbative control, since the gauge bosons are massive, and the gauge theory is spontaneously broken. Thus we can compute the low-energy matrix elements, which are finite and free of infrared divergences. This provides additional insight into the structure of gauge theory amplitudes. In the usual QCD analysis, the low energy matrix elements are infrared divergent, and infinite in perturbation theory.

Using the results in this paper, one can compute radiative corrections in the  effective theory to \emph{all} partonic hard scattering processes, and thus to any hard scattering process at the LHC with an arbitrary number of external particles, if the high scale matching (step (2) below) is known. The high-scale matching is suppressed by two powers of a large logarithm relative to the EFT corrections. In the examples we have studied, the high scale matching (including QCD) is much smaller (by a factor of four or more) than the electroweak EFT corrections, so one can profitably use our EFT results even for cases where the high scale matching is not known. The only exception is the cross-section for $q \bar q \to W^+_T W^-_T$, where the high-scale matching is almost 20\%. The radiative corrections are also large for this case, and exceed 20\% below 1~TeV.

\begin{figure}
\begin{center}
\includegraphics[bb=120 370 305 660,width=8cm]{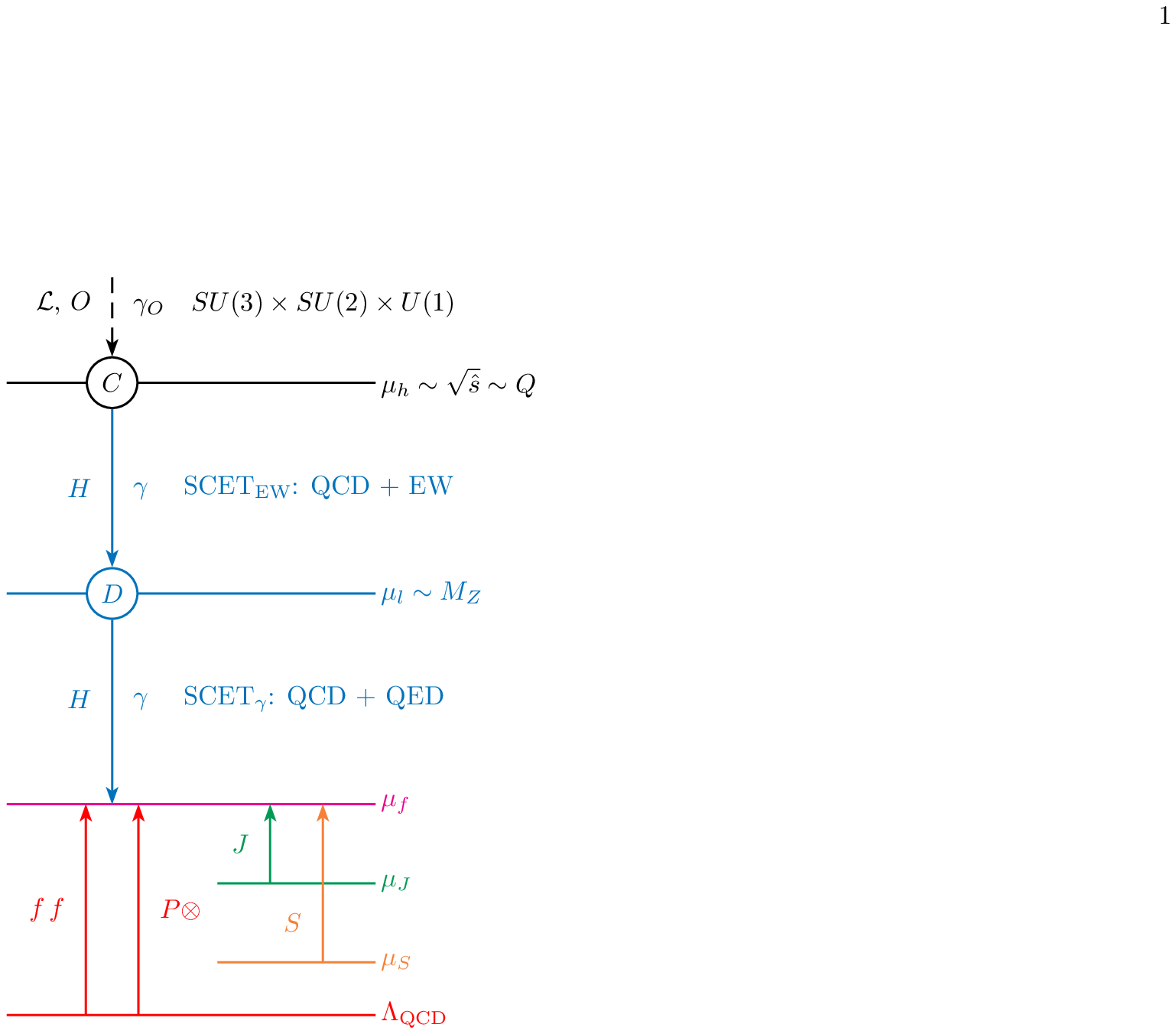} 
\end{center}
\caption{\label{fig:schematic} Steps in the computation of a hadronic cross-section at high energy. The effective theory can be used for everything except the hard matching correction $C$, and the running $\gamma_O$ in the full theory.}
\end{figure}

The computation of the complete hadronic cross-section can be done using the following steps, shown schematically in Fig.~\ref{fig:schematic}:\\
%%%----FIGURE--------------------------------------------------------------------------------------
%%
%------------------------------------------------------------------------------------------------------
(1) If the amplitude is the matrix element of a full theory operator, such as the Sudakov form factor for the scattering of a gluon by the operator $O(\mu_0)=G_{\mu\nu}^AG^{A\,\mu\nu}$, scale the operator using the standard model anomalous dimension $\gamma_O$ to the scale $\mu_h$ (high-scale matching). If the amplitude is a scattering amplitude computed using vertices in the standard model Lagrangian,  this step is not necessary, since the Lagrangian is $\mu$ independent. 

(2) Compute the hard matching\footnote{pun intended} coefficient $C$ at a scale $ \mu_h$. This is a standard example of a matching computation to an effective field theory. $C$ is computed from graphs in the full theory, setting all small scales such as the gauge boson and fermion masses to zero. The resulting graphs are single-scale graphs, and have ultraviolet and infrared divergences, which are regulated using dimensional regularization, leading to $1/\epsilon$ terms. The ultraviolet divergences are cancelled by renormalization counterterms, and the infrared divergences are reproduced by corresponding infrared divergences in the effective theory. Thus $C$ is given by keeping the finite part of the full theory diagrams, and throwing away all singular terms in $1/\epsilon$.\footnote{The above procedure only works if dimensional regularization is used to regulate both the ultraviolet and infrared divergences. A check on the matching computation is that the $1/\epsilon$ infrared divergent terms must agree with the anomalous dimension in the effective theory. See, for example, Refs.~\cite{amhqet,ameft} for a more extensive discussion.} The hard matching is independent of low scales such electroweak symmetry  breaking and fermion masses, and so can be computed in the unbroken theory using gluons, $W$ and $B$ gauge bosons, without worrying about electroweak gauge boson mixing, even for processes such as $Z$ production.  All mass effects are included  in the EFT computation in steps (4)--(6).

All scattering amplitudes are independent of the choice of $\mu_h$. However, in perturbation theory, there is residual $\mu_h$ dependence from higher order terms that are not included in the computation. $\mu_h$ is chosen to be of order the hard scale in the scattering problem so that there are no parametrically large logarithms in $C$, and all large logarithms are summed by renormalization group evolution. We will choose $\mu_h = \sqrt{\hat s}$, the partonic center of mass energy in the collision.

The high-scale matching for processes involving a small number of external partons is known and is included in the examples considered in Ref.~\cite{CKM} and in this paper. The high scale matching, in general, does not obey the factorization structure of the effective theory amplitude. This is the only piece of the computation which cannot be computed using the effective theory results.

(3) Run the amplitude from the high scale $\mu_h$ to a low scale $\mu_l$ of order $M_Z$ using the SCET anomalous dimension computed in \sceth. (\sceth\ is SCET with dynamical $SU(3) \times SU(2) \times U(1)$ gauge bosons). The SCET anomalous dimension is linear in $\log Q^2/\mu^2$ to all orders~\cite{dis}, and the renormalization group evolution sums the Sudakov double-logarithms. For $Q \sim 1$~TeV, $\LL^2 = \log^2 Q^2/M_Z^2 \sim 21$ so the Sudakov corrections can be very large, and lead to a suppression of the cross-section. It is well-known that the QCD Sudakov corrections can have a huge effect on the cross-sections. The electroweak Sudakov corrections are also significant, ranging from 10\%--50\% depending on the process at energies of a few TeV, increasing rapidly with energy. The Sudakov double logarithms which are summed by the LL series are two powers of $\LL$ more important than the high-scale matching in step (1). The anomalous dimension is independent of the electroweak scale, and like the high-scale matching, can be computed in the unbroken theory.

(4) Compute the low-scale matching at $\mu_l$ by integrating out the $W$, $Z$,  Higgs and $t$-quark, onto a theory containing only QCD and electromagnetism, which we call \scetl. The new feature found in Ref.~\cite{CGKM1,CGKM2,CKM} is that this low-scale matching in SCET contains a $\log Q^2/\mu_l^2$ term of order $\LL$. This term is related to the cusp anomalous dimension. No higher powers of a log appear to all orders in perturbation theory. For most processes, the low-scale matching correction is purely electroweak, since the gluon graphs are continuous through the electroweak threshold. The low-scale matching $D$ can be comparable to or slightly larger than the QCD contribution to  the high scale matching $C$ because of the $\log Q^2/\mu_l^2$ enhancement. There are low-scale QCD corrections for processes involving external $t$-quarks, and here the corrections are much larger, of order 35\%. The low-scale matching is one power of $\LL$ more important than the high-scale matching.

(5) Run the amplitude to $\mu_f$ (the factorization scale) using the anomalous dimension in \scetl. (\scetl is SCET with dynamical gluons and photons). This gives the parton level scattering amplitude at the scale $\mu_f$, i.e.\ the hard amplitude at $\mu=\mu_f$, denoted by $H$ in Fig.~\ref{fig:schematic}.

(6) The last step depends on the experimental observable being computed. To compute a hadronic cross-section, one would first square the parton scattering amplitude and multiply by the partonic phase space to get the hard partonic cross-section $\hat \sigma(\mu_f) \propto \abs{H(\mu_f)}^2$. An SCET amplitude with two incoming collinear fields and $n$ outgoing collinear fields can be used to compute the cross-section for $pp \rightarrow n$ jets. This is an inclusive cross-section at the level of partons, i.e.\ each jet can contain any number of collinear partons, but it is exclusive at the level of jets, i.e.\ one only includes exactly $n$ jets, and no additional hard radiation. For such exclusive jet rates, one in general can have parton distribution functions (PDFs) describing the parton distributions inside the proton, as well as jet and soft functions~\cite{css}. These must all be scaled to the common scale $\mu_f$, and then convoluted with the partonic cross-section $\hat \sigma(\mu_f)$,  to get the experimentally measurable hadronic cross-section. The $\mu_f$ dependence of $\hat \sigma(\mu_f)$ is cancelled by that in $f \otimes f \otimes J \otimes S$. This provides non-trivial constraints on the anomalous dimension, since the convolution running of $f$, $J$ and $S$ must be cancelled by the multiplicative running of $H$~\cite{top1,top2}. For complete consistency, the parton distribution functions need to be extracted including QED corrections. The QED running below $M_Z$ is very small, so one can simply drop this and match to PDFs obtained neglecting electromagnetic corrections.

The final step (6) above is common to all hadronic computations, and has been studied extensively. We have nothing new to say here. We therefore stop our computation at the end of (5). In the numerical results we present, we will choose $\mu_f=M_Z$.

Various software packages such as {\sc BlackHat}~\cite{blackhat}, {\sc Rocket}~\cite{rocket}  and {\sc CutTools}~\cite{cuttools} are being developed to streamline the computation of QCD radiative corrections to experimentally relevant cross-sections at the LHC. These calculations typically include the QCD effects discussed in (1)--(4) above, but not the electroweak radiative corrections. They also perform the difficult task of taking partonic cross-sections and computing infrared safe observables (such as jet rates) in proton scattering. The electroweak corrections are independent of the QCD corrections to one-loop, and the mixed QCD-electroweak terms from renormalization group running at two-loops are tiny. One can then incorporate our results in the existing methods by using the results in this paper with $\alpha_s \to 0$ as a multiplicative correction factor to the scattering amplitudes. If the partonic amplitude is needed at a scale $\mu_f > M_Z$, it can be obtained by using our results for $\alpha_s \to 0$ and $\mu_f=M_Z$, and then scaling back up to $\mu_f$ using the pure QCD expressions.

The outline of the paper is as follows. In Sec.~\ref{sec:plots}, we give plots of a few representative processes to illustrate our results. We consider $\bar u_L u_L \to \bar c_L c_L$ as an example of fermion scattering, $\bar u_L u_L \to W_T W_T, Z_T Z_T$ for transverse electroweak gauge boson production,   $\bar u_L u_L \to W_L W_L$ for longitudinal gauge boson production, $\bar u_L u_L \to H Z_L$ for associated Higgs boson production, and $gg \to W_T W_T$. The kinematics and notation is given in Sec.~\ref{sec:kinematics}, and the structure of the logarithmic series in Sec.~\ref{sec:log}. Sec.~\ref{sec:eft} discusses the EFT result that we need for our computations, which have been checked by explicit computation. There follows a long discussion in Sec.~\ref{sec:regulator}-Sec.~\ref{sec:hq} on the general form of the radiative corrections to all orders in perturbation theory, which can be skipped by readers not interested in technical details. The topics discussed are only relevant beyond three loops, and away from the $N \to \infty$ limit. Sec.~\ref{sec:regulator} discusses the $\Delta$-regulator, and Sec.~\ref{sec:fac} discusses the factorization structure of the amplitude. The general form of the scattering amplitude is given in Sec.~\ref{sec:scatamp}. Casimir scaling, the sum-on-pairs form, and $K$-factors are given in Sec.~\ref{sec:casimir}. The decomposition into collinear and soft functions is given in Sec.~\ref{sec:csfn}, and the extension to heavy quarks in Sec.~\ref{sec:hq}. The Goldstone boson equivalence theorem, and the one-loop radiative corrections, are summarized in Sec.~\ref{sec:equivthm}. Sec.~\ref{sec:results} gives the results for the collinear and soft functions in an $SU(2)$ gauge theory. The Magnea-Sterman result for the Sudakov form factor is given in Sec.~\ref{sec:sudakovff}, and conclusions in Sec.~\ref{sec:concl}.

\section{Plots}\label{sec:plots}

The numerical results for a few representative standard model computations are shown here. The details of how these plots were computed will be discussed later in this article. All the computations have been done using QCD plus electroweak one-loop matching at $\mu_h$ and $\mu_l$, and renormalization group evolution using the SCET QCD plus electroweak non-cusp anomalous dimension to two loops, and the QCD plus electroweak cusp anomalous dimension to three loops. The electroweak three-loop cusp anomalous dimension is only known neglecting the Higgs scalar contribution.

The results have a very small dependence on the Higgs mass. All the numerical results are for $m_H=200$~GeV. Changing $m_H$ to 500~GeV changes the rates by less than one part in $10^4$.

The full standard model is matched onto \sceth\ at $\mu_h$, and the electroweak gauge bosons, Higgs, and top quark are integrated out at $\mu_l$. We will choose $\mu_h=\sqrt{\hat s}$ and $\mu_l=M_Z$ in computing scattering rates. It is convenient to integrate out the electroweak gauge bosons, Higgs, and top quark at a common scale, so that one does not have to treat an effective theory with electroweak  gauge bosons but broken gauge symmetry.

\begin{figure}
\begin{center}
\includegraphics[bb=40 152 473 713,width=7cm]{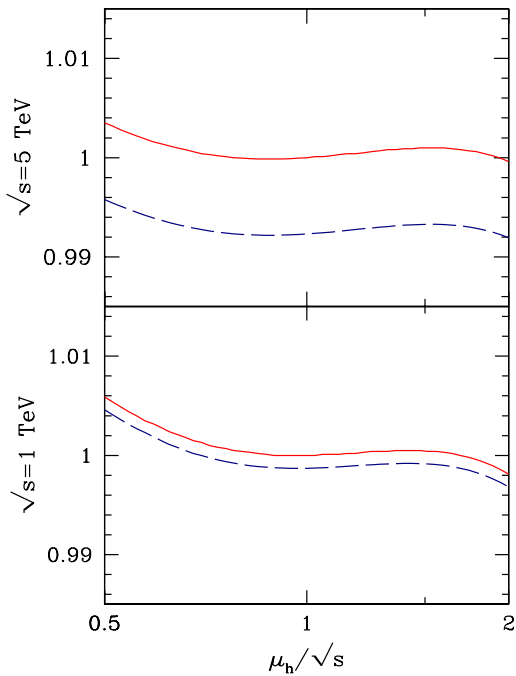} 
\end{center}
\caption{\label{fig:varymu} Variation of the $\bar u_L u_L \to \bar c_L c_L$ cross-section as a function of $\mu_h$ for $\mu_l=M_Z$ (solid red) and $\mu_l=m_t$ (dashed blue), for $\sqrt{s}=1$~TeV (lower panel) and $\sqrt{s}=5$~TeV (upper panel). The cross-section has been normalized to that at $\mu_h=\sqrt{s}$ and $\mu_l=M_Z$.}
\end{figure}

\begin{figure}
\begin{center}
\includegraphics[bb=40 152 473 713,width=7cm]{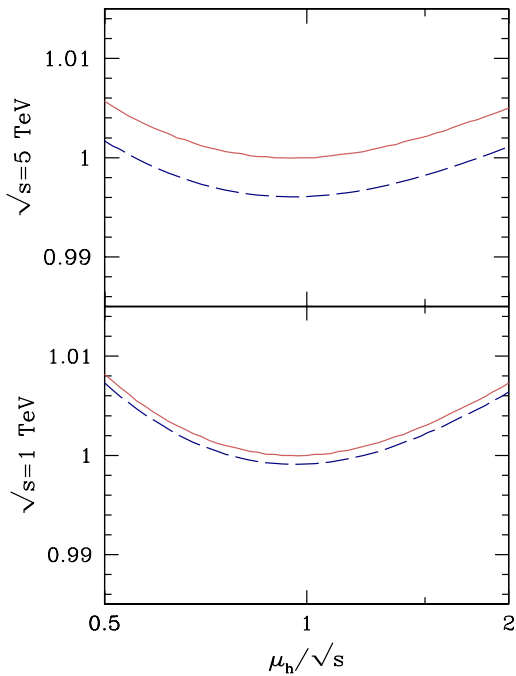} 
\end{center}
\caption{\label{fig:varymuZH} Variation of the $\bar u_L u_L \to Z_L H$ cross-section. See Fig.~\ref{fig:varymu} caption.}
\end{figure}

The $\mu_{h,l}$ dependence is very small in almost all the processes we have studied. Fig.~\ref{fig:varymu} shows the variation in the $\bar u_L u_L \to \bar c_L c_L$ cross-section as these scales are varied. $\mu_h$ is varied between $\sqrt{\hat s}/2$ and $2\sqrt{\hat s}$, and $\mu_l$ is varied between the two natural choices, $M_Z$ and $m_t$, which is approximately a factor of two variation. A similar plot of the variation for $\bar u_L u_L \to Z_L H$ is shown in Fig.~\ref{fig:varymuZH}.%
%%%---FIGURE--------------------------------------------------------------------------------------
%%
%%%---FIGURE--------------------------------------------------------------------------------------
%%
%%%---FIGURE--------------------------------------------------------------------------------------
\begin{figure}
\begin{center}
\includegraphics[bb=40 152 473 713,width=7cm]{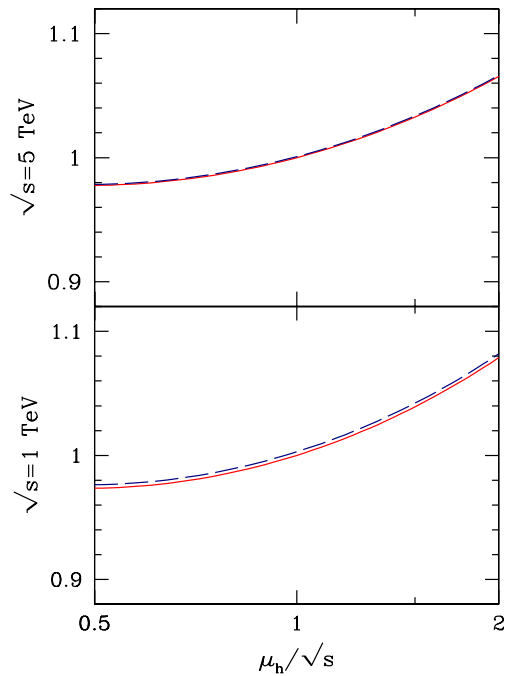} 
\end{center}
\caption{\label{fig:varymuWW} Variation of the $\bar u_L u_L \to W^+_T W^-_T$ cross-section. See Fig.~\ref{fig:varymu} caption. Note the tenfold change in vertical scale.}
\end{figure}%
The only example with significant $\mu_h$ variation is transverse gauge boson production. Fig.~\ref{fig:varymuWW} shows the $\mu_h$ variation for $\bar u_L u_L \to W^+_T W^-_T$. The $\mu_l$ dependence is still very small, but the $\mu_h$ variation is nearly 10\%. This is being investigated further. The high-scale matching is large for this process, almost 20\%. We have checked that the $\mu_h$ dependence of the high-scale matching agrees with the EFT anomalous dimension.

The scale dependence is below 1\% in all cases except $W_T$ production, so $\mu_h$ and $\mu_l$ will be held fixed in subsequent plots. The main contribution to the scale dependence is from the two-loop cusp anomalous dimension in the running, which is not cancelled by logarithms in the matching, since that has only been computed to one-loop. The scale-dependence would be even smaller if only the one-loop SCET anomalous dimension was used. The $\mu_l$ dependence increases with energy because of the $\log Q/\mu_l$ term in the low-scale matching. The total radiative correction increases even faster.

\begin{figure}
\begin{center}
\includegraphics[bb=52 152 473 713,width=8.5cm]{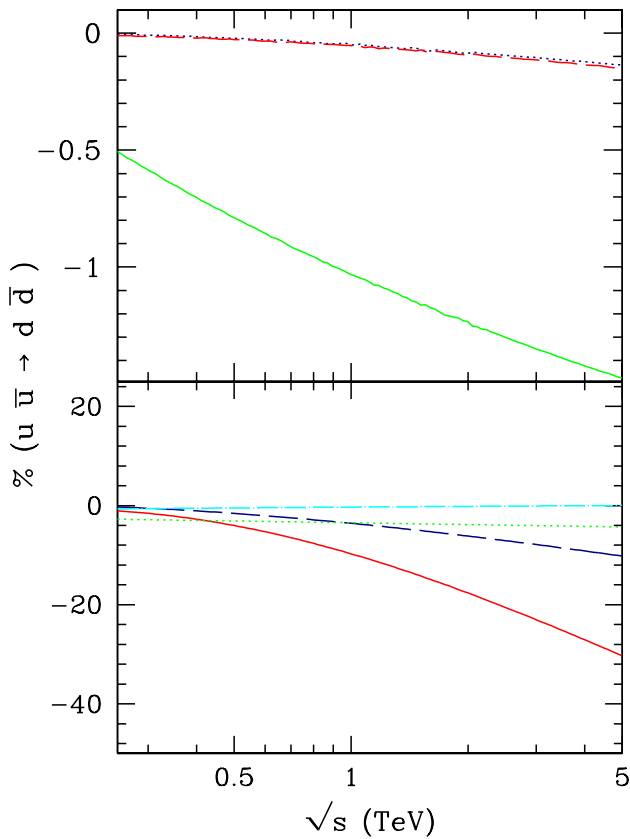} 
\end{center}
\caption{\label{fig:UUDDSize} Different contributions to the $\bar u_L u_L \to \bar c_L c_L$ cross-sections as a percentage of the total rate. Lower panel: electroweak corrections (solid red), two-loop QCD cusp anomalous dimension (dashed blue), low-scale matching (dotted green), and the high-scale matching (dot-dashed cyan). Upper panel: two-loop non-cusp QCD anomalous dimension (solid green), three-loop QCD cusp anomalous dimension (dotted blue), and two-loop electroweak cusp anomalous dimension (dashed red).}
\end{figure}

\begin{figure}
\begin{center}
\includegraphics[bb=52 152 473 713,width=8.5cm]{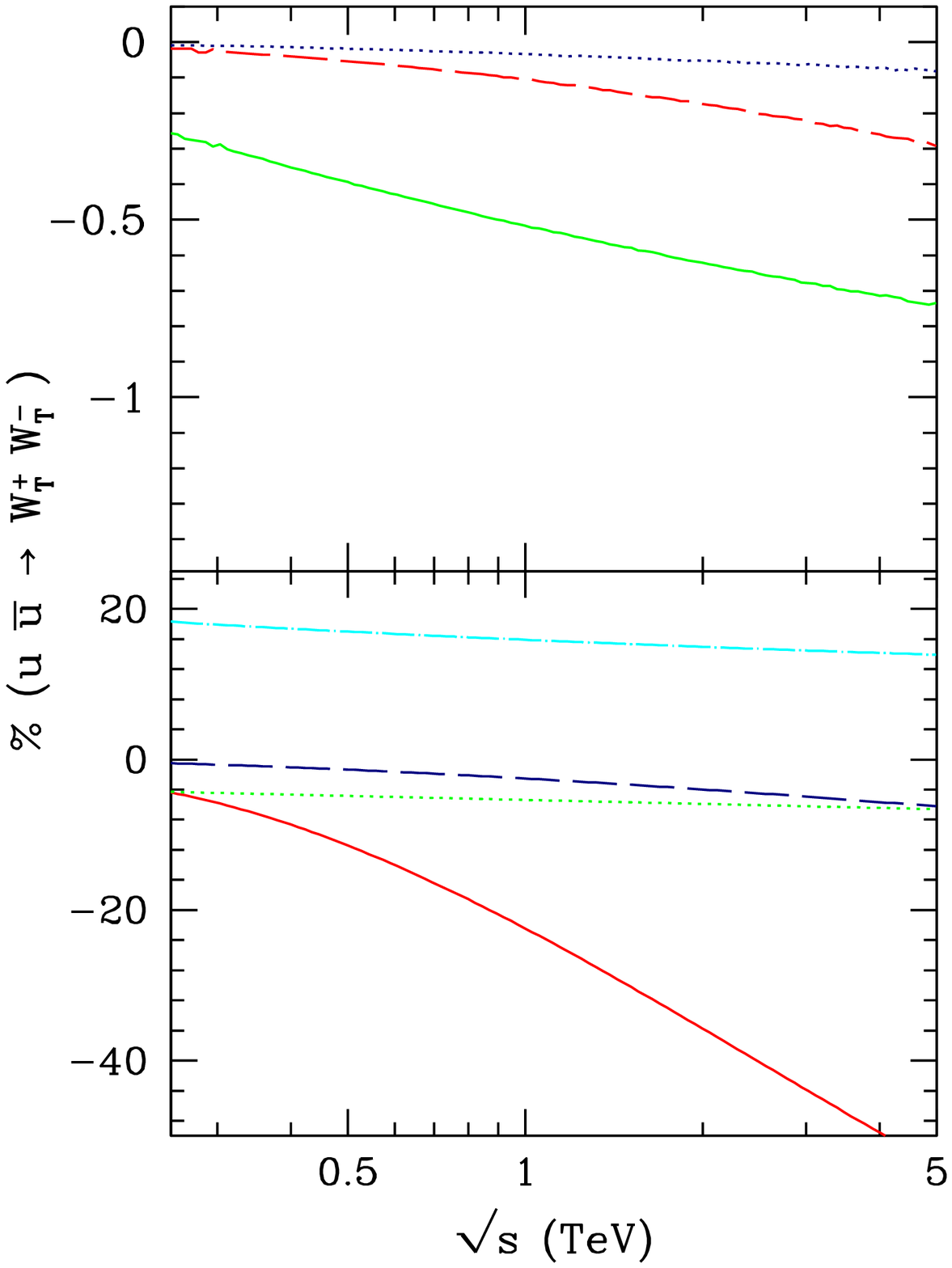} 
\end{center}
\caption{\label{fig:WWSize} Different contributions to the $\bar u_L u_L \to W^+_T W^-_T$ cross-sections as a percentage of the total rate. See
Fig.~\ref{fig:UUDDSize} caption.}
\end{figure}

\begin{figure}
\begin{center}
\includegraphics[bb=52 152 473 713,width=8.5cm]{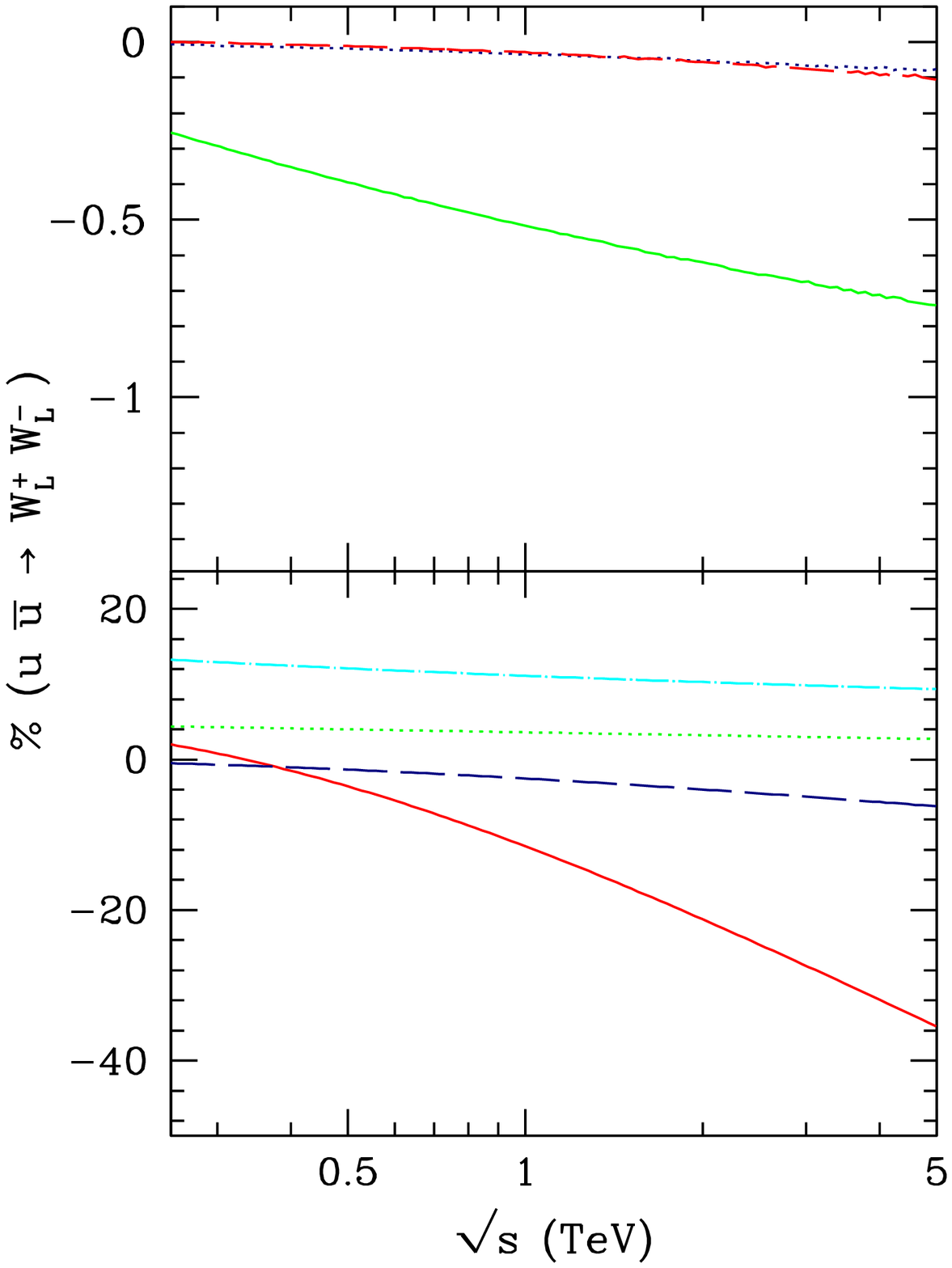} 
\end{center}
\caption{\label{fig:WLSize} Different contributions to the $\bar u_L u_L \to W_L W_L$ cross-sections as a percentage of the total rate. See
Fig.~\ref{fig:UUDDSize} caption.}
\end{figure}

\begin{figure}
\begin{center}
\includegraphics[bb=52 152 473 713,width=8.5cm]{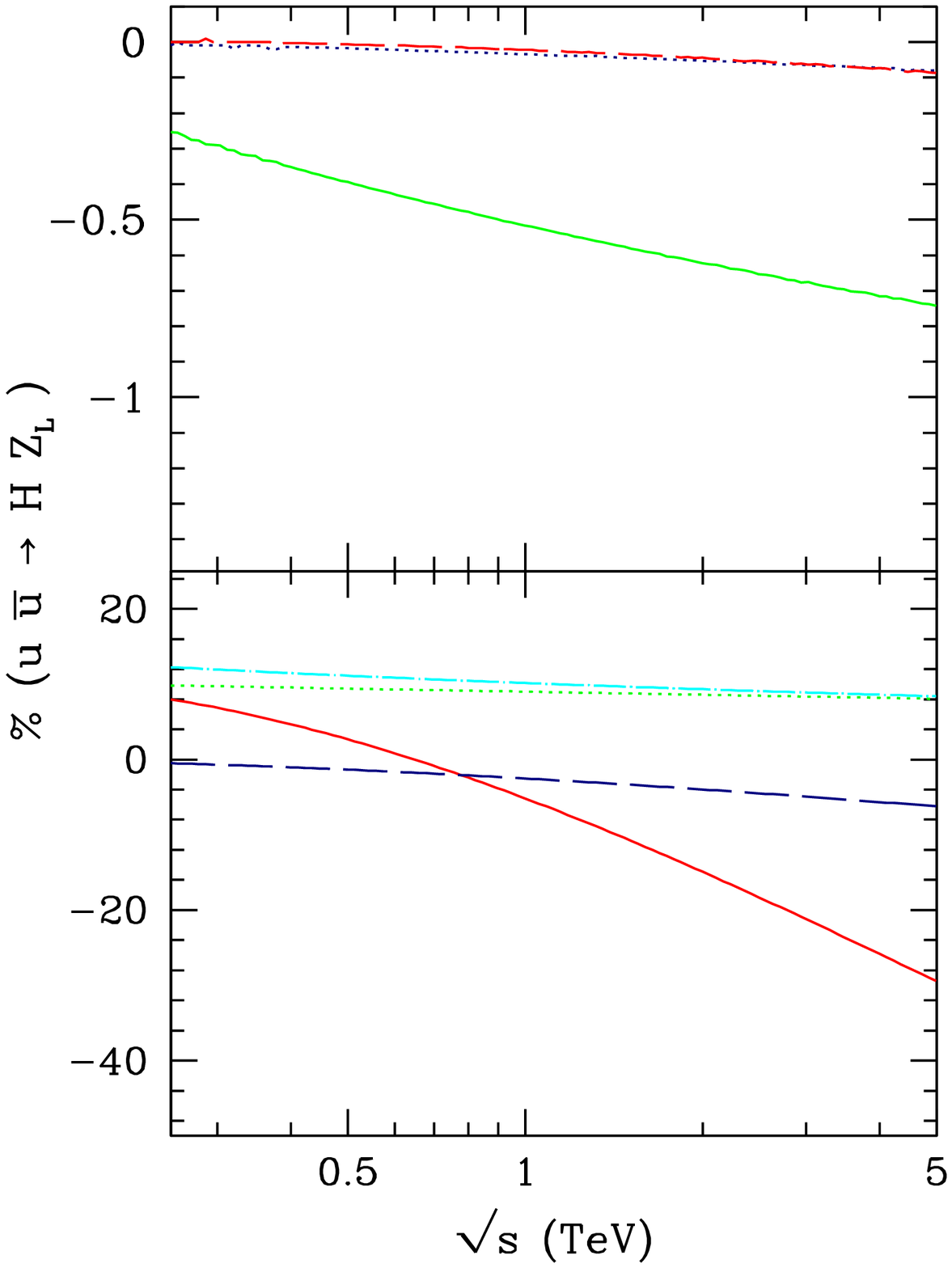} 
\end{center}
\caption{\label{fig:ZHSize} Different contributions to the $\bar u_L u_L \to H Z_L $ cross-sections as a percentage of the total rate. See
Fig.~\ref{fig:UUDDSize} caption.}
\end{figure}

The relative size of different contributions to the radiative corrections, such as the two and three-loop cusp and non-cusp anomalous dimensions, high-scale and low-scale matching, are shown for $\bar u_L u_L \to \bar c_L c_L$ in Fig.~\ref{fig:UUDDSize}, $\bar u_L u_L \to W^+_T W^-_T$ in Fig.~\ref{fig:WWSize}, $\bar u_L u_L \to W_L W_L$ in Fig.~\ref{fig:WLSize} and $\bar u_L u_L \to H Z_L $ in Fig.~\ref{fig:ZHSize}. We have determined the size of the electroweak corrections by turning them on and off in the effective theory, while keeping the full one-loop matching. 
%%%---FIGURE--------------------------------------------------------------------------------------
%%%
%%%%---FIGURE--------------------------------------------------------------------------------------
%%%
%%%%---FIGURE--------------------------------------------------------------------------------------
%%%
%%%%---FIGURE--------------------------------------------------------------------------------------

In processes such as $q \bar q \to q \bar q$, the tree-level matching has terms of order $\alpha_i$, $i=1,2,3$ from $B$, $W$ and gluon exchange, and the one-loop matching has terms of order $\alpha_i \alpha_j$ from box graphs. One has to decide whether to treat the $\alpha_{1,2}$ tree-level matching as an electroweak correction to the $\alpha_3$ term, and whether $\alpha_{1,2} \alpha_3$ terms in the one-loop matching should be treated as electroweak corrections to the tree-level QCD matching, or QCD corrections to the tree-level electroweak matching. To avoid this, we have used the full one-loop matching in all comparisons. The EFT graphs are one-loop diagrams, so it is clear whether to call them QCD or electroweak corrections.

The electroweak radiative corrections are substantial, and increase rapidly with energy. For transverse $W$ production, the electroweak corrections exceed 20\% at $\sqrt{\hat s}=1$~TeV. The corrections for transversely polarized gauge bosons are larger than those for longitudinally polarized gauge bosons and the Higgs. At high energies, longitudinally polarized gauge bosons and Higgs bosons behave like members of the scalar doublet, and their electroweak radiative corrections depend on the $SU(2)$ Casimir of the fundamental representation $C_F=3/4$, whereas transverse gauge bosons depend on the Casimir of the adjoint $C_A=2$, which is larger. The high-scale matching for transverse $W$ production is almost 20\%, and both QCD and electroweak matching is important. The large  $\mu_h$ dependence is an indication that the two-loop high-scale matching might be important for this problem.  

The electroweak corrections are more important than the two-loop QCD cusp anomalous dimension.\footnote{Note that $\alpha_s/\alpha_2 \sim 2.7$ at 1~TeV and $\sim 2.4$ at 5~TeV.} The only corrections that are necessary for sub-1\% accuracy are the one-loop QCD and electroweak corrections, the one-loop matching, and the two-loop QCD anomalous dimension. The two-loop electroweak and three-loop QCD corrections are completely negligible; even the two-loop QCD non-cusp anomalous dimension barely exceeds one percent in a few cases. The one-loop high scale matching can be as large as 10\%.  Higgs corrections proportional to the top-quark Yukawa coupling are a few percent. These arise even for processes not involving top quarks via virtual  top quark  loops in the anomalous dimensions for the longitudinal gauge bosons and Higgs (see Ref.~\cite{p2}).

%%%---FIGURE--------------------------------------------------------------------------------------
\begin{figure}
\begin{center}
\includegraphics[bb=60 152 473 713,width=8.5cm]{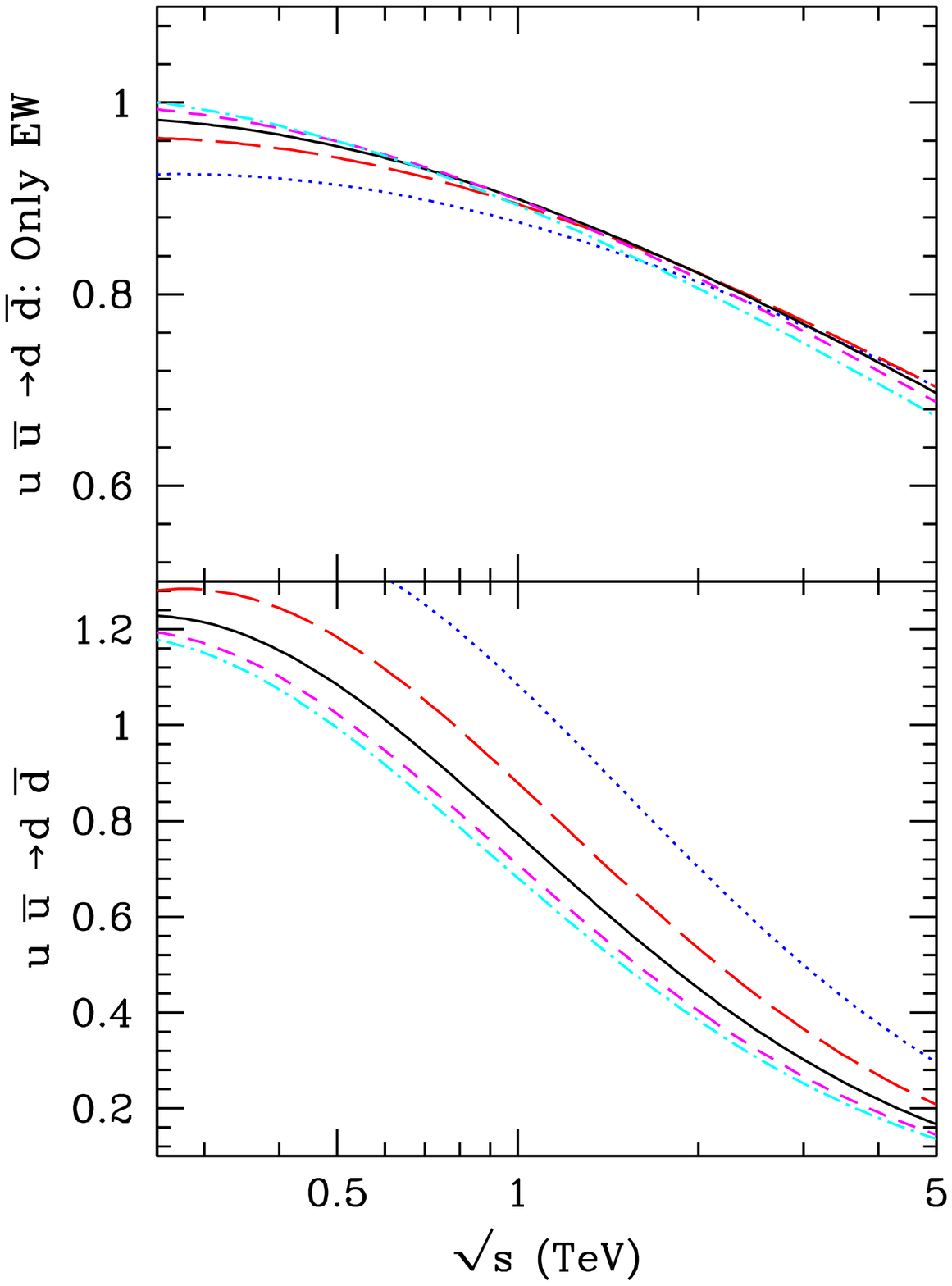} 
\end{center}
\caption{\label{fig:UUDDrate}Plot of the $\bar u_L  u_L \to \bar c_L c_L$ cross-section normalized to the tree-level rate, for $t/s=-0.2$ (dotted blue), $-0.35$ (dashed red), $-0.5$ (solid black), $-0.65$ (dashed magenta) and $-0.8$ (dot-dashed cyan). The lower panel shows the rate with QCD and electroweak corrections, and the upper panel with only the electroweak corrections.  }
\end{figure}
%%
%%%---FIGURE--------------------------------------------------------------------------------------
\begin{figure}
\begin{center}
\includegraphics[bb=60 152 473 713,width=8.5cm]{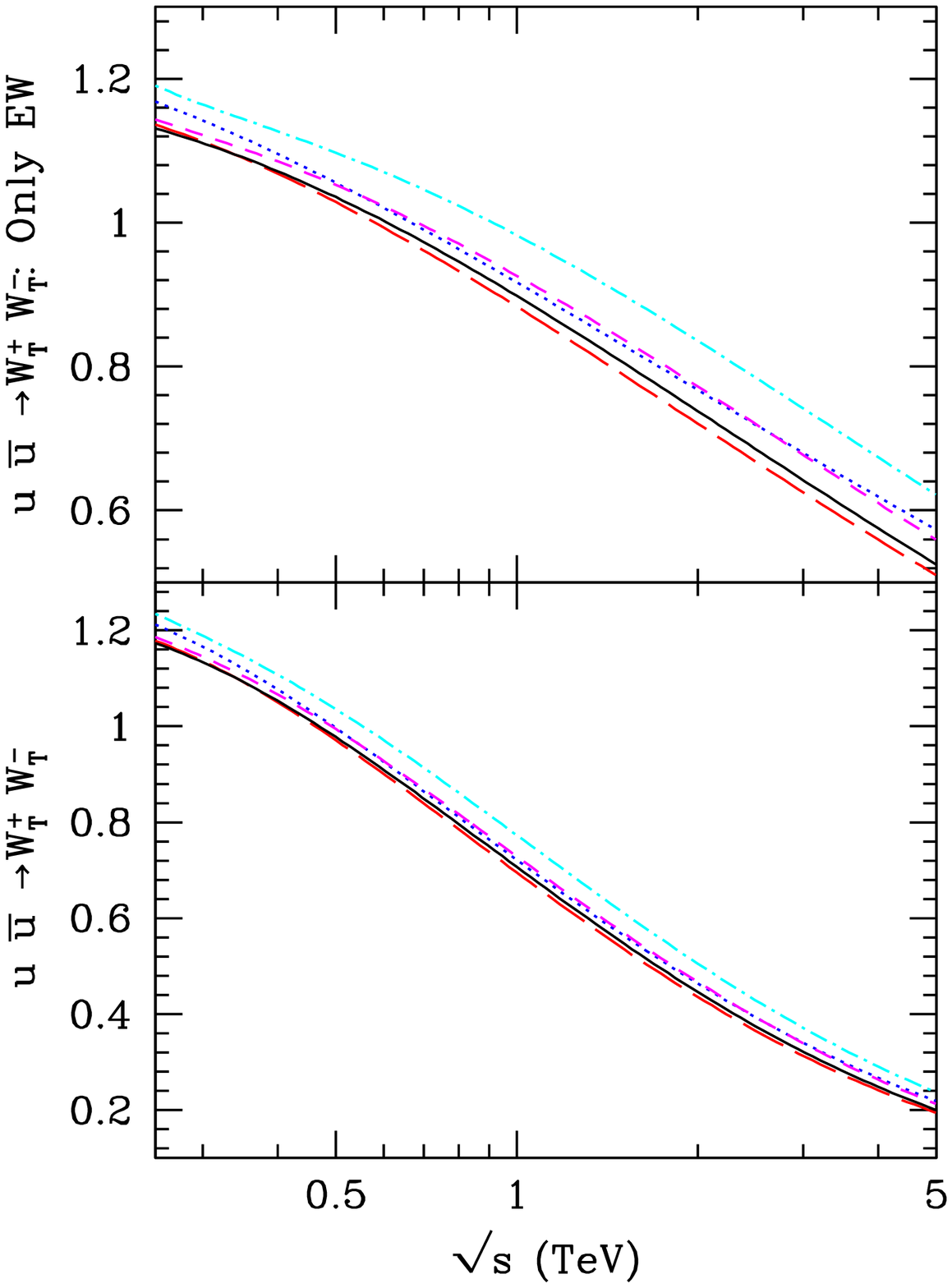} 
\end{center}
\caption{\label{fig:UUWWrate}Plot of the $\bar u_L  u_L \to W^+_T W^-_T$ cross-section normalized to the tree-level rate. See Fig.~\ref{fig:UUDDrate} caption.  }
\end{figure}
%%
%%%---FIGURE--------------------------------------------------------------------------------------
\begin{figure}
\begin{center}
\includegraphics[bb=60 152 473 713,width=8.5cm]{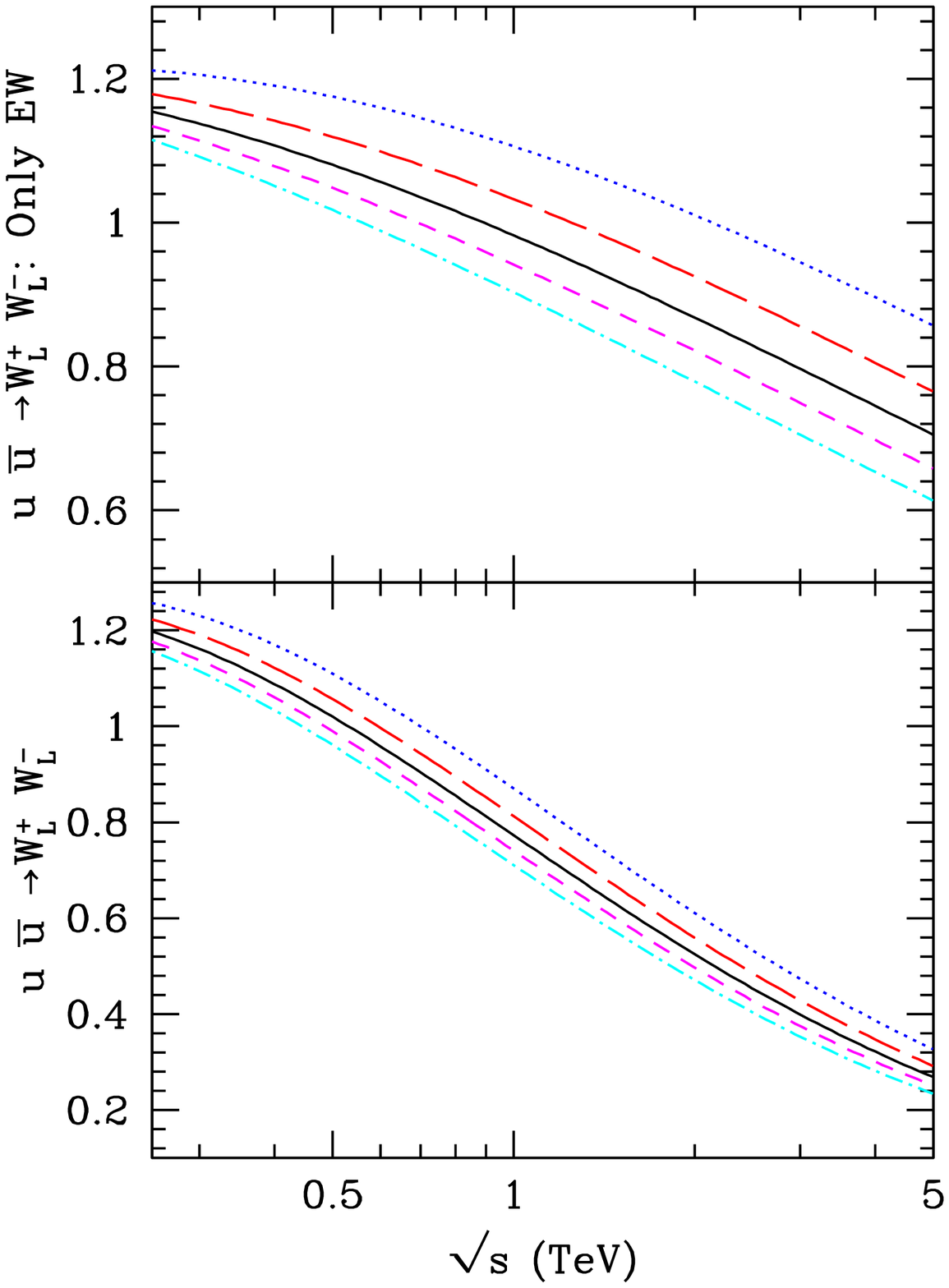} 
\end{center}
\caption{\label{fig:UUPhirate}Plot of the $\bar u_L  u_L \to W^+_L W^-_L$ cross-section normalized to the tree-level rate. See Fig.~\ref{fig:UUDDrate} caption.  }
\end{figure}
%%
%%%---FIGURE--------------------------------------------------------------------------------------
\begin{figure}
\begin{center}
\includegraphics[bb=60 152 473 713,width=8.5cm]{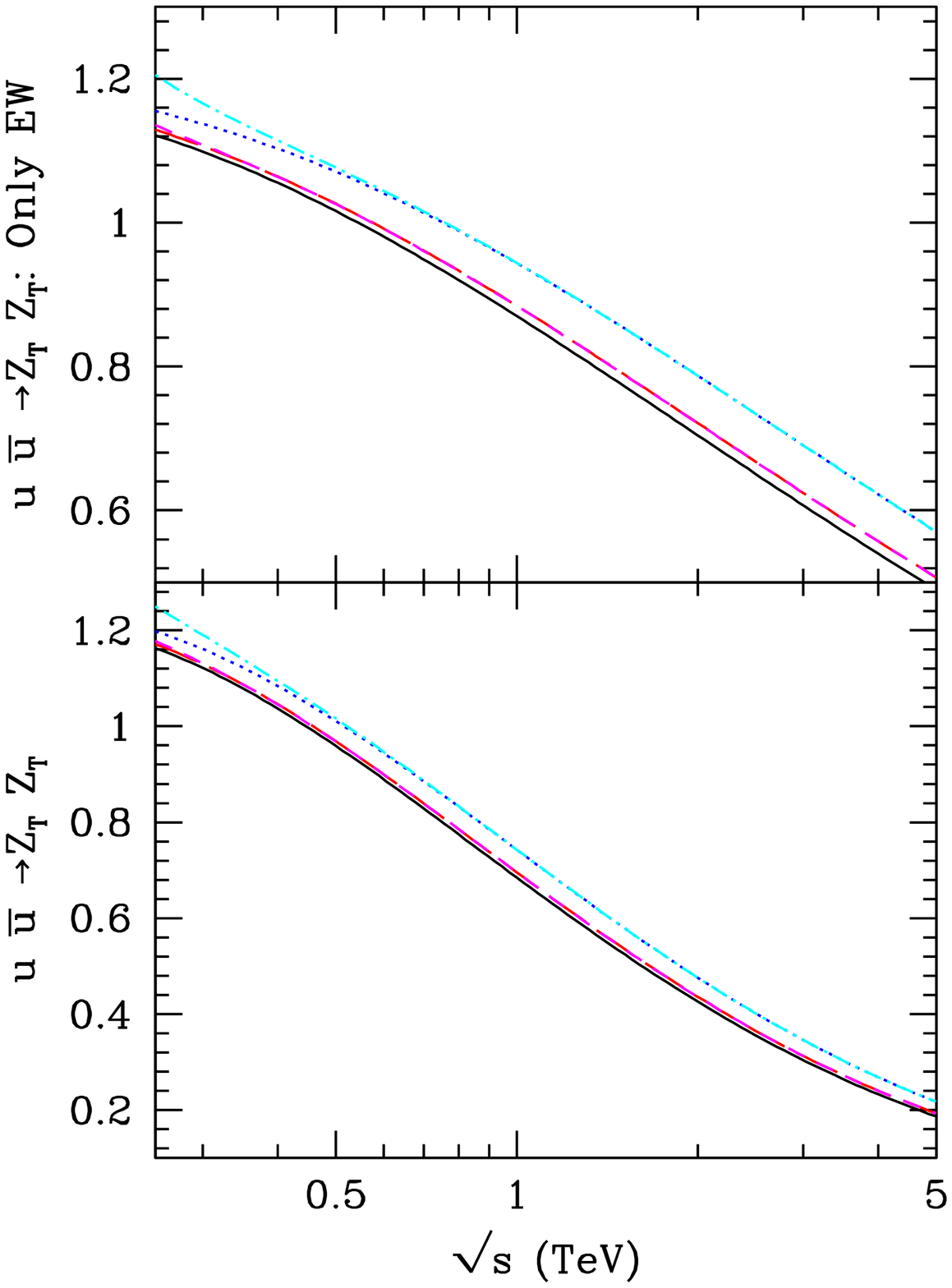} 
\end{center}
\caption{\label{fig:UUZZrate}Plot of the $\bar u_L  u_L \to Z_T Z_T$ cross-section normalized to the tree-level rate. See Fig.~\ref{fig:UUDDrate} caption. }
\end{figure}
%%
%%%---FIGURE--------------------------------------------------------------------------------------
\begin{figure}
\begin{center}
\includegraphics[bb=60 152 473 713,width=8.5cm]{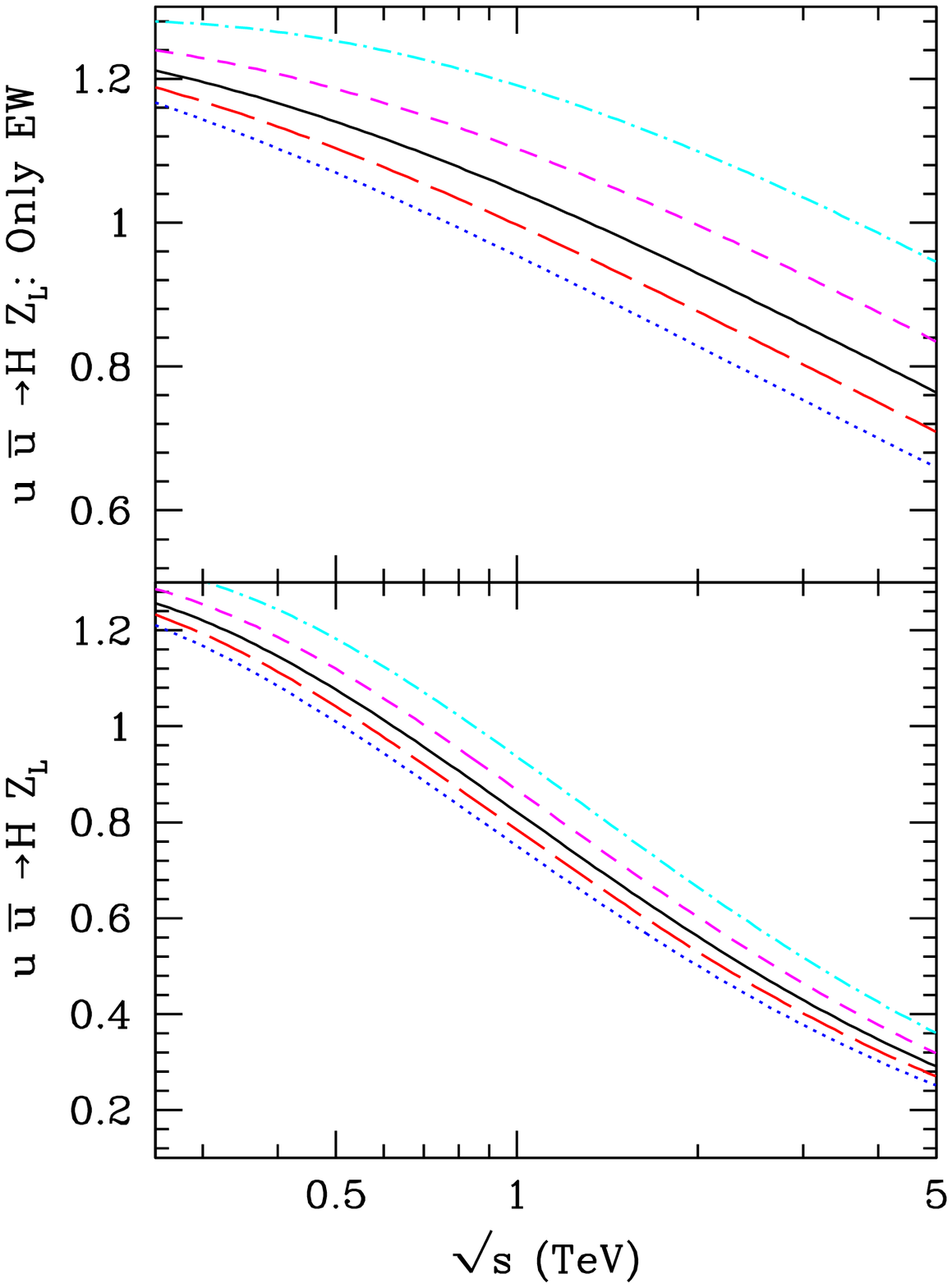} 
\end{center}
\caption{\label{fig:UUZHrate}Plot of the $\bar u_L  u_L \to H Z_L$ cross-section normalized to the tree-level rate. See Fig.~\ref{fig:UUDDrate} caption.  }
\end{figure}
%%
%%%---FIGURE--------------------------------------------------------------------------------------
\begin{figure}
\begin{center}
\includegraphics[bb=60 152 473 713,width=8.5cm]{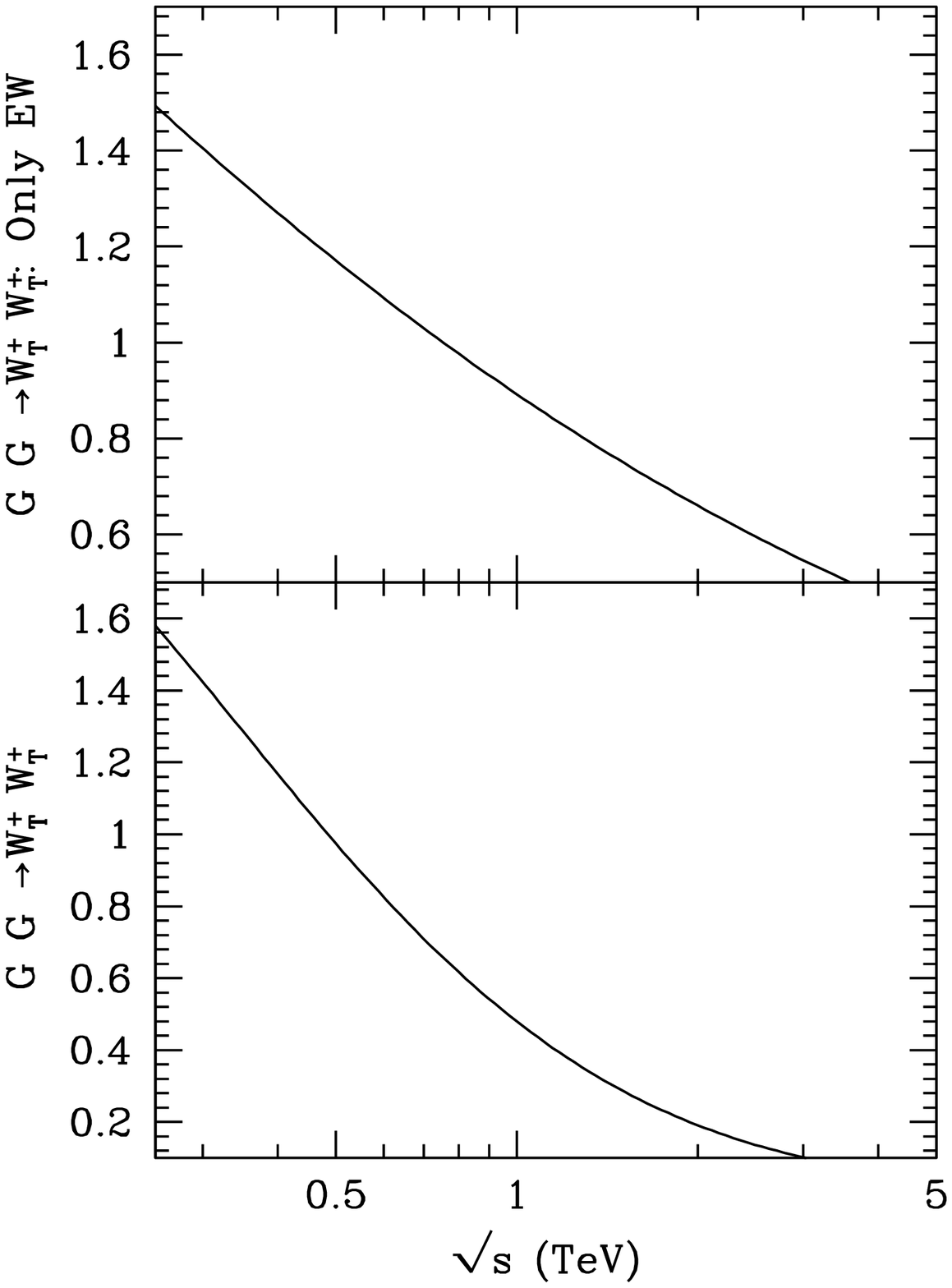} 
\end{center}
\caption{\label{fig:GGWWrate} Plot of the radiative corrections to the $GG \to W^+_T W^-_T$ cross-section. The lower panel shows the rate with QCD and electroweak corrections, and the upper panel with only the electroweak corrections. The radiative corrections are independent of $\hat t$.  }
\end{figure}

Finally, in Figs.~\ref{fig:UUDDrate}--\ref{fig:GGWWrate},  we show the complete radiative corrections to the scattering rate for some representative processes involving quarks, transverse and longitudinal gauge bosons, and the Higgs, normalized to the tree-level rate. At high energies, the radiative corrections suppress the scattering rate, but at lower energies, the radiative corrections lead to an enhancement. The rates are shown as a function of $\sqrt{\hat s}$ for different values of $\hat t$. The lower figure shows the total correction to the rate, whereas the upper panel shows only the electroweak contribution. The numerical values of the electroweak contribution are different from the previous plots, because they have been normalized to the tree-level rate. The previous plots computed the electroweak contribution using the ratio of the total rate including electroweak corrections, to the total rate without electroweak corrections.

Fig.~\ref{fig:GGWWrate} shows the radiative corrections to $g g \to W_T W_T$, which is a background for the Higgs search at the LHC. This process first occurs at one-loop in the standard model. We have not included the radiative corrections to the high scale matching $C$, which would involve computing two-loop diagrams, but all the other radiative corrections have been included using the effective theory.

\section{Kinematics and Notation}\label{sec:kinematics}

We will compute amplitudes for an arbitrary scattering process with $r$ external  legs in the high energy limit. The large energy scale will be denoted by $Q$, and can be chosen to be the center of mass energy in the collision. The effective theory computation is valid in the limit that the kinematic invariants $p_i \cdot p_j$ are of order $Q^2$, and invariant masses are small, $m_i \ll Q$, $M_{W,Z}\ll Q$, where $m_i$ is the mass of particle $i$. The resulting amplitudes can be used, for example, to compute the rate for $W + \text{jets}$ as long as the jet invariant masses are small compared with $Q$. The SCET computation neglects power corrections of the form $M_{W,Z}^2/Q^2$ and $m_i^2/Q^2$ in the one-loop and higher radiative corrections, and these are estimated in Sec.~\ref{sec:power}. The full dependence on mass ratios such as $m_t^2/M_{W,Z}^2, M_H^2/M_{W,Z}^2$, which are order unity in the power counting, is included in the EFT computation.

It is convenient to introduce unit three-vectors $\mathbf{n}_i$,  $\abs{\mathbf{n}_i}=1$, which point along the direction of motion of the scattering particles. For incoming particles $i$, define the null vectors $n_i=(1,\mathbf{n}_i)$ and $\bn_i =(1,-\mathbf{n}_i)$, with $n_i \cdot \bn_i=2$.  For outgoing particles $j$, $n_j$ and $\bn_j$ are defined by $n_j=-(1,\mathbf{n}_j)$ and $\bn_j =-(1,-\mathbf{n}_j)$.

In SCET, fast moving particles in the $n_i$ direction are referred to as $n_i$-collinear particles, and are described by fields $\xi_{n_i}$. A four-vector $p$ can be decomposed as
\begin{eqnarray}
p^\mu &=& \frac12 \left(\bn_i \cdot p\right) n^\mu_i + \frac12 \left(n_i \cdot p \right) 
\bn^\mu_i
+ \mathbf{p}^\mu_\perp\,.
\label{pdef}
\end{eqnarray}
If $p$ is the momentum of an $n_i$ collinear particle, and $p_i^2=\lambda^2 Q^2$, $\lambda\ll1$, then $\bar n \cdot p \sim \mathcal{O}(Q)$, $n \cdot p \sim \mathcal{O}(\lambda^2 Q)$, $\mathbf{p}_\perp \sim \mathcal{O}(\lambda Q)$. If $n_i$ is chosen to point precisely along the direction of $p_i$, so that $\mathbf{p}_\perp=0$, then $\bar n_i \cdot p_i = E_i + \abs{\mathbf{p}_i} \sim 2 E_i$ and $n_i \cdot p_i =  E_i - \abs{\mathbf{p}_i} \sim m_i^2/(2 E_i)$.

In an $r$-leg scattering amplitude, all momenta $p_i$  are chosen to be incoming, so that for an outgoing particle $j$, $p_j$ is the negative of the momentum of the particle. $\bar n_j \cdot p_j \approx 2E_j >0$ is still positive for outgoing particles, because of our definition of $n_j$ for outgoing particles.

A massive particle such as the top quark is described by  a velocity four-vector $v^\mu$, with $v \cdot v=1$, where $v^\mu = \gamma\left(1,\bm{\beta}\right)$, $\gamma=1/\sqrt{1-\bm{\beta}\cdot\bm{\beta}}$. For energetic top-quarks, it is sometimes convenient to use the four-vector $\beta^\mu=(1,\bm{\beta})$, with $\beta^2=1/\gamma^2 \to 0$ in the high-energy limit. This allows for a smooth transition in the high energy limit to a massless description, with $\beta \to n$.

\begin{figure}
\begin{center}
\includegraphics[width=6cm]{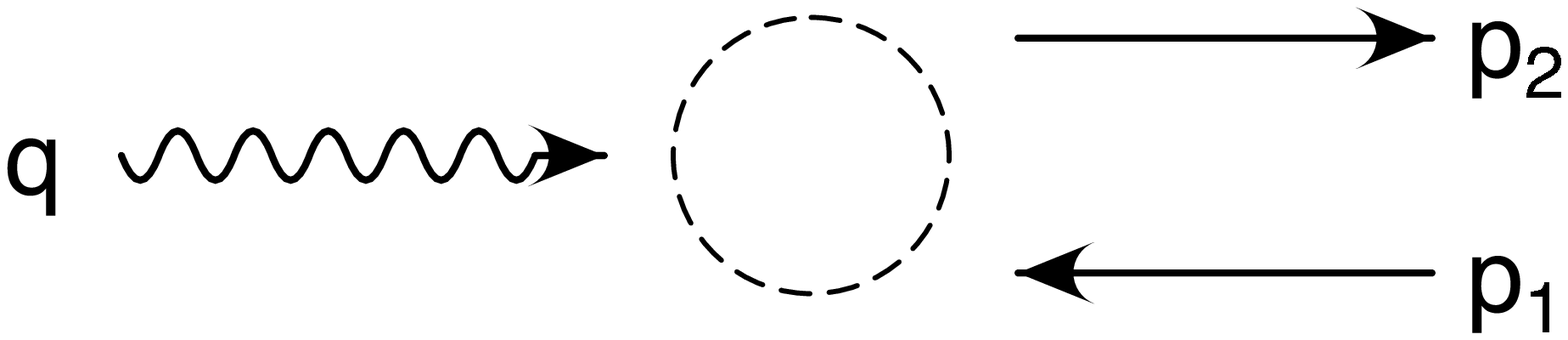} 
\end{center}
\caption{\label{fig:breit} The Breit frame, where the incoming particle is 
backscattered by an external current.}
\end{figure}

The Sudakov form factor will play an important role in this paper. The spacelike Sudakov form factor $F(Q^2)$   is defined as the particle scattering amplitude by an external current, with momentum transfer $Q^2=-q^2 >0$. It is convenient to compute the form factor in the Breit frame (see Fig.~\ref{fig:breit}), where the particle is back-scattered, and the momentum transfer $q$ has $q^0=0$.
%%%----FIGURE--------------------------------------------------------------------------------------
%%
%------------------------------------------------------------------------------------------------------
The Sudakov form factor is an $r=2$ scattering amplitude, where the incoming and outgoing particle are identical. In the Breit frame, $n_1=(1,\mathbf{n})$, $n_2=-(1,-\mathbf{n})$ so that $\bar n_1=-n_2$ and $\bar n_2=-n_1$.

The labelling convention chosen for the Higgs doublet is
\begin{eqnarray}
\phi &=& \frac{1}{\sqrt 2}\left[ \begin{array}{c}
\varphi^2 + i \varphi^1 \\
v + H - i \varphi_3 \end{array} \right]\,,
\end{eqnarray}
so that $\varphi^a \propto iT^a \vev{\phi}$. The charged gauge and Goldstone bosons are
\begin{eqnarray}
W^\pm &=& \frac1{\sqrt2}\left(W^1 \mp iW^2 \right)\,,\nn
\varphi^\pm &=& \frac1{\sqrt2}\left(\varphi^1 \mp i\varphi^2 \right)\,,
\end{eqnarray}
and the sign convention for the $Z$ and photon fields is
\begin{eqnarray}
Z &=& \cos \theta_W W^3-\sin\theta_W B\,,\nn
A &=& \sin \theta_W W^3+\cos\theta_W B\,.
\end{eqnarray}
The $SU(2)$ and $U(1)$ fine structure constants are $\alpha_2$ and $\alpha_1$ respectively,
and the QED fine structure constant is $\aem$, with
\begin{eqnarray}
\frac{1}{\aem} &=& \frac{1}{\alpha_2}+\frac{1}{\alpha_1}\,.
\label{6}
\end{eqnarray}
We will use the abbreviations
\begin{eqnarray}
 \alpha_W &=& \frac{\aem}{\sin^2 \theta_W}=\alpha_2\,,\nn
\alpha_Z &=& \frac{\aem}{\sin^2 \theta_W \cos^2 \theta_W}\,,\nn
c_W &=& \cos \theta_W\,,\nn
s_W &=& \sin \theta_W\,,\nn
\lM &=& \log \frac{M^2}{\mu^2}\,,\nn
\lQ &=& \log \frac{Q^2}{\mu^2}\,,\nn
\lQM &=& \log \frac{Q^2}{M^2}\,.
\label{logdef}
\end{eqnarray}

All couplings are in the $\overline{\text{MS}}$ scheme so that Eqs.~(\ref{6},\ref{logdef}) provide a definition of $\sin^2\theta_W(\mu)$ in terms of $\alpha_{1,2}(\mu)$.

A generic gauge coupling is denoted by $\alpha$, and the electromagnetic coupling constant is $\aem$.

Anomalous dimensions, form-factors and matching coefficients will be expanded in powers of $\alpha/(4\pi)$,
\begin{eqnarray}
\gamma &=& \gamma^{(1)} \frac{\alpha}{4\pi}+\gamma^{(2)} \left(\frac{\alpha}{4\pi}\right)+\ldots\,.
\end{eqnarray}

\section{Structure of the Logarithms}\label{sec:log}

The schematic structure of the radiative corrections is
\begin{eqnarray}
\amp &=& \left( \begin{array}{cccccccccc} 1 \\[5pt]
\alpha \LL^2 & \alpha \LL & \alpha  \\[5pt]
\alpha^2 \LL^4 & \alpha^2 \LL^3 & \alpha^2 \LL^2 & \alpha^2 \LL & \alpha^2  \\[5pt]
\alpha^3 \LL^6 & \multicolumn{4}{c}{\ldots} \\[5pt]
\vdots
\end{array}\right)
\label{x1}
\end{eqnarray}
where the first row is the tree-level result, the second row gives the one-loop corrections, etc., and all the terms are added to give the total amplitude. The standard terminology used in the literature is to call the first row leading order, the second row NLO, the third row NNLO, etc.\ and refer to the $\alpha^n \LL^{2n}$ term as LL, the $\alpha^n \LL^{2n-1}$ term as NLL, etc. Thus the $\alpha^2 \LL^2$ term is referred to as the NNLL term at NNLO.

In the leading-log regime, where $\alpha \LL \sim 1$, the terms have the form
\begin{eqnarray}
\amp &=& \left( \begin{array}{cccccccccc} 1 \\[5pt]
\frac{1}{\alpha} & 1 & \alpha  \\[5pt]
\frac{1}{\alpha^2} & \frac{1}{\alpha} & 1 & \alpha & \alpha^2  \\[5pt]
\frac{1}{\alpha^3} & \multicolumn{4}{c}{\ldots} \\
\vdots
\end{array}\right).
\label{x3}
\end{eqnarray}
and terms of high order become important.\footnote{The leading double-log series exponentiates, and is convergent, i.e.\ $e^z$ has a series expansion with infinite radius of convergence.} Clearly fixed order perturbation theory breaks down completely when $\LL$ get comparable to $1/\alpha$, and one needs to include all terms along and below the $45^\circ$ diagonal in Eq.~(\ref{x3}).

The SCET formalism gives a result for the logarithm of the amplitude in the form
\begin{eqnarray}
\log \amp &=& \left( \begin{array}{cccccccccc} 
\alpha \LL^2 & \alpha \LL & \alpha  \\[5pt]
\alpha^2 \LL^3 & \alpha^2 \LL^2 &  \alpha^2 \LL & \alpha^2  \\[5pt]
\alpha^3 \LL^4 & \alpha^3 \LL^3 & \alpha^3 \LL^2 &  \alpha^3 \LL & \alpha^3  \\[5pt]
\alpha^4 \LL^5 & \multicolumn{4}{c}{\ldots} \\[5pt]
\vdots
\end{array}\right)
\label{x2}
\end{eqnarray}
which is the known form of the perturbation series for the Sudakov form factor~\cite{collinslog,mueller,sen}. Eq.~(\ref{x2}) implies many non-trivial relations between the coefficients in Eq.~(\ref{x1}). In the leading-log regime, the exponentiated series becomes
\begin{eqnarray}
\log \amp &=& \left( \begin{array}{cccccccccc} 
\frac{1}{\alpha} & 1 & \alpha  \\[5pt]
\frac{1}{\alpha} & 1 &  \alpha & \alpha^2  \\[5pt]
\frac{1}{\alpha} & 1 & \alpha &  \alpha^2 & \alpha^3  \\[5pt]
\frac{1}{\alpha} & \multicolumn{4}{c}{\ldots} \\
\vdots
\end{array}\right).
\label{x4}
\end{eqnarray}
To get a reliable value for the amplitude in the leading-log regime, it is essential to sum the first two columns of Eq.~(\ref{x4}). In this case, \emph{the higher order terms are small and under perturbative control, even though the total radiative correction can be very large} (order $1/\alpha$). We will use the exponentiated form of the amplitude. The first column in Eqs.~(\ref{x3},\ref{x4}) is the LL series, the second column the NLL series, etc.

The SCET analysis automatically gives the amplitude in the exponentiated form Eq.~(\ref{x2}). By expanding the SCET  result in a series in $\alpha$, one can recover the form Eq.~(\ref{x1}). The exponentiated form must be used for the QCD corrections. It also needs to be used for electroweak corrections in $W$ production at several TeV, where the electroweak corrections become order unity. At lower energies ($\sim 1-2$~TeV) it is not necessary to sum the entire set of terms along and below the $45^\circ$ diagonal in Eq.~(\ref{x3}). It is still necessary to include the two-loop  $\LL^4,\LL^3,\LL^2$ corrections, and in some cases, the three-loop $\LL^6,\LL^5$
terms. The SCET formalism automatically sums all these terms (as well as higher order $\LL$ enhanced terms). The two-loop $\alpha^2 (\LL^4,\LL^3,\LL^2)$ terms can be obtained using a one-loop renormalization-group improved computation. We have checked that the SCET formalism reproduced the known two-loop $\LL^n$, $n >1$ terms~\cite{fadin,kps,fkps,jkps,jkps4,beccaria,dp1,dp2,hori,beenakker,dmp,pozzorini,js,melles1,melles2,melles3,kuhnW}. 

It is worth emphasizing that with the EFT theory method, no additional work needs to be done to sum the higher order terms. This is done automatically by the renormalization group. Instead one needs to work harder to drop these terms by expanding the SCET expressions in $\alpha$, and then truncating the series. By contrast, in a fixed order computation, one first computes the one- and two-loop graphs, matches the resulting amplitude to the infrared evolution equation and then exponentiates the resulting amplitude, so summing the higher order terms requires additional work.

\section{Estimate of Neglected Power Corrections}\label{sec:power}

The EFT computation neglects $1/Q^2$ power corrections in the one-loop and higher radiative corrections. There are also power corrections at tree-level, which arise from $W,Z,m_t$ masses in the propagators and particle phase space. These are easy to include, and do not limit the computation. Thus the neglected power corrections are small, because they only arise from loop graphs.

To estimate the size of power corrections in the loop graphs, we compare the exact and EFT results for the one-loop Sudakov form factor with a massive gauge boson. The exact one-loop result is
\begin{eqnarray}
&&F(Q^2/M^2)\nn
&=&1+\frac{\alpha C_F}{4\pi}\Biggl[-\frac{7}{2} -\frac{2\pi^2}{3}
+ \frac{4\pi^2}{3} \frac{M^2}{Q^2}
- 3 \log \frac{M^2}{Q^2}\nn
&&-
\log^2 \frac{M^2}{Q^2}+2 \frac{M^2}{Q^2}\log^2 \frac{M^2}{Q^2}\nn
&&+2 \log \frac{M^2}{Q^2} \log\left(1-\frac{M^2}{Q^2}\right)\nn
&&-4\frac{M^2}{Q^2} \log \frac{M^2}{Q^2} \log\left(1-\frac{M^2}{Q^2}\right)\nn
&&+2 \text{Li}_2\left( \frac{M^2}{Q^2}\right)-4\frac{M^2}{Q^2}\text{Li}_2\left( \frac{M^2}{Q^2}\right)\Biggr]
\end{eqnarray}
In the $Q^2 \gg M^2$ limit, this becomes
\begin{eqnarray}
F(Q^2/M^2)&=&F_{\text{EFT}}(Q^2/M^2)\nn
&&+ \frac{\alpha C_F}{4\pi}\frac{M^2}{Q^2} \left[2 \log^2 \frac{M^2}{Q^2} 
-2 \log \frac{M^2}{Q^2}+\frac{4\pi^2}{3}+2 \right]\nn
&&+\mathcal{O}\left(\frac{M^4}{Q^4} \right)
\end{eqnarray}
which is the EFT result, 
\begin{eqnarray}
&&F_{\text{EFT}}(Q^2/M^2)\nn
&=&1+\frac{\alpha C_F}{4\pi}\Biggl[ -\frac{7}{2} -\frac{2\pi^2}{3}
- 3 \log \frac{M^2}{Q^2}-
\log^2 \frac{M^2}{Q^2}\Biggr]\nn
\end{eqnarray}
plus $M^2/Q^2$ suppressed corrections. We define the power corrections $\delta F(Q^2/M^2)$ as the difference $F(Q^2/M^2)-F_{\text{EFT}}(Q^2/M^2)$. At $Q^2=M^2$,
\begin{eqnarray}
\delta F(Q^2=M^2) &=&\frac{\alpha C_F}{4\pi} \pi^2\,.
\end{eqnarray}

\begin{figure}
\begin{center}
\includegraphics[bb=65 200 572 576,width=8.5cm]{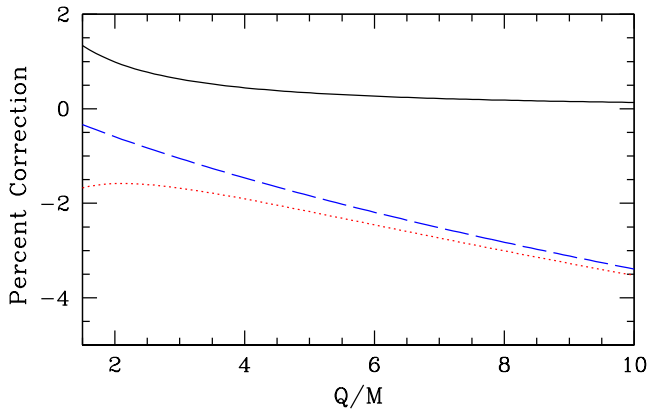} 
\end{center}
\caption{\label{fig:power} Plot of the one-loop power corrections to the Sudakov form factor (solid black), the EFT one-loop correction (dotted red), and their sum, which is the total one-loop correction (dashed blue) as a percentage of the total form factor. The gauge coupling constant has been chosen to be the standard model weak $SU(2)$ value.}
\end{figure}

In Fig.~\ref{fig:power}, we have plotted the power corrections as a percentage of the tree-level value $F(Q^2/M^2)=1$, using an $SU(2)$ gauge theory with coupling $\alpha=\aem/(4 \pi \sin^2 \theta_W)$, the value of the standard model $SU(2)$ gauge coupling at $M_Z$. The power corrections shown by the solid black line are less than 2\% even at $Q=M$, and fall below 1\% by the time $Q/M > 2$. The EFT radiative corrections, shown by the dotted red line, grow with energy.
%%%---FIGURE--------------------------------------------------------------------------------------

\begin{figure}
\begin{center}
\includegraphics[bb=65 200 572 576,width=8.5cm]{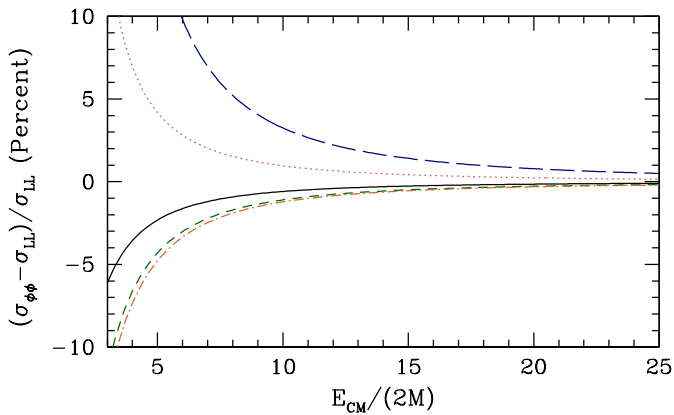} 
\end{center}
\caption{\label{fig:et} Percentage difference between the tree-level $q \bar q \to W_L W_L$ cross-section computed using the equivalence theorem and the tree-level cross-section computed directly, as a function of the center-of-mass energy for center-of-mass scattering angles of $45^\circ$  (long-dashed blue), $60^\circ$ (dotted red), $90^\circ$ (solid black),  $120^\circ$ (short-dashed green) and $135^\circ$ (dot-dashed brown).}
\end{figure}

The other place where power corrections enter is through the use of the Goldstone boson equivalence theorem to compute the cross-section for longitudinally polarized gauge bosons, since the polarization vector $\epsilon_L^\mu$ is replaced by the momentum $k^\mu$. Here the power corrections exist even in the tree-level amplitude. In Fig.~\ref{fig:et} is plotted the ratio of the
%%%---FIGURE--------------------------------------------------------------------------------------
$q \bar q \to W_LW_L$ tree-level cross-section $\sigma_{\text{ET}}$ computed using the equivalence theorem, to the exact expression $\sigma_{\text{exact}}$, as a function of the center-of-mass energy $E_{\text{CM}}$ for different center-of-mass scattering angles $\theta_{\text{CM}}$. The ratio $\sigma_{\text{ET}}/\sigma_{\text{exact}}$ asymptotes to one, but the corrections do not go below 1\% until $E_{\text{CM}}/(2M) \agt 15$. The correction is $\sim 8$\% for $E_{\text{CM}}/(2M) = 5$. Production threshold is $E_{\text{CM}}/(2M) = 1$.

\begin{figure}
\begin{center}
\includegraphics[bb=65 200 572 576,width=8.5cm]{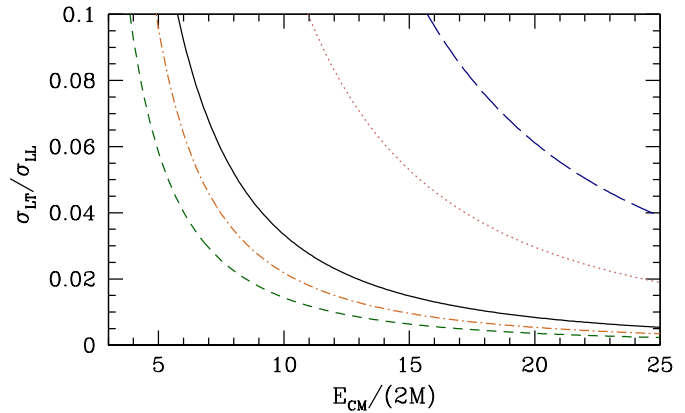} 
\end{center}
\caption{\label{fig:lt} Ratio of the  $q \bar q \to W_L W_T$ cross-section to $q \bar q \to W_L W_L$ at tree-level, as a function of the center-of-mass energy for center-of-mass scattering angles of $45^\circ$  (long-dashed blue), $60^\circ$ (dotted red), $90^\circ$ (solid black),  $120^\circ$ (short-dashed green) and $135^\circ$ (dot-dashed brown).}
\end{figure}

The $q \bar q \to W_LW_T$ cross-section is suppressed by $M^2/E_{\text{CM}}^2$ relative to the $q \bar q \to W_T W_T$ and $q \bar q \to W_LW_L$ cross-sections. The $W_L W_T$ cross-section has been omitted in our analysis. In Fig.~\ref{fig:lt}, we
%%%---FIGURE--------------------------------------------------------------------------------------
plot the ratio of the $W_L W_T$ to the $W_L W_L$ cross-section. The $W_L W_T$ rate falls with energy, and is sub-1\% of the total cross-section at $E_{\text{CM}}/(2M) \agt 15$.

The parton-level results in this paper have an accuracy better than 1\%, with the exception of $W_T$ production. From the estimates in this section, we see that they can be used for processes other than those involving external longitudinal $W$ and $Z$ bosons at relatively low energies, $2-3$ times $M_Z$, while retaining their sub-1\% accuracy. For $W_L,Z_L$ processes, the use of the equivalence theorem introduces power corrections at tree-level, so the corrections are not sub-1\% till much higher energies. One can achieve sub-1\% accuracy at much lower energies by including the tree-level power corrections in the final result, and by including the tree-level rate for $W_L W_T$ production. Then only $\alpha/(4\pi)$ suppressed power corrections are omitted, as for the Sudakov form factor, and the results can be used close to threshold with high accuracy.

\section{The Effective Theory Amplitude}\label{sec:eft}

The next few sections contain a lengthy discussion of the general form of the scattering amplitude to all orders in perturbation theory. We have already seen that for practical purposes, a one-loop analysis augmented by the two-loop cusp anomalous dimension is sufficient. The general form of the amplitude to this order is known explicitly, and the subtleties discussed in the subsequent sections are not relevant. We therefore provide a brief summary of the subsequent sections, and the form of the amplitude.

A generic $r$-particle scattering amplitude can have several different gauge invariant contributions. One writes the amplitudes and anomalous dimensions as matrices in a standard operator basis. For example, the $q_1 \bar q_2 \to q_3 \bar q_4$ amplitude can have the two possible color structures,\footnote{\emph{The term color refers to the $SU(3) \times SU(2) \times U(1)$ gauge quantum  numbers, not just to QCD.}} written schematically as
\begin{eqnarray}
\bar \psi_3 \psi_4\ \bar \psi_2 \psi_1,\qquad \bar \psi_3 T^A \psi_4\ \bar \psi_2 T^A ,
\psi_1
\label{eq7}
\end{eqnarray}
where the color of the initial particles is combined to a color singlet, or color octet, respectively.  The amplitude and anomalous dimensions are matrices in this operator basis. A more useful notation is to use operators in color space acting on the various particles~\cite{catani}, rather than matrices in a particular basis. $\mathbf{T}^A_i$ is a color operator acting on particle $i$. The action of the color operators on the particle fields is
\begin{eqnarray}
\mathbf{T}^A_j \psi_{i\alpha} &=& - \left(T^A\right)_{\alpha \beta} \psi_{i\beta}
\delta_{ij} =-\left(T^A \psi_i\right)_\alpha\, \delta_{ij} \nn
\mathbf{T}^A_j \bar \psi_{i\alpha} &=& \bar \psi_{i\beta} \left(T^A\right)_{\beta \alpha}\delta_{ij} = \left(\bar \psi_i T^A\right)_\alpha\delta_{ij}
\label{eq28}
\end{eqnarray}
where $T^A$ is the color matrix in the representation of particle $i$. The minus sign in the first of Eq.~(\ref{eq28}) is because $\psi_i$ annihilates particle $i$. One can think of $\mathbf{T}^A_i$ as a field-theory operator $\hat T^A_i$ acting by commutation, so that
\begin{eqnarray}
\mathbf{T}^A_j \psi_{i\alpha} &\equiv& \left[ \hat T^A_j, \psi_{i\alpha} \right]=- \left(T^A\right)_{\alpha \beta} \psi_{i\beta}\, \delta_{ij}\,.
\end{eqnarray}
Taking the hermitian conjugate then gives the second of Eq.~(\ref{eq28}). Then, for example,
\begin{eqnarray}
&& \mathbf{T}^B_1 \cdot \mathbf{T}^B_2\ \bar \psi_3 \psi_4\ \bar \psi_2 \psi_1
= - \bar \psi_3 \psi_4\ \bar \psi_2 T^BT^B \psi_1\nn
&=& - C_F \bar \psi_3 \psi_4\ \bar \psi_2  \psi_1\,,\nn[5pt]
&& \mathbf{T}^B_1 \cdot \mathbf{T}^B_2\ \bar \psi_3 T^A \psi_4\ \bar \psi_2 T^A \psi_1 
=  - \bar \psi_3 T^A \psi_4\ \bar \psi_2 T^B T^A T^B \psi_1\nn
&=&  -\left(C_F -\frac12 C_A\right) \bar \psi_3 T^A \psi_4\ \bar \psi_2  T^A \psi_1\,.
\end{eqnarray}
Thus $\mathbf{T}_1 \cdot 
\mathbf{T}_2$ is equivalent to the matrix structure
\begin{eqnarray}
\left[ \begin{array}{cc} 
-C_F & 0 \\
0 & -\left(C_F -\frac12 C_A\right)
\end{array}\right]
\end{eqnarray}
in the basis Eq.~(\ref{eq7}).

The number of color invariants grows rapidly with the number of external legs, and the color operator method provides a compact way of writing the group theoretic form of the amplitude. The soft anomalous dimension and matching has universal form when written using color operators.

\subsection{The Renormalization Group Improved Amplitude}\label{sec:sum}

At high energies, fixed order perturbation theory breaks down because of large logarithms of the form $\log( p_i \cdot p_j/M^2)$, which are of order $\LL \sim \log Q^2/M^2$ in the log-counting, since all invariants $p_i \cdot p_j$ ($i \not=j$) are of order $Q^2$. From the point of view of the high-energy theory, these logarithms arise from infrared divergences as $M \to 0$. The same logarithms arise from ultraviolet divergences as $Q \to \infty$ in the EFT, and can be summed using the EFT renormalization group equations. 

The renormalization group improved on-shell scattering amplitude  (i.e.\ an $S$-matrix element) has the schematic form~\cite{CGKM1,CGKM2}
\begin{eqnarray}
&& \bamp = \exp \Biggl[ \mathbf{D}\left(\alpha(\mu_l),\log\frac{M^2}
{\mu_l^2},*\right) \Biggr] \nn
&&\times  P \exp\left[ - \int_{\mu_l}^{\mu_h}\frac{{\rm d}\mu}{\mu} \bm{\gamma}\left(\alpha(\mu),*\right) \right]\nn
&& \times \mathbf{c}\left(\alpha(\mu_h),\left\{\log\frac{p_i \cdot p_j}{\mu_h^2}\right\}
\right)
\label{13a}
\end{eqnarray}
where boldface quantities are matrices, and $P$ denotes path ordering so that higher values of $\mu$ are to the right. The $\mu_h$ and $\mu_l$ scales are of order $Q$ and $M$, respectively, and the $\mu_{h,l}$ dependence cancels between the matching conditions and the renormalization evolution, so that $\bamp$ is  $\mu_{h,l}$ independent. The invariants $p_i \cdot p_j$ are of order $Q^2$. $*$ denotes a possible dependence on kinematic variables that is explained below. Further details on the form Eq.~(\ref{13a}) are given in Secs.~\ref{sec:fac}, \ref{sec:casimir}.

In Eq.~(\ref{13a}), $\mathbf{c}$ is the high scale matching coefficient at $\mu_h$, and can be a function of all the invariants $\left\{\log (p_i \cdot p_j)/\mu_h^2\right\}$. It is independent of low-energy scales such as the gauge boson mass $M$ or fermion masses $m \ll Q$. The matching scale $\mu_h$ is chosen of order $Q$, so that $\left\{\log (p_i \cdot p_j)/\mu_h^2\right\}$ are all of order unity in the log-counting. 

The anomalous dimension  $\bm{\gamma}\left(\alpha(\mu),*\right) $ is the SCET anomalous dimension used to evolve the operators from $\mu_h$ to $\mu_l$. $\bm{\gamma}$ is given by the ultraviolet divergence structure of the effective theory, and is independent of low-energy scales such as $M$ and $m$.  The $*$ denotes \emph{linear} dependence on $\log (p_i \cdot p_j)/\mu^2$ and a possible dependence on cross-ratios (also referred to as conformal ratios) of the form
\begin{eqnarray}
\mathcal{P}_{ijkl}&=&\log \frac{(p_i \cdot p_j)(p_k \cdot p_l)}{(p_i \cdot p_k)(p_j \cdot 
p_l)}
\end{eqnarray}
which are order unity in the log-counting, can potentially arise in scattering amplitudes with four or more external particles beyond two-loop order\cite{armoni,alday1,alday2,dms,gardi,becherneubert,alday3}, and are discussed further in Sec.~\ref{sec:casimir}.

In deep inelastic scattering and $B \to X_s \gamma$, the SCET anomalous dimension is linear in $\log Q^2/\mu^2$ to all orders in perturbation theory~\cite{dis,Bauer:2003pi}, and can be written as $\bm{\gamma}(\mu)=A\left(\alpha(\mu)\right) \log \mu^2/Q^2+\mathbf{B}\left(\alpha(\mu)\right)$. $A$ is known as the cusp anomalous dimension, and $B$ will be referred to as the non-cusp anomalous dimension. The $A$ term is order $\LL$ in the log-counting, and is proportional to the unit matrix. The integral of this term is order $\LL^2$, and gives the exponentiated Sudakov double-logarithms in high-energy scattering. The $\mathbf{B}$ term is order unity in log-counting, and its integral gives an order $\LL$ contribution to $\log \bamp$. $\mathbf{B}$ can have a non-trivial matrix structure in color space. 

The analysis in Sec.~\ref{sec:regulator}--\ref{sec:csfn} is concerned with determining the structure of the anomalous dimension and low-scale matching to all orders in perturbation theory, and determining what possible variables can occur in $*$. For the applications discussed in Sec.~\ref{sec:plots}, it is sufficient to know the form of the anomalous dimension and matching to two-loop order. This is known by explicit computation, and many of the complications discussed in later sections do not occur to two-loop order. We summarize here the results to two-loop order. The reader not interested in the technical details of their derivation can skip directly from the end of this section to Sec.~\ref{sec:equivthm}.

The matching $D$ and anomalous dimension $\gamma$ can both be written as the sum of collinear and soft contributions, $\mathbf{D}=D_C \openone +\mathbf{D}_S$ and $\bm{\gamma}=\gamma_C \openone+\bm{\gamma}_S$, where the collinear contributions are unit matrices in color space. The collinear running and matching can be written as the sum of one-particle contributions,
\begin{eqnarray}
\gamma_C &=& \sum_i \gamma_{C,i}(\bar n_i \cdot p_i/\mu)\,,\nn
D_C &=& \sum_i D_{C,i}(\bar n_i \cdot p_i/\mu,M/\mu)\,,
\end{eqnarray}
where the sum is over the incoming and outgoing external particles in the hard scattering process. The collinear contributions are process-independent, so that, for example, $\gamma_{C}$ and $D_C$ for a $u_L$ quark have the same values in $u_L \bar u_L \to d_L \bar d_L$ and in $u_L \bar d_L \to W^+ Z$.
The collinear anomalous dimensions have the form
\begin{eqnarray}
 \gamma_{C,i}(\bar n_i \cdot p_i/\mu) &=& \Gamma(\alpha(\mu))\mathbf{T}_i \cdot \mathbf{T}_i \log \frac{\bar  n_i \cdot p_i}{\mu}
 + B_i(\alpha(\mu))\,.\nn
 \end{eqnarray}
 Recall that $\bar  n_i\cdot p_i$ is twice the energy of the particle. The first term is  order $\LL$ in the power counting, and the second term is order unity. $\Gamma$ is the cusp anomalous dimension. At one-loop,
\begin{eqnarray}
\Gamma(\alpha(\mu)) &=& \frac{\alpha(\mu)}{\pi} 
\end{eqnarray}
and the two-loop value is given by  multiplying by the $K$-factor, Eq.~(\ref{kfactor}).\footnote{See Sec.~\ref{sec:higherorder} for the definition and references.} $\Gamma$ has no subscript $i$ because it does not depend on the particle.

The collinear-matching has the form
\begin{eqnarray}
D_{C,i}(\bar n_i \cdot p_i/\mu) &=& J(\alpha(\mu)) \log \frac{\bar  n_i \cdot p_i}{\mu} \mathbf{T}_i \cdot \mathbf{T}_i 
 + E_i(\alpha(\mu))\,,\nn
 \end{eqnarray}
and at one-loop,
\begin{eqnarray}
J(\alpha(\mu)) &=& \frac{\alpha(\mu)}{2\pi} \ \log \frac{M^2}{\mu^2}\,,
\end{eqnarray}
and does not depend on the particle.
The collinear matching at the low scale $\mu_l$ has a dependence on the high scale $Q$ through the $\log (\bar n_i \cdot p_i)$ term. $J$ is related to the cusp anomalous dimension, and is required for consistency  of the low-scale matching, see Sec.~\ref{sec:sudakovff}.

The matrix dependence is in the soft anomalous dimension and matching. These have a universal form when written using the color operator notation,
\begin{eqnarray}
\bm{\gamma}_S &=& -  \Gamma(\alpha(\mu)) \sum_{\vev{ij}}  \mathbf{T}_i \cdot 
\mathbf{T}_j \log \frac{(-n_i \cdot n_j-i0^+)}{2}\,,\nn
 \mathbf{D}_S &=&- J(\alpha(\mu),\lM) \sum_{\vev{ij}}  \mathbf{T}_i \cdot 
\mathbf{T}_j \log \frac{(-n_i \cdot n_j-i0^+)}{2}\nn
\end{eqnarray}
where the sum is over all pairs of external particles in the scattering process. The two-loop values are given by multiplying by the $K$-factor, as shown in Refs.~\cite{aybat1,aybat2} for the soft anomalous dimension in QCD.

The soft-collinear split divides the amplitude into a collinear contribution which depends only on the energies $\bar n_i \cdot p_i$ of the individual particles and is independent of their directions, and a soft contribtion which depends only on the vectors $\{n_i\}$ and hence on the directions of the external particles, but not on their energies. The full amplitude depends only on the Lorentz invariant dot products $p_i \cdot p_j$, and the dependence on the vectors $\{n_i\}$
introduced by the effective theory cancels between the soft and collinear terms by reparametrization invariance~\cite{rpi1,rpi2}.

The anomalous dimension has a cusp contribution $\Gamma$, which is order $\LL$, and the rest, which will be referred to as the non-cusp anomalous dimension, given by $B_i$ and $\bm{\gamma}_S$, collectively referred to as $\mathbf{B}$.

The low scale matching $\mathbf{D}=D_L \openone + \mathbf{D}_0$ has $D_L$ proportional to the unit matrix which contributes at order $\LL$,  and $\mathbf{D}_0$  with non-trivial matrix structure which contributes at order unity. $D_L$ is given by the $J$ term in the collinear matching, and $\mathbf{D}_0$ by the $E_i$ terms in the collinear matching, and the soft matching $\mathbf{D}_S$.  The low-scale matching does depend on the low-energy scales $M$ and $m$. In the standard model, it depends on the electroweak symmetry breaking scale, and includes contributions from the top-quark mass and $\gamma-Z$ mixing.
The new feature of the high-energy scattering problem first found in Refs.~\cite{CGKM1,CGKM2,CKM} is the presence of the single-log term $D_L$ in the low-scale matching. This term can be shown to exist by explicit computation, and is required by consistency of the effective theory --- the $\mu_l$ dependence in the running from the cusp anomalous dimension $\Gamma$, which is of order $\LL$, is cancelled by the $\mu_l$ dependence of $D_1$. One can prove that the low scale  matching is order $\LL$ in the log-counting to all orders in perturbation theory, so that terms such as $\log^2 Q^2/\mu_l^2$ do not appear~\cite{CGKM1,CGKM2,CKM}. This is discussed further in Sec.~\ref{sec:fac}.

The leading-log  (LL) series is given by the one-loop cusp anomalous dimension, the next-to-leading-log (NLL) series by the two-loop cusp anomalous dimension and the one-loop values of $\mathbf{B}$ and $D_1$, the NNLL series by the three-loop cusp, two-loop $\mathbf{B}$ and $D_1$, and one-loop $\mathbf{D}_0$ and $\mathbf{c}$, and the N${}^n$LL series by the $n+1$ loop cusp, the $n$-loop $\mathbf{B}$ and $D_1$, and the $n-1$ loop $\mathbf{D}_0$ and $\mathbf{c}$. 

Eq.~(\ref{13a}) for the standard model sums the QCD and electroweak corrections, including cross terms such as $\alpha_s \alpha_{1,2}$, $\alpha_s y_t^2$, or $\alpha_{1,2} y_t^2$ which depend on mixed products of the Yukawa, strong and electroweak coupling constants. 
\begin{table}
\begin{eqnarray*}
\renewcommand{\arraystretch}{2.0} 
\begin{array}{c|ccccc}
\text{Series} & \Gamma & \mathbf{B} & D_L & \mathbf{D}_0 & \mathbf{c} \\
\hline
\text{LL} & 1\surd & - & - & - & - \\
\text{NLL} & 2\surd & 1\surd & 1\surd & 0\surd & 0\surd\\
\text{NNLL} & 3\surd* & 2\surd* & 2 & 1\surd & 1\surd
\end{array}
\end{eqnarray*}
\caption{\label{tab:orders} The loop-order of various quantities needed to sum the LL, NLL, etc.\ series. $\Gamma$ and $\mathbf{B}$  are the cusp and non-cusp anomalous dimensions, $D_L$ and $\mathbf{D}_0$ are the log and non-log parts of the low-scale matching, and $\mathbf{c}$ is the high-scale matching. $\surd$ means the term has been included in our numerical results. $\surd*$ means the term has been included, except for the scalar contribution. }
\end{table}
Table~\ref{tab:orders} shows the order that the various quantities are needed to sum the different series. The $\surd$ shows which terms have been included in our numerical results, and $\surd*$ shows the terms which have been included except for the scalar contribution. The only quantities remaining to sum the NNLL series are the log term in the low-scale matching $D_L$, and the scalar terms in the three-loop cusp and two-loop non-cusp anomalous dimension. The two-loop value of $D_L$ has been computed in Ref.\cite{jkps4} for fermions when $m_H=M_W$, and is given in Eq.~(\ref{37.12}).

\subsection{The UV/IR Correspondence}\label{sec:uvir}

The gauge theory under consideration is a spontaneously broken $SU(2)$ gauge theory with a Higgs in the fundamental representation, so that all the gauge bosons are massive. Our analysis is also valid in the more general case of an arbitrary broken gauge theory with massive gauge bosons. The advantage of studying this theory instead of a massless gauge theory such as QCD is that the infrared structure of the theory is under complete perturbative control, with the divergences being regulated by the gauge boson mass $M$, if the gauge coupling at $M$ is weak.

The high energy behavior of a spontaneously broken theory is the same as that of the unbroken theory, since spontaneous symmetry breaking is soft, and can be neglected in the high energy limit. Thus our computations also gives the high energy behavior of gauge theory amplitudes in an arbitrary unbroken gauge theory with massless gauge bosons. The ultraviolet behavior of an unbroken gauge theory is identical to the infrared behavior in perturbation theory, so we also obtain information on the infrared structure of gauge theory amplitudes in perturbation theory, which is a topic of current interest~\cite{armoni,alday1,alday2,dms,gardi,becherneubert,alday3}.

In the broken gauge theory infrared divergences are regulated by the gauge boson mass, so the renormalized amplitudes are functions of $\alpha$, $\lQ$ and $\lM$, $A(\alpha(\mu),\lQ,\lM)$, and the scattering amplitudes are IR finite. In a massless gauge theory, the scattering amplitudes are infrared divergent, and can be regulated in perturbation theory using dimensional regularization, so they have the form $A(\alpha(\mu),\lQ,\epsilon)$ with $1/\epsilon$ playing the role of $\lM$. The form Eq.~(\ref{13a}) holds for the massless gauge theory, with the low-scale matching replaced by a low-energy parton matrix element computed in perturbation theory. This is the same as Eq.~(\ref{13a}) with $\lM \to \epsilon$  in the $D$ terms:
\begin{eqnarray}
&& \bamp = \exp \Biggl[ \mathbf{D}\left(\alpha(\mu_l),\epsilon,*\right) \Biggr] \nn
&&\times  P \exp\left[ - \int_{\mu_l}^{\mu_h}\frac{{\rm d}\mu}{\mu} \bm{\gamma}\left(\alpha(\mu),*\right) \right]\nn
&& \times \mathbf{c}\left(\alpha(\mu_h),\left\{\log\frac{p_i \cdot p_j}{\mu_h^2}\right\}
\right)
\label{13b}
\end{eqnarray}
where $\bm{\gamma}$, and $\mathbf{c}$ are unchanged from their values in the broken gauge theory. The $1/\epsilon$ terms in Eq.~(\ref{13b}) are  infrared divergences, and arise because one is computing on-shell scattering amplitudes with a fixed number of external particles. They cancel between real and virtual graphs if one computes experimentally measurable quantities such as jet production rates, which are infrared safe.

We will illustrate the relation between the massive amplitude Eq.~(\ref{13a}) and the massless amplitude Eq.~(\ref{13b}) in Sec.~\ref{sec:sudakovff} by deriving the Magnea-Sterman equations for the Sudakov form factor~\cite{ms} for both theories, and comparing the results.

\section{Factorization Regulator}\label{sec:regulator}

In SCET, the collinear and soft integrals need an infrared regulator, which cancels in the total amplitude. A regulator is necessary for all  factorization calculations --- otherwise one can show that the $\log Q^2/\mu^2$ term in the anomalous dimension, which leads to the Sudakov double logarithms, would be absent.

A convenient regulator is the $\Delta$-regulator introduced in Ref.~\cite{Delta}. The $\Delta$-regulator for particle~$i$ is given by replacing the  propagator denominators by
\begin{eqnarray}
\frac{1}{(p_i+k)^2-m_i^2} \to \frac{1}{(p_i+k)^2-m_i^2-\Delta_i}.
\label{deltareg}
\end{eqnarray}
$\Delta_i$ has mass dimension 2. This regulator can be implemented at the level of the Lagrangian, since it corresponds to  a shift in the particle mass. The on-shell condition remains $p_i^2=m_i^2$. The form Eq.~(\ref{deltareg}) determines the regulator to be used for soft and collinear graphs in the effective theory.

In SCET, the $n_i$-collinear propagator denominators have the replacement Eq.~(\ref{deltareg}). Collinear gauge invariance requires a collinear Wilson line $W_{n_i}$, so that $n_i$ collinear fields only appear in the combination $\left[ W_{n_i}^\dagger \xi_{n_i}\right]$. As shown in Ref.~\cite{Delta,Chiu:2009yz}, this is true even in the presence of the $\Delta$-regulator, since the collinear Wilson lines do not require a regulator after zero-bin subtraction~\cite{zerobin,leesterman,Delta,Chiu:2009yz,messenger}.\footnote{The necessity of subtractions has been shown previously using other regulators~\cite{hautmann}.} $n_i$-collinear graphs only depend on the regulator $\Delta_i$, and are independent of $\Delta_j$ for $j\not=i$.

If particle $i$ interactions with a soft-gluon ($k^\mu \to 0$), the interaction with a $\Delta$-regulator becomes
\begin{eqnarray}
\frac{2 p_i^\mu}{(p_i+k)^2-\Delta_i} &\approx& \frac{2 p_i^\mu}{2p_i \cdot k-
\Delta_i} \,.
\label{eq11}
\end{eqnarray}
Using Eq.~(\ref{pdef}), and keeping only the leading term gives
\begin{eqnarray}
\frac{\left( \bar n_i \cdot p_i \right) n_i^\mu}{\left(\bar n_i \cdot p_i\right) \left(n_i 
\cdot k\right)-\Delta_i}&=& \frac{n_i^\mu}{ \left(n_i \cdot k\right)-\delta_i}\,,
\label{eq4}
\end{eqnarray}
where
\begin{eqnarray}
\delta_i &=& \frac{\Delta_i}{\bar n_i \cdot p_i}\,.
\label{5}
\end{eqnarray}
It is convenient to use $\delta_i$ as the regulator for the soft gluon interactions, even though it is determined from $\Delta_i$ by Eq.~(\ref{5}). $\delta_i$ has mass dimension 1.

The soft interaction Eq.~(\ref{eq4}) is invariant under the rescaling $n_i \to s_i n_i$, $\delta _i \to s_i n_i$, so that soft graphs can only depend on  the combination $n_i^\mu/\delta_i$. This can also be seen directly from
\begin{eqnarray}
\frac{n_i^\mu}{ \left(n_i \cdot k\right)-\delta_i}&=& \frac{n_i^\mu/\delta_i}{ \left(n_i /
\delta_i\cdot k\right)-1}\,,
\end{eqnarray}
so that only the ratio $n^\mu_i/\delta_i$ appears in soft graphs.

We also need the $\Delta$-regulated form for graphs containing heavy particles to study processes such as top quark production. We will use the boosted HQET (bHQET) formalism of Refs~\cite{top1,top2}, in which heavy particles are described by HQET fields $W_{n_I}^\dagger h_{v_I}$, where $v_I$ is the velocity four-vector $v_I^2=1$ and $n_I $ is a null vector $n_I^2=0$ along the direction of particle $I$.\footnote{We will use lower case subscripts $i$ for light particles and upper case subscripts $I$ for heavy particles.} The momentum is $p_I = m_I v_I$, and $\gamma_I=E_I/m_I=n_I \cdot v_I/2$ is the boost. $W_{n_I}$ contains ultracollinear fields that couple to $h_{v_I}$ analogous to the  collinear Wilson line $W_{n_i}$ for collinear fields $\xi_{n_i}$. $W_{n_I}$ does not require a regulator after zero-bin subtraction~\cite{Delta}.

For the $h_v$ propagator and vertex, Eq.~(\ref{eq11},\ref{eq4}) become
\begin{eqnarray}
\frac{2 p_I^\mu}{(p_I+k)^2-\Delta_I} &\to & \frac{v_I^\mu}{v_I \cdot k-\Delta^\prime_{I}} \,,
\label{eq11h}
\end{eqnarray}
where
\begin{eqnarray}
\Delta^\prime_{I} &=& \frac{\Delta_I}{2m_I}\,,
\label{5h}
\end{eqnarray}
so that the HQET propagator is regulated by $\Delta_I^\prime$, and amplitude only depends on the ratio $v_I/\Delta_{I}^\prime$.

The soft interactions of the heavy quark are given in terms of a soft Wilson line 
$Y_{n_I}$, 
for which Eq.~(\ref{eq11},\ref{eq4}) become
\begin{eqnarray}
\frac{\left( \bar n_I \cdot p_I \right) n_I^\mu}{\left(\bar n_I \cdot p_I\right) \left(n_I 
\cdot k\right)-\Delta_I}
&=& \frac{n_I^\mu}{\left(n_I \cdot k\right)-\delta_I}\,,\nn
\label{eq4hh}
\end{eqnarray}
where
\begin{eqnarray}
\delta_I &=& \frac{\Delta_I}{\bar n_I \cdot p_I} = \frac{\Delta_I}{m_I (\bar n_I \cdot 
v_I)}=\frac{2\Delta_I^\prime}{\bar n_I \cdot p_I}\,,
\label{5hh}
\end{eqnarray}
and the amplitude only depends on the ratio $n_I^\mu/\delta_I$. $\Delta_I^\prime$ and $\delta_I$ for heavy quarks both have mass dimension 1.

In previous papers, we have used the analytic regulator~\cite{beneke,smirnov}, defined by letting
\begin{eqnarray}
\frac{1}{p_i^2-m^2} &\to& \frac{1}{\left(p_i^2-m^2\right)^{1+\widetilde\delta_i}}\,,
\end{eqnarray}
where $\widetilde \delta_i$ are dimensionless. The same regulator is also used in Feynman graph computations using the method of regions (see, e.g.\ Ref.~\cite{jkps}). The analytic regulator is scaleless, so that SCET soft-graphs vanish on-shell. However, the analytic regulator breaks the color Ward identities necessary to write the collinear fields in the form $[W_n^\dagger \xi_n]$~\cite{HardScattering,CKM} if there is more than one gauge invariant amplitude. It can used for the Sudakov form factor (where there is only one color invariant), while maintaining the $[W_n^\dagger \xi_n]$ form of the collinear fields.

\section{Factorization Structure of the Amplitude}\label{sec:fac}

The scattering amplitude $\bamp$ computed in the effective theory is computed from soft and collinear diagrams. $\bamp$ will refer to the amplitude, or to the  anomalous dimension, depending on the context. We will use $m_i$ to denote the masses and any other quantum numbers of particle $i$ (such as whether it is a fermion or scalar), and $T^A_i$ to denote its gauge quantum numbers.The argument $\left\{ T^A_i \right\}$ denotes dependence on the gauge quantum numbers of all the external particles, whereas  $T^A_i$ denotes dependence on the gauge quantum numbers only of particle $i$, etc.\ Note that all amplitudes depend on the total particle content of the gauge theory, for example through the $\beta$-function and vacuum polarization diagrams. What $\left\{ T^A_i\right\}$  and $T^A_i$ denote is dependence on the quantum numbers of the particular external particle(s) for a given scattering amplitude. 

In the general $r$-particle scattering amplitude, all the momenta and gauge indices will be treated as flowing into the vertex. Thus an outgoing particle with momentum $p$ in representation $\mathfrak{R}$ is treated as an incoming particle with momentum $-p$ in  representation $\bar\mathfrak{R}$. The analysis below holds for an operator at arbitrary  momentum, so we do not assume that the total incoming momentum $\sum_i p_i$ vanishes, e.g.\ for the Sudakov form factor ($r=2$), momentum transfer $q$ is inserted at the vertex.

\begin{figure}
\begin{center}
\includegraphics[width=4cm]{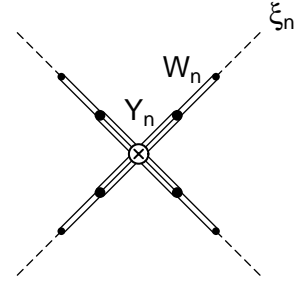} 
\end{center}
\caption{\label{fig:op} Structure of the $r$-particle operator in the effective theory. The dashed lines are collinear fields $\xi_n$, the double lines are collinear Wilson lines $W_{n}^\dagger$, and the triple lines are ultrasoft Wilson lines $Y_n$. The gauge indices on the $r$ ultrasoft Wilson lines are combined into a color singlet at the vertex.}
\end{figure}

\begin{figure}
\begin{center}
\includegraphics[width=4cm]{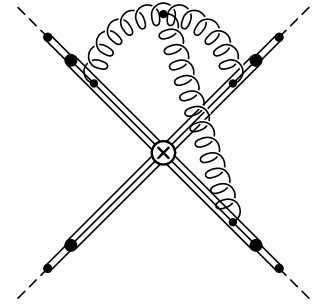} 
\end{center}
\caption{\label{fig:opsoft} Soft gluons can interact with all the $Y_n$s (triple lines).}
\end{figure}

A generic $r$-particle operator has the form
\begin{eqnarray}
f(\left\{b_i\right\})\prod_{i=1}^r \left(Y_{n_i}\right)_{b_i}{}^{a_i}\left[  W_{n_i}^\dagger 
\xi_{n_i}\right]_{a_i}
\label{eq9}
\end{eqnarray}
where $a_i$ are gauge labels, and $f$ is a Clebsch-Gordan  coefficient to  make a gauge singlet operator. Here $\xi_{n_i}$ is an $n_i$-collinear external line which couples to $n_i$-collinear gauge fields, but not to soft gauge fields. $W_{n_i}$ is the $n_i$-collinear Wilson line in the representation of particle $i$ necessary for $n_i$-collinear gauge invariance, and $Y_{n_i}$ is a soft Wilson line in the $n_i$ direction in the  representation of particle $i$ and contains soft gauge fields.  The structure of the interaction is shown in Fig.~\ref{fig:op}, and shows how
the gauge indices are contracted.
%%%---FIGURE--------------------------------------------------------------------------------------
%%
%------------------------------------------------------------------------------------------------------
Soft gluons couple only to the $Y_n$s, the triple lines in Fig.~\ref{fig:op}, and they can interact with any of the triple lines, as shown in Fig.~\ref{fig:opsoft}.
%%%----FIGURE--------------------------------------------------------------------------------------
%%
%------------------------------------------------------------------------------------------------------
$n_i$-collinear gluons can interact with $W_{n_i}$ and $\xi_{n_i}$, so they interact with the double line and dashed line for a \emph{given} direction $n_i$. They do not communicate between two different directions.
%%%----FIGURE--------------------------------------------------------------------------------------
\begin{figure}
\begin{center}
\includegraphics[width=4cm]{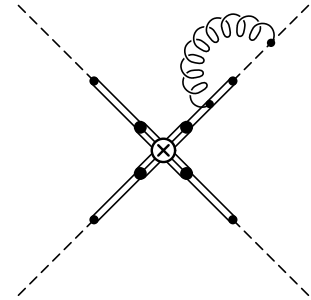} 
\end{center}
\caption{\label{fig:opcoll} $n$-collinear interactions couple $W_n$ (double line) and $\xi_n$ (dashed line).}
\end{figure}
%%
%------------------------------------------------------------------------------------------------------

The $n_i$ collinear interactions involve only the  $\left[W_{n_i}^\dagger \xi_{n_i} \right]$ part of the operator. Thus they only depend on the interactions of particle $i $, and are independent of all the other particles in the process. Note that the Wilson line $W_{n_i}$ is in the gauge representation of particle $i$, and does not need a $\Delta$-regulator. Thus collinear gauge invariance implies that each collinear amplitude has the form (to all orders)
\begin{eqnarray}
\exp I_i\left(\alpha(\mu),\lM,T^A_i,\Delta_i/\mu^2,m_i/\mu \right)\openone\,,
\label{eqcoll}
\end{eqnarray}
where $m_i$ is an abbreviation for all the quantum numbers of particle $i$, such as its mass, and whether it is a fermion, scalar, or gauge boson. The amplitude is a unit matrix in gauge space since $\left[W_{n_i}^\dagger \xi_{n_i}\right]_{a_i}$ gets corrections which do not change the index $a_i$. The index cannot mix with the other $a_{j\not=i}$ indices, since the other indices do not transform under $n_i$-collinear gauge transformations. The exponential form is convenient for the subsequent analysis. The  collinear amplitude depends on the $\Delta$ regulator only for particle $i$, which enters into the particle-$i$ collinear propagator. 

The soft amplitude is given by graphs in which the gauge bosons couple to the soft Wilson lines $Y_{n_i}$ via eikonal interactions. The soft amplitude can be written (to all orders) as
\begin{eqnarray}
\exp\mathbf{S}\left(\alpha(\mu),\lM,\left\{T^A_i\right\},\left\{\mu n_i^\mu/\delta_i\right\}
\right).
\label{eqsoft}
\end{eqnarray}
The amplitude is a matrix in gauge space, since the soft interactions can mix the gauge labels in Eq.~(\ref{eq9}), so $\exp\mathbf{S}$ is a matrix exponential. The amplitude is written as $\exp \mathbf{S}$ rather than $\mathbf{S}$ for future convenience. $\mathbf{S}$ depends on the gauge quantum numbers and directions $n_i$ of all the particles. The argument in Sec.~\ref{sec:regulator} shows that it only depends on $n_i$ and $\delta_i$ in the combination $n_i^\mu/\delta_i$. Note that the soft graphs do not depend on the masses of the external  particles, or whether they are fermions, scalars, or gauge bosons. Thus the amplitude does not depend on $\left\{m_i\right\}$. The soft interactions do not generate $\epsilon_{\alpha \beta \lambda \sigma}$, and so can only depend on the invariants $(n_i \cdot n_j)/(\delta_i \delta_j)$.

\begin{figure}
\begin{center}
\includegraphics[width=4cm]{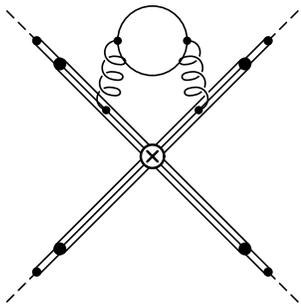} 
\end{center}
\caption{\label{fig:mass} The soft function can depend on $m_t$ through vacuum polarization graphs.}
\end{figure}

The soft amplitude can depend on particle masses through vacuum  polarization graphs such as those in Fig.~\ref{fig:mass}. They arise from the mass-modes discussed in Ref.~\cite{top1,top2}. This dependence is omitted from the argument of Eq.~(\ref{eqsoft})
%%%----FIGURE--------------------------------------------------------------------------------------
%%
%------------------------------------------------------------------------------------------------------
for simplicity, and does not change the following discussion. The dependence on particle masses $\{m_i\}$ is an invariant dependence, i.e.\ it does not depend on the particular external legs in the scattering amplitude.

It is well-known that the Sudakov form factor depends on $Q^2=2p_1 \cdot p_2$, and that the $r$-particle scattering amplitude depends on $\log p_i \cdot p_j$. This can be seen explicitly by a one-loop computation. If the soft and collinear sectors were truly independent, then the scattering amplitude would be completely independent of $p_i \cdot p_j$. This is because each collinear sector is Lorentz invariant and can only depend on one momentum at a time, whereas the soft sector does not depend on $\left\{p_i\right\}$, so there is no way to generate $p_i \cdot p_j
$. The flaw in the argument is that the collinear and soft sectors need a regulator to be separately defined, and communicate via the regulator. The $p_i \cdot p_j$ dependence is generated through this regulator dependence. This is not specific to SCET, but is a universal feature present in any method which factors the total amplitude into soft and collinear contributions. All such factorization methods must necessarily introduce a 
factorization scale that acts as a regulator and enters both the soft and collinear functions. The $p_i \cdot p_j$ dependence enters through this regulator dependence, as we will see explicitly below for the $\Delta$ regulator.

The total amplitude is
\begin{eqnarray}
\bamp&=&\exp\mathbf{S}\ \prod_{i=1}^r 
\exp \left[ I_i\openone \right]\,,
\end{eqnarray}
and the matrix ordering between $I_i$ and $\mathbf{S}$ is irrelevant since the collinear $I_i$ terms are proportional to $\openone$. The amplitude is the product of the collinear and soft contributions since the different sectors are independent. The log of the scattering amplitude is
\begin{eqnarray}
\log \bamp&=& \mathbf{S}\left(\alpha(\mu),\lM,\left\{T^A_i\right\},\left\{\frac{\mu^2 n_i 
\cdot n_j}{\delta_i \delta_j}\right\}\right)\nn
&& + \sum_{i=1}^r  I_i\left(\alpha(\mu),\lM,T^A_i,\Delta_i/\mu^2,m_i/\mu \right)\openone\nn
\label{eq19}
\end{eqnarray}
where matrix ordering between collinear and soft is again irrelevant because the collinear terms are proportional to $\openone$.

The total amplitude is independent of the infrared regulator, since the SCET scattering amplitude in a massive gauge theory is not infrared divergent. This is because the effective theory has the same infrared behavior as the full theory, and the full theory has no infrared divergences since they are all regulated by the gauge boson mass. This implies that the regulator dependence must cancel between the soft and collinear terms.\footnote{One can write down the integrals for the soft and collinear amplitudes, and first combine the integrands before evaluating the integral. This allows one to compute the total amplitude without the necessity of introducing a regulator such as the $\Delta$ regulator~\cite{idilbimehen1,idilbimehen2}.} The only reason for the $\Delta$-regulator is that divergences are introduced in the effective theory by the 
split into collinear and soft sectors.

The cancellation of the regulator in Eq.~(\ref{eq19}) puts strong constraints on the form of the effective theory amplitude. The $\delta$ dependence must cancel between the soft function and the $r$ collinear functions. The collinear amplitudes are a sum of one-particle terms, each depending on a single $\delta_i$, whereas the soft amplitude depends on the products $\delta_i \delta_j$ of all possible pairs of $\delta$s.

The soft function depends on $r(r-1)/2$ variables $(n_i \cdot n_j)/(\delta_i \delta_j)$. [Remember that $n_i^2=0$.] It is convenient to rewrite these in terms of $r(r-3)/2$ cross-ratios which are independent of $\left\{\delta_i\right\}$ and $r$ variables which depend on $\delta_i$. Define
\begin{eqnarray}
\conf(ij|r-2,r-1,r) &\equiv& \log \frac{( n_i\cdot n_j)( n_{r-2} \cdot n_r)( n_{r-1} \cdot 
n_r)}
{(n_i \cdot n_r)( n_j \cdot n_r)( n_{r-2}\cdot n_{r-1})} \nn
\label{eq20}
\end{eqnarray}
for $1 \le i < j  < r$ and
\begin{eqnarray}
\conf(ir|r-2,r-1,r) &\equiv& 0
\label{eq21}
\end{eqnarray}
for $i < r$. In Eqs.~(\ref{eq20},\ref{eq21}), the particles $r-2$, $r-1$ and $r$ play a special role, and have been included explicitly in the indices of $\conf$. Most of the time, we will keep these reference particles fixed, and denote the cross-ratios by $\conf(ij)$. $\conf(r-2,r-1)=0$, so there are $r(r-3)/2$ non-trivial independent cross-ratios. The cross-ratios Eq.~(\ref{eq20}) are homogeneous in $\left\{n_i\right\}$ and so have the same values if $n_i \to n_i /\delta_i$. 

The cross-ratios are invariant under rescaling $n_i \to \lambda_i n_i$, and so one can make the replacement $n_i \to (\bar n_i \cdot p_i) n_i$, i.e.\ $n_i \to p_i$, and write them in terms of ratios of dot products of momenta,
\begin{eqnarray}
\conf(ij|r-2,r-1,r) &\equiv& \log \frac{ (p_i\cdot p_j )( p_{r-2} \cdot p_r)( p_{r-1} \cdot 
p_r)}
{(p_i \cdot p_r)( p_j \cdot p_r)( p_{r-2}\cdot p_{r-1})}\,. \nn
\label{eq20p}
\end{eqnarray}
The cross-ratios first appear for $r=4$, and so can enter into the $2 \to 2$ scattering amplitude. They are absent for the Sudkaov form factor ($r=2$). For two-particle scattering $1 + 2 \to 3 + 4$, the only cross-ratios are
\begin{eqnarray}
\mathcal{P}_{1243}=\conf(12 | 2 3 4) &=& \log \frac{ (n_1\cdot n_2 )(  n_{3} \cdot 
n_4)}
{(n_1 \cdot n_4)( n_{2}\cdot n_{3})}\nn
&=& \log \frac{ (p_1\cdot p_2)(  p_{3} \cdot p_4)}
{(p_1 \cdot p_4)(p_{2}\cdot p_{3})} =  \log \frac{s^2}{u^2}\nn[10pt]
\mathcal{P}_{1342}=\conf(13 | 2 3 4) &=& \log \frac{ (n_1\cdot n_3 )(  n_{2} \cdot 
n_4)}
{(n_1 \cdot n_4)(n_{2}\cdot n_{3})}\nn
&=& \log \frac{ (p_1\cdot p_3 )(  p_{2} \cdot p_4)}
{(p_1 \cdot p_4)(p_{2}\cdot p_{3})} = \log \frac{t^2}{u^2}\nn
\end{eqnarray}
in terms of the Mandelstam variables.

Define the remaining $r$ variables by
\begin{eqnarray}
\mathcal{Y}(i|r-2,r-1,r)  &=&\frac12 \log \frac{ n_{r-2} \cdot n_{r-1}}{(n_{r-2}\cdot 
n_r)(n_{r-1}\cdot n_r)}\nn
&&+\log (n_i \cdot n_r),\quad  i < r\nn[10pt]
\mathcal{Y}(r|r-2,r-1,r) &=& -\frac12 \log \frac{ n_{r-2} \cdot n_{r-1}}{(n_{r-2}\cdot 
n_r)(n_{r-1}\cdot n_r)}\,.\nn
\label{eq24}
\end{eqnarray}
In terms of these variables
\begin{eqnarray}
\log (n_i \cdot n_j) &=& \conf(ij) + \mathcal{Y}(i)+\mathcal{Y}(j), \qquad i \not = j
\label{eq23}
\end{eqnarray}
if we let $\conf(ij) \equiv \conf(ji)$ if $i>j$. Thus any cross-ratio of the $\left\{n_i\right\}$ which is invariant under rescalings of the $n_i$ can be written in terms of $\conf$:
\begin{eqnarray}
\log \frac{ (n_i \cdot n_j)(n_k \cdot n_l)}{(n_i \cdot n_k)(n_j \cdot n_l)}
&=& \conf(ij)+\conf(kl)-\conf(ik)-\conf(jl)\nn
\end{eqnarray}
since $\mathcal{Y}(i)$ cancel on substituting Eq.~(\ref{eq23}).

Then the $r(r-1)/2$ variables $\log (\mu^2 n_i \cdot n_j)/(\delta_i \delta_j)$ can be written as a linear combination of $\conf(ij)$ and $\mathcal{Y}(i)-\log (\delta_i/\mu)$, so the soft function $\mathbf{S}$ can be considered to be a function of $\conf(ij)$ and $\mathcal{X}(i)=\mathcal{Y}(i)-\log (\delta_i/\mu)$.\footnote{It is useful to use $\delta_i/\mu$ so that the variables are dimensionless.} The collinear function $I_i$ is a function of $\Delta_i/\mu^2 = \delta_i (\bar n_i \cdot p_i)/\mu^2$ (see Eq.~(\ref{5})), so it can be regarded as a function of $\mathcal{Z}(i)=\log (\delta_i/
\mu) + \log (\bar n_i \cdot p_i/\mu)$. The total amplitude can then be written as
\begin{eqnarray}
\log \bamp&=& \mathbf{S}\left(\left\{\conf(ij),\mathcal{X}(i)\right\}\right) + 
\sum_{i=1}^r  I_i\left(\mathcal{Z}_i \right)\openone 
\label{eq25}
\end{eqnarray}
where only the momentum and regulator dependence has been shown explicitly. Since $\bamp$ is independent of $\delta_i$, differentiating it w.r.t.\ $\delta_i$ gives
\begin{eqnarray}
\frac{\partial \mathbf{S}}{\partial \mathcal{X}(i)} &=&  \frac{\partial I_i}{\partial 
\mathcal{Z}(i)}\openone \,.
\label{eq26}
\end{eqnarray}
Since $I_i$ only depends on $\mathcal{Z}(i)$, differentiating w.r.t. $\delta_j$ gives
\begin{eqnarray}
0 &=& \frac{\partial^2 \mathbf{S}}{\partial \mathcal{X}(i) \partial \mathcal{X}(j)} 
\qquad i \not = j 
\label{eq27}
\end{eqnarray}
and all mixed partial derivatives of $\mathbf{S}$ vanish. This implies that $\mathbf{S}$ must have the form
\begin{eqnarray}
\mathbf{S} &=& \sum_{i=1}^r \mathbf{S}_i(\mathcal{X}(i)),
\label{eq28alt}
\end{eqnarray}
a sum of matrix functions each depending on a single $\mathcal{X}(i)$ variable. Substituting Eq.~(\ref{eq28alt}) in Eq.~(\ref{eq26}) gives
\begin{eqnarray}
\frac{\partial \mathbf{S}_i}{\partial \mathcal{X}(i)} &=&  \frac{\partial I_i}{\partial 
\mathcal{Z}(i)}\openone \,,
\label{eq26a}
\end{eqnarray}
from which it follows that both sides are a constant since $\mathcal{X}(i)$ and $\mathcal{Z}(i)$ can be varied independently. The constant must be proportional to the unit matrix since the r.h.s.\ is. It can only depend on variables other than $\mathcal{X}(i)$ and $\mathcal{Z}(i)$ which are common to both $\mathbf{S}$ and $I_i$ and have been suppressed in Eq.~(\ref{eq26a}). Thus one gets
\begin{eqnarray}
\frac{\partial \mathbf{S}_i}{\partial \mathcal{X}(i)} &=&  \frac{\partial I_i}{\partial 
\mathcal{Z}(i)}\openone = J\left(\alpha(\mu),\lM,T_i^A\right)\openone
\label{eq26aa}
\end{eqnarray}
so that $J$ only depends on the color of particle $i$ and the gauge boson mass, and is independent of particle masses, and whether they are scalars, fermions, or gauge bosons. Thus $J$ is a universal function independent of $i$, since the color dependence is explicit through the argument $T_i^A$.

Integrating Eq.~(\ref{eq26aa}) gives
\begin{eqnarray}
\mathbf{S}_i &=&  J\left(\alpha(\mu),\lM,T_i^A\right)\openone \mathcal{X}(i)\nn
&& + \mathbf{G}_i\left(\alpha(\mu),\lM,\left\{T_i^A\right\},\left\{\conf(ij)\right\}
\right)\,,\nn[10pt]
 I_i &=&  J \left(\alpha(\mu),\lM,T_i^A\right)\openone \mathcal{Z}(i) + E
\left(\alpha(\mu),\lM,T_i^A,m_i/\mu\right) \,,\nn
\label{eq26b}
\end{eqnarray}
and
\begin{eqnarray}
\log \bamp &=& \sum_{i=1}^r
J \left(\alpha(\mu),\lM,T_i^A\right)\openone \left[\mathcal{Y}(i)+\log 
\left(\frac{\bar n_i \cdot p_i}{\mu}\right)\right]\nn
&&+ \sum_{i=1}^r  E\left(\alpha(\mu),\lM,T_i^A,m_i/\mu\right)\openone \nn
&& + \mathbf{G}\left(\alpha(\mu),\lM,\left\{T_i^A\right\},\left\{\conf(ij)\right\}\right)\,,
\label{eq:form}
\end{eqnarray}
where $\mathbf{G}=\sum_i \mathbf{G}_i$. The $\left\{\delta_i\right\}$ dependence has cancelled, and the amplitude depends  linearly on the combination $\mathcal{X}(i)+\mathcal{Z}(i)=\mathcal{Y}(i)+\log (\bar n_i \cdot p_i)$.  The final result Eq.~(\ref{eq:form}) is completely symmetric under the exchange of particle labels. This is not manifest in Eq.~(\ref{eq:form}) because the split of kinematic variables into $\conf(ij)$ and $\mathcal{Y}(i)$ is not symmetric.  The final expressions we need for our computation given in Sec.~\ref{sec:results} are manifestly symmetric.

The form Eq.~(\ref{eq:form}) is valid to all orders in perturbation theory. $J$ is related to the cusp anomalous dimension. The factorized form Eq.~(\ref{eq:form}) is valid for the renormalized amplitude $\log \bamp$ and for the anomalous dimension $\bm{\gamma}$ in the effective theory, both of which  have the additive structure Eq.~(\ref{eq25}). The EFT anomalous dimension is given by the UV divergences in the EFT, and does not depend on IR scales such as $M$ and $m_i$, so these variables can be dropped for $\bm{\gamma}$. The anomalous dimension can depend on $\bar n_i \cdot p_i$ and on the directions $n_i$, since these appear explicitly as labels on the EFT operator. The amplitude Eq.~(\ref{eq:form}) is linear in $\log \mu$ to all orders in perturbation theory, a result first shown in SCET for deep-inelastic scattering in Ref.~\cite{dis}. For the multiparticle case, the $\log \mu$ dependence is proportional to the unit matrix, and commutes with the other contributions.

\subsection{The Light-Light Sudakov Form Factor}\label{sec:r=2}

The above discussion works for four or more external legs, $r \ge 4$, where one can use the variables $\conf(ij|r-2,r-1,r)$ and $\mathcal{Y}(i|r-2,r-1,r)$ defined by Eqs.~(\ref{eq20},\ref{eq24}). For $r=3$, there are no cross-ratios, but one can still use Eq.~(\ref{eq24}) and Eq.~(\ref{eq23}), dropping $\conf(ij)$ in the final result Eq.~(\ref{eq:form}). 

For $r=2$, the soft function depends only on the single variable $(n_1 \cdot n_2)/(\delta_1 \delta_2)$, and there is only a single color invariant, so that there is no matrix structure. The derivation leading to Eq.~(\ref{eq:sform}) for the two-particle case has already been given  earlier in several different ways~\cite{dis,CGKM1,CGKM2,CKM}.  We repeat the derivation because we will use the results for $r=2$ (the Sudakov form factor) extensively in the remainder of this article. In Eq.~(\ref{eq25}), the soft function can be chosen to be a function of $\mathcal{X} = \log(\mu^2 n_1 \cdot n_2)/(2 \delta_1 \delta_2)$. Since $\bamp$ is independent of $\delta_{1,2}$, differentiating w.r.t.\ $\delta_{1,2}$ gives
\begin{eqnarray}
\frac{\partial S}{\partial \mathcal{X}} = \frac{\partial I^{(1)}}{\partial \mathcal{Z}(1)}=
\frac{\partial I^{(2)}}{\partial \mathcal{Z}(2)}.
\label{eq35}
\end{eqnarray}
Differentiating again w.r.t.\ $\delta_{1,2}$ gives
\begin{eqnarray}
\frac{\partial^2 S}{\partial \mathcal{X}^2} = 0\,,
\end{eqnarray}
so that
\begin{eqnarray}
S &=& J(\alpha(\mu),\lM,T_i^A) \mathcal{X} + G(\alpha(\mu),\lM,T_i^A) \,,\nn
I_i &=& J(\alpha(\mu),\lM,T_i^A) \mathcal{Z}(i) + E(\alpha(\mu),\lM,T^A_i,m_i/
\mu)\,. \nn
\end{eqnarray}
Note that because the operator is a gauge singlet, $C_F=\mathbf{T}_1 \cdot \mathbf{T}_1 = \mathbf{T}_2 \cdot \mathbf{T}_2=-\mathbf{T}_1 \cdot \mathbf{T}_2$ and $\mathbf{T}_1^A+\mathbf{T}_2^A=0$, so that there is only one independent color operator. We will drop the subscript $i$ on $T^A$ when we write the color dependence of the Sudakov amplitude.

Eq.~(\ref{eq35}) implies $J_1=J_2$, so there is no subscript $i$ on $J$. This also follows from the discussion below Eq.~(\ref{eq26a}). The collinear function $I_i$ has the same value whether particle $i$ scatters via a two-particle operator, or via an $r$-particle operator, so $J$ and $E$ are the same as in Eq.~(\ref{eq26b}).

Adding the pieces gives
\begin{eqnarray}
\log \bamp &=& 
J  \left(\alpha(\mu),\lM,T^A\right) \log \left(\frac{-Q^2}{\mu^2}\right)\nn
&&+ \sum_{i=1}^2  E\left(\alpha(\mu),\lM,T^A,m_i/\mu\right) \nn
&& + G\left(\alpha(\mu),\lM,T^A\right)\,.
\label{eq:sform}
\end{eqnarray}
In deriving Eq.~(\ref{eq:sform}) we have used $J_1=J_2=J$ and
\begin{eqnarray}
&&
J \left[ \log \frac{n_1 \cdot n_2}{2} + \log \frac{\bar n_1 \cdot p_1}{\mu}+ \log 
\frac{\bar n_2 \cdot p_2}{\mu} \right]\nn
&=&
J \log \left(\frac{-Q^2}{\mu^2}\right)\,,
\end{eqnarray}
where $-Q^2=q^2=(p_1+p_2)^2=(\bar n_1 \cdot p_1)(\bar n_2 \cdot p_2)(n_1 \cdot 
n_2)/2$, since all momenta are incoming.

%The final form of the result, which we will need later, is given by Eq.~(\ref{eq:srun}) and Eq.~(\ref{eq:slowD}).

\subsection{Change of Basis Transformation}\label{sec:basis}
 
The basis for cross-ratios Eq.~(\ref{eq20}) picked out particles $r-2$, $r-1$ and $r$. One can use a different reference set of particles for the cross-ratios,   $r \to s_0$,  $r-1 \to s_{1}$ and $r-2 \to s_{2}$ in Eq.~(\ref{eq20}). The cross-ratios in the new basis are given in terms of the old basis by
\begin{eqnarray}
&& \conf(ij|s_0s_1s_2) =  \conf(ij|r-2,r-1,r)\nn
&&\qquad +\conf(s_1 s_0 |r-2,r-1,r) -  \conf(i s_1 |r-2,r-1,r)\nn
&& \qquad -  \conf(j s_0 |r-2,r-1,r)\,,\nn[10pt]
&& \mathcal{Y}(i \not = s_0 |s_0s_1s_2)=\mathcal{Y}(i| r-2,r-1,r) \nn
&&\quad + \frac12 \conf(s_2 s_1|r-2,r-1,r)-\frac12 \conf(s_2 s_0|r-2,r-1,r)\nn
&&\quad-\frac12 \conf(s_1 s_0| r-2,r-1,r)+ \conf(i s_0 | r-2,r-1,r)\,,\nn[5pt]
&& \mathcal{Y}(s_0 |s_0s_1s_2)= -\frac12 \conf(s_2 s_1|r-2,r-1,r)\nn
&&\quad+\frac12 \conf(s_2 s_0|r-2,r-1,r)+\frac12 \conf(s_1 s_0| r-2,r-1,r)\,.\nn
\label{eq34}
\end{eqnarray}
The transformation is linear in $\mathcal{Y}$, and $\conf$ does not involve $\mathcal{Y}$, so linearity of Eq.~(\ref{eq:form}) in $\mathcal{Y}$ is maintained, and changing the basis does not provide any additional constraints. If $\mathbf{G}$ is linear in $\left\{\conf(ij)\right\}$ in one basis, then it remains linear in $\left\{\conf(ij)\right\}$ in any other basis. Linearity in $\conf$ is connected with Casimir scaling, and will be important in Sec.~\ref{sec:casimir}.

\section{The Scattering Amplitude}\label{sec:scatamp}

One can now write down the general expression for a hard scattering amplitude using the EFT method.

{(1)} One matches from the full theory on to SCET at a scale $\mu_h$ of order $Q$. This gives a column vector of coefficients $\mathbf{c}$ of the allowed SCET operators. We will normalize $\mathbf{c}$ so that it is dimensionless. $\mathbf{c}$ is a column vector, and has the form
\begin{eqnarray}
\mathbf{c}\left(  \alpha(\mu_h), \left\{p_i \cdot p_j \right\},\mu_h \right)
\label{eq:high}
\end{eqnarray}
and can depend on the invariants $p_i \cdot p_j$ which are order $Q^2$, and logarithmically on $\mu_h$, but not on low scales parametrically smaller than $Q$, such as $M$. Since $\mathbf{c}$ is dimensionless, it can depend on ratios such as $(p_i \cdot p_j)/(p_k \cdot p_l)$, which are invariant under an overall rescaling $p_i \to \lambda p_i$ of the momenta, but are not invariant under individual rescalings $p_i \to \lambda_i p_i$. The invariance under $p_i \to \lambda p_i$ is broken by the $\mu_h$ dependence, i.e.\ by the anomalous dimension of $\mathbf{c}$. $\mathbf{c}$ is computed from the hard part of the diagram, i.e.\ by using pure dimensional regularization with all infrared scales set to zero. This is the usual procedure for computing a matching condition, and the EFT factorization methods do not apply to this part of the computation. $\mu_h$ is chosen to be of order $Q$ to minimize the logarithms in $\mathbf{c}$.

(2) Run the coefficients from $\mu_h$ down to a low scale $\mu_l$ of order $M$ using the SCET anomalous dimension. The SCET anomalous dimension has the form Eq.~(\ref{eq:form}), and can be written as
\begin{eqnarray}
\bm{\gamma}(\mu) &=& \openone \sum_{i=1}^r
\Gamma \left(\alpha(\mu),T_i^A\right) \left[\mathcal{Y}(i)+\log 
\left(\frac{\bar n_i \cdot p_i}{\mu}\right)\right]\nn
&&+\openone \sum_{i=1}^r  \Omega\left(\alpha(\mu),T_i^A,s_i\right) \nn
&& + \bm{\Sigma}\left(\alpha(\mu),\left\{T_i^A\right\},\left\{\conf(ij)\right\}\right)\,.
\label{eq:run}
\end{eqnarray}
$\Gamma, \Omega, \bm{\Sigma}$ are the analogs of $J$, $E$ and $\mathbf{G}$ for the anomalous dimension. Since $\gamma$ does not depend on low scales such as $M$ or $m_i$, these variables are dropped in Eq.~(\ref{eq:run}). $\Gamma$, $\Omega_i$ and $\bm{\Sigma}$ are obtained from the $1/\epsilon$ ultraviolet divergent terms in Eq.~(\ref{eq:form}), and obey the factorization structure of the EFT amplitude. $\Gamma$, which is the cusp anomalous dimension, was shown above to only depend on the color of particle $i$, but not on whether it is a fermion, gauge boson, or scalar. Thus we have written it as $\Gamma(\alpha(\mu),T_i^A)$ where the color dependence is explicitly through the second argument. The function $\Gamma$ is then universal, i.e.\ there is no need to introduce different functions $\Gamma_i$ for each particle. While $\Omega$ does not depend on the mass of particle $i$, it can depend on whether the particle is a fermion or scalar or gauge boson, so we have included an argument  $s_i$ (for statistics) in $\Omega$; again there is no need for different functions $\Omega_i$.

For the Sudakov problem, the scattering of particle~$i$ by an external current, $r=2$, there is no color matrix structure, and
\begin{eqnarray}
 \bm{\gamma}_i^{(r=2)}(\mu) &=& \Gamma \left(\alpha(\mu),T_i^A\right) \left[
\log \frac{Q^2}{\mu^2}\right]+ \sum_{i=1}^2  \Omega\left(\alpha(\mu),T_i^A,s_i\right)
\nn
&& + \sigma\left(\alpha(\mu),\left\{T_i^A\right\}\right)
\label{eq:srun}
\end{eqnarray}
using the results of Sec.~\ref{sec:r=2} and the form Eq.~(\ref{eq:sform}). $\bm{\Sigma}$ for the Sudakov problem, which we require later, has been called $\sigma$, and the $\log(-1)\Gamma $ term has been absorbed into  $\sigma$. The branch cuts are determined by analytically continuing the form factor from Euclidean to Minkowski space, or by an explicit graphical computation. While there is no color matrix structure, the anomalous dimensions till depend on the color of the  particles (through $C_F$), which is why the $T_i$ dependence is retained in Eq.~(\ref{eq:srun}).

The evolution of the amplitude from $\mu_h \to \mu_l$ is given by
\begin{eqnarray}
\mathbf{H} &=& P \exp -\int_{\mu_l}^{\mu_h} \frac{\rd \mu}{\mu} \bm{\gamma}(\mu)
\label{eq:H}
\end{eqnarray}
where $P$ denotes $\mu$-ordering with larger values of $\mu$ on the right.

(3) Compute the low-scale matching at $\mu=\mu_l \sim M$. The matching is also constrained by the EFT form Eq.~(\ref{eq:form}),
\begin{eqnarray}
\mathbf{D} &=& \log \mathbf{d} \nn
&=&\openone  \sum_{i=1}^r
J \left(\alpha(\mu),\lM,T_i^A\right) \left[\mathcal{Y}(i)+\log 
\left(\frac{\bar n_i \cdot p_i}{\mu}\right)\right]\nn
&&+ \openone \sum_{i=1}^r  E\left(\alpha(\mu),\lM,T_i^A,m_i/\mu\right)\nn
&& + \mathbf{G}\left(\alpha(\mu),\lM,\left\{T_i^A\right\},\left\{\conf(ij)\right\}
\right)\,.
\label{eq:low}
\end{eqnarray}
$\mathbf{D}$ is linear in $\log Q$ (the $\log (\bar n_i \cdot p_i)$ terms) to all orders in perturbation theory. The existence of this logarithm in the low-scale matching, and linearity to all orders, was first pointed out in Ref.~\cite{CGKM1,CGKM2}. $\mathbf{D}$ and $\mathbf{d}$ are square matrices. Previous studies of factorization have all been for massless gauge theories. Eq.~(\ref{eq:low}) is the factorization form for the low-scale matching in a massive gauge theory, which has not been studied previously.

We break $\mathbf{D}$ into the single-log term and the rest,
\begin{eqnarray}
\mathbf{D} &=& \mathbf{D}_0 + \openone D_L\,,\nn
D_L &=& \sum_{i=1}^r
J \left(\alpha(\mu),\lM,T_i^A\right)\log \left(\frac{\bar n_i \cdot p_i}{\mu}
\right)\,,\nn
 \mathbf{D}_0 &=&\openone \sum_{i=1}^r
J \left(\alpha(\mu),\lM,T_i^A\right) \mathcal{Y}(i)\nn
&&+\openone \sum_{i=1}^r  E\left(\alpha(\mu),\lM,T_i^A,m_i/\mu\right) \nn
&& + \mathbf{G}\left(\alpha(\mu),\lM,\left\{T_i^A\right\},\left\{\conf(ij)\right\}
\right)\,,
\label{eq:lowD}
\end{eqnarray}
so the low-scale matching can be written as
\begin{eqnarray}
\mathbf{d} &=& \exp{\mathbf{D_0}} \exp({D_L} \openone)
\end{eqnarray}
where the two terms commute.

For the Sudakov form-factor ($r=2$) for particle $i$,
\begin{eqnarray}
D_i &=& D_{0,i} + D_{1,i}\log \frac{Q^2}{\mu^2}\,,\nn
D_{1,i} &=&  J\left(\alpha(\mu),\lM,T_i^A\right)\,,\nn
D_{0,i} &=& \sum_{i=1}^2  E\left(\alpha(\mu),\lM,T_i^A,m_i/\mu\right) \nn
&& + g \left(\alpha(\mu),\lM,\left\{T_i^A\right\}\right)\,,
\label{eq:slowD1}
\end{eqnarray}
where $\mathbf{G}$ for the Sudakov problem is denoted by $g$, the $\log(-1)J$ term has been absorbed into $g$, and there is no matrix structure.

The final result for the amplitude is (showing the scales at which the various pieces are computed)
\begin{eqnarray}
\bamp &=& \mathbf{d}(\mu_l)\ \mathbf{H}(\mu_l \leftarrow \mu_h)\ \mathbf{c}
(\mu_h)
\label{eq:total}
\end{eqnarray}
and  is independent of $\mu_h$ and $\mu_l$. This is the amplitude given in Eq.~(\ref{13a}).

It is worth emphasizing that the total amplitude $\bamp$ in the gauge theory does not have the factorization properties studied here. The factorization properties only hold for the effective theory amplitude, and hence for $\mathbf{d}$ and $\mathbf{H}$ in Eq.~(\ref{eq:total}) but not for the hard matching correction $\mathbf{c}$. Thus the factorization properties are obscured in an explicit diagrammatic computation. The effective theory reorganizes the computation to make the factorization structure manifest, analogous to the way in which heavy quark effective theory (HQET) makes heavy quark spin-flavor symmetry manifest~\cite{hqetbook}.

\subsection{Consistency Conditions}\label{sec:consistency}

$\mu_h$ independence of Eq.~(\ref{eq:total}) gives
\begin{eqnarray}
\mu \frac{\rd \mathbf{c}}{\rd \mu} &=& \bm{\gamma}(\mu)\mathbf{c}(\mu)
\end{eqnarray}
and  $\mu_l$ independence gives
\begin{eqnarray}
\mu \frac{\rd \mathbf{d}}{\rd \mu} &=& -\mathbf{d} \bm{\gamma}(\mu)
\end{eqnarray}
so
\begin{eqnarray}
e^{-\mathbf{D}_0}\mu \frac{\rd }{\rd \mu} e^{\mathbf{D}_0} 
+ \openone  \mu \frac{\rd D_L}{\rd \mu}&=& - \bm{\gamma}(\mu)\,.
\end{eqnarray}
Comparing the $\log(\bar n_i \cdot p_i)$ terms on both sides gives
\begin{eqnarray}
\mu \frac{\rd J}{\rd \mu} &=&  -\Gamma
\end{eqnarray}
which relates the $\log (\bar n_i \cdot  p_i)$ term  in the low-scale matching to the cusp anomalous dimension. Thus the presence of the $\log (\bar n_i \cdot  p_i)$ term in the low-scale matching follows from consistency of the effective theory~\cite{CKM}.

\section{Casimir Scaling}\label{sec:casimir}

The results of the previous section simplify dramatically at one-loop. The collinear graphs are proportional to the color factor $\mathbf{T}_i \cdot \mathbf{T}_i$ and the soft graphs are a sum over pairs of two-particle terms proportional to $\mathbf{T}_i \cdot \mathbf{T}_j$. 
The collinear and soft amplitude can be written as
\begin{eqnarray}
I_i &=& \mathbf{T}_i \cdot \mathbf{T}_i\left[I(\alpha(\mu),\lM,\Delta_i/\mu^2,m_i/\mu) 
\right]\nn
\mathbf{S} &=& \sum_{\vev{ij}} \mathbf{T}_i \cdot \mathbf{T}_j\ S
\left(\alpha(\mu),\lM,\frac{n_i \cdot n_j}{\delta_i \delta_j}\right)
\label{eq79}
\end{eqnarray}
where $\vev{ij}$ is a sum over pairs of particles with $i\not=j$. This is a special case of Eq.~(\ref{eq19}), where the color dependence of the collinear and soft amplitudes has been explicitly written, so that $I_i$ and $S$ do not depend on color, and $S$ has no matrix structure. Furthermore, the $S$ function  multiplying $ \mathbf{T}_i \cdot \mathbf{T}_j$ can only depend on the single variable $(n_i \cdot n_j)/(\delta_i  \delta_j)$, rather than all possible dot products, and so the EFT amplitude cannot depend on cross-ratios, which involve at least four particles.

The EFT anomalous dimension and matching Eqs.~(\ref{eq72C},\ref{eq73C},\ref{eq:runsoft},\ref{eq:lowsoft}) simplify considerably because of the restricted form Eq.~(\ref{eq79}). It is simplest to rederive the anomalous dimensions and matching conditions which follow from the $\delta$-cancellation conditions starting from Eq.~(\ref{eq79}), rather than imposing the restriction of Eq.~(\ref{eq79}) on the results of the previous section. The answer is
\begin{eqnarray}
\log \bamp &=&J(\alpha(\mu),\lM) \sum_i \mathbf{T}_i \cdot 
\mathbf{T}_i \, \log \frac{\bar  n_i \cdot p_i}{\mu} \nn
&&-J(\alpha(\mu),\lM) \sum_{\vev{ij}}  \mathbf{T}_i \cdot \mathbf{T}_j \log \frac{-n_i \cdot n_j-i0^+}{2}\nn
&&\hspace{-1cm}+\sum_i  \mathbf{T}_i \cdot \mathbf{T}_i\left[ E(\alpha(\mu),\lM,m_i/\mu)
+\frac12 g(\alpha(\mu),\lM)\right]\nn
\label{eq86a}
\end{eqnarray}
and the derivation is given in Appendix~\ref{app:sudform}. The functions on the r.h.s.\ no longer depend on the color quantum numbers, and the matrix function $\mathbf{G}$ has been replaced by a single function $g$.

\subsection{Anomalous Dimension}\label{sec:anomdim}

Assuming Casimir scaling, Eq.~(\ref{eq79}), the anomalous dimension Eq.~(\ref{eq:run}) simplifies to
\begin{eqnarray}
\bm{\gamma}(\mu) &=& \Gamma(\alpha(\mu))  \sum_i \mathbf{T}_i \cdot 
\mathbf{T}_i \, \log \frac{\bar  n_i \cdot p_i}{\mu} \nn
&&-\Gamma \left(\alpha(\mu)\right) \sum_{\vev{ij}}  \mathbf{T}_i \cdot \mathbf{T}_j \log \frac{-n_i \cdot n_j-i0^+}{2}\nn
&&+\sum_i  \mathbf{T}_i \cdot \mathbf{T}_i\left[ \Omega(\alpha(\mu),s_i)
+\frac12 \sigma(\alpha(\mu))\right]
\label{eq:srunz1}
\end{eqnarray}
and the Sudakov anomalous dimension Eq.~(\ref{eq:srun})  becomes
\begin{eqnarray}
 \bm{\gamma}_i^{(r=2)}(\mu) &=& \mathbf{T}_i \cdot 
\mathbf{T}_i  \biggl[ \Gamma \left(\alpha(\mu)\right) 
\log \frac{Q^2}{\mu^2} +  \Omega\left(\alpha(\mu),s_1\right)\nn
&&+\Omega\left(\alpha(\mu),s_2\right)
 + \sigma\left(\alpha(\mu)\right)\biggr]
\label{eq:srunz}
\end{eqnarray}
using $Q^2= (\bar n_1 \cdot p_1)(\bar n_2 \cdot p_2)(-n_1 \cdot n_2)/2$.

It is convenient to break the anomalous dimension into a collinear and soft anomalous dimension, the precise definition of which is given in Sec.~\ref{sec:csfn} for the general case. The collinear anomalous dimension is
\begin{eqnarray}
\bm{\gamma}_C &=& \sum_i \mathbf{T}_i \cdot \mathbf{T}_i 
\Biggl[ \Gamma(\alpha(\mu))  \log \frac{\bar n_i \cdot p_i}{\mu}\nn
&&+\Omega(\alpha(\mu),s_i)
+\frac12 \sigma(\alpha(\mu))\Biggr]\,,
\label{eq86ca}
\end{eqnarray}
the soft anomalous dimension is
\begin{eqnarray}
\bm{\gamma}_S &=& -  \Gamma(\alpha(\mu)) \sum_{\vev{ij}}  \mathbf{T}_i \cdot 
\mathbf{T}_j \log \frac{(-n_i \cdot n_j-i0^+)}{2}\,,\nn
\label{eq86c}
\end{eqnarray}
and the total  anomalous dimension is $\bm{\gamma}=\bm{\gamma}_C+\bm{\gamma}_S$. \emph{The soft anomalous dimension for an $r$-leg amplitude has a universal form, and is completely determined in terms of the cusp anomalous dimension.} At one-loop,
\begin{eqnarray}
\Gamma(\alpha(\mu)) &=& \frac{\alpha(\mu)}{\pi}\,.
\label{eq86d}
\end{eqnarray}
Eq.~(\ref{eq86c}) is also valid at two-loops.

\subsection{Low-Scale Matching $D$}\label{sec:lowscale}

Assuming Eq.~(\ref{eq79}), the low scale matching Eq.~(\ref{eq:low}) simplifies to
\begin{eqnarray}
\mathbf{D} &=& \log \mathbf{d} \nn
&&= \sum_{i=1}^r
J \left(\alpha(\mu),\lM\right) \mathbf{T}_i \cdot 
\mathbf{T}_i \  \log \frac{\bar n_i \cdot p_i}{\mu}\nn
&&- J \left(\alpha(\mu),\lM\right) \sum_{\vev{ij}}  \mathbf{T}_i \cdot \mathbf{T}_j \log \frac{-n_i \cdot n_j-i0^+}{2}\nn
&&+ \sum_{i=1}^r  \mathbf{T}_i \cdot 
\mathbf{T}_i \ E\left(\alpha(\mu),\lM,m_i/\mu\right)\nn
&&+\frac12\sum_i  \mathbf{T}_i \cdot \mathbf{T}_i\
 g(\alpha(\mu),\lM)\,.
\label{eq:lowz}
\end{eqnarray}
$\mathbf{D}$ is linear in $\log Q$ (the $\log (\bar n_i \cdot p_i)$ terms)  and in $\log\mu$ to all orders  in perturbation theory. 

We break $\mathbf{D}$ into the single-log term and the rest, so Eq.~(\ref{eq:lowD}) becomes
\begin{eqnarray}
\mathbf{D} &=& \mathbf{D}_0 + \openone D_L\,,\nn
D_L &=& \sum_{i=1}^r
J \left(\alpha(\mu),\lM\right) \mathbf{T}_i \cdot 
\mathbf{T}_i \ \log \frac{\bar n_i \cdot p_i}{\mu}\,,\nn
 \mathbf{D}_0 &=&- J \left(\alpha(\mu),\lM\right) \sum_{\vev{ij}}  \mathbf{T}_i \cdot \mathbf{T}_j \log \frac{-n_i \cdot n_j-i0^+}{2}\nn
&&+ \sum_{i=1}^r  \mathbf{T}_i \cdot 
\mathbf{T}_i \ E\left(\alpha(\mu),\lM,m_i/\mu\right)\nn
&&+\frac12\sum_i  \mathbf{T}_i \cdot \mathbf{T}_i\
 g(\alpha(\mu),\lM)\,,
\label{eq:lowDz}
\end{eqnarray}
so the low-scale matching can be written as
\begin{eqnarray}
\mathbf{d} &=& \exp{\mathbf{D_0}} \exp({D_L} \openone)
\end{eqnarray}
where the two terms commute.

For the Sudakov problem
\begin{eqnarray}
D_i &=& D_{0,i} + D_{1,i}\log \frac{Q^2}{\mu^2}\nn
D_{1,i} &=&  J \left(\alpha(\mu),\lM\right)\mathbf{T}_i \cdot 
\mathbf{T}_i \nn
D_{0,i} &=&\mathbf{T}_i \cdot 
\mathbf{T}_i\biggl[ \sum_{i=1}^2  E\left(\alpha(\mu),\lM,m_i/\mu\right) \nn
&& + g \left(\alpha(\mu),\lM\right)\biggr]
\label{eq:slowD}
\end{eqnarray}

The matching can also be written as the sum of  a collinear contribution
\begin{eqnarray}
\mathbf{D}_C &=& \sum_i \mathbf{T}_i \cdot \mathbf{T}_i  \Biggl[ J(\alpha(\mu),\lM) \log \frac{\bar n_i \cdot p_i}{\mu}\nn
&&+  E(\alpha(\mu),\lM,m_i/\mu)+\frac12 g(\alpha(\mu),\lM)\Biggr]
\label{eq86b}
\end{eqnarray}
and a soft contribution
\begin{eqnarray}
 \mathbf{D}_S &=&- J(\alpha(\mu),\lM) \sum_{\vev{ij}}  \mathbf{T}_i \cdot 
\mathbf{T}_j \log \frac{(-n_i \cdot n_j-i0^+)}{2}\nn
\label{eq86e}
\end{eqnarray}
with $\mathbf{D}  = \mathbf{D}_C + \mathbf{D}_S $.
\emph{The soft matching for an $r$-leg amplitude has a universal form, and is completely determined in terms of the single-log term in the matching, which is related to the cusp anomalous dimension.} At one-loop,
\begin{eqnarray}
J(\alpha(\mu),\lM) &=&  \frac{\alpha(\mu)}{2\pi} \lM =\frac{\alpha(\mu)}{2\pi}\log \frac{M^2}{\mu^2}.
\label{eq86f}
\end{eqnarray}

The above expressions give a remarkable simplification in computing scattering amplitudes. The collinear contributions are a sum over one-particle terms which do not depend on the scattering process, and the soft contributions are universal. The soft functions have the minimal form necessary to restore reparametrization invariance~\cite{rpi1,rpi2} (i.e.\ boost invariance) of the EFT amplitudes. For back-to-back particles $n_i \cdot n_j=-2$, and $\log(-n_i \cdot n_j)/2=0$.

\subsection{Form  of the EFT Functions}\label{sec:functions}

Comparing the formul\ae\ in this section with Eqs.~(\ref{eq:run},\ref{eq:srun},\ref{eq:lowD},\ref{eq:slowD}), we see that
\begin{eqnarray}
\Gamma_i(\alpha(\mu),T_i^A) &=& \mathbf{T}_i \cdot \mathbf{T}_i\ 
\Gamma(\alpha(\mu))\nn
\Omega_i(\alpha(\mu),T_i^A,s_i) &=&  \mathbf{T}_i \cdot \mathbf{T}_i\ 
\Omega(\alpha(\mu),s_i)\nn
\sigma_i(\alpha(\mu),T_i^A) &=& \mathbf{T}_i \cdot \mathbf{T}_i\ 
\sigma(\alpha(\mu))\nn
J(\alpha(\mu),\lM,T_i^A) &=&\mathbf{T}_i \cdot 
\mathbf{T}_i\ J(\alpha(\mu),\lM) \nn
E\left(\alpha(\mu),\lM,T_i^A,m_i/\mu\right) &=&  \mathbf{T}_i \cdot \mathbf{T}_i\  
E\left(\alpha(\mu),\lM,m_i/\mu\right)\nn
g_i(\alpha(\mu),\lM,T_i^A) &=&  \mathbf{T}_i \cdot \mathbf{T}_i\  
g(\alpha(\mu),\lM)
\label{eq71}
\end{eqnarray}
so that the color dependence is factored out into a prefactor. In addition, the non-trivial matrix structures $\bm{\Sigma}$ and $\mathbf{G}$ in the soft contribution are determined in terms of scalar functions $\sigma$ and $g$ which appear in the Sudakov problem.

\subsection{Higher Order}\label{sec:higherorder}

The  one-loop functions Eq.~(\ref{eq71}) and the anomalous dimension Eq.~(\ref{eq:srunz}) and amplitude Eq.~(\ref{eq:lowz}) have two very special properties.  The first is proportionality to $\mathbf{T}_i \cdot \mathbf{T}_j$, which is referred to as Casimir scaling. Obviously, this is a trivial result at one-loop, since the one-loop graphs all explicitly have this form, but it is non-trivial at higher orders. The cusp anomalous dimension $\Gamma(\alpha(\mu),T_i^A)$ has been computed to three-loop order~\cite{moch:ns}, and obeys Casimir scaling to this order. The cusp anomalous dimension is
\begin{eqnarray}
\Gamma(\alpha(\mu),T_i^A) &=& \sum_{n}
\left(\frac{\alpha}{4\pi}\right)^n  \Gamma^{(n)}(T_i^A)\,,\nn
\Gamma^{(1)} &=& 4 C_F\,,\nn
\Gamma^{(2)}  & =& K_1 \Gamma^{(1)}\,,\nn
\Gamma^{(3)}&=&  K_2 \Gamma^{(1)}\,,
\end{eqnarray}
where the $K$-factors are
\begin{eqnarray}
K_1 &=& \left(\frac{67}{9} - \frac13\pi^2\right) C_A  - \frac{20}{9}n_F T_F
-\frac89 t_S n_S \nn
K_2 &=& \left({245\over 6}-{134\over 27}\pi^2
+{22\over 3}\zeta(3)+{11\over 45}\pi^4\right)C_A^2\nn
&&+\left(-{418\over 27}+{40\over 27}\pi^2
-{56\over 3}\zeta(3)\right)C_AT_Fn_F
\nn
&&+\left(-{55\over 3}+{16}\zeta(3)\right)C_FT_Fn_F-
{16\over 27}(T_Fn_F)^2\nn
\label{kfactor}
\end{eqnarray}
The scalar contributions to $K_1$ is known~\cite{jkps,jkps4}, but not to $K_2$. 

Casimir scaling can be seen from the proportionality of $\Gamma^{(2,3)}$ to $\Gamma^{(1)}$, so that all three terms in $\Gamma$ are proportional to $C_F$.
%%-FIGURE--------------------------------------------------------------------------------------
\begin{figure}
\begin{center}
\includegraphics[width=4cm]{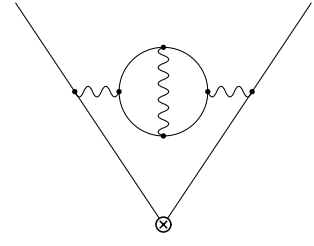} 
\end{center}
\caption{\label{fig:cusp} Three loop correction to the cusp anomalous dimension.}
\end{figure}
%%
%------------------------------------------------------------------------------------------------------
It appears that the $C_F$ term in $K_2$ violates Casimir scaling; however, it originates from the two-loop vacuum polarization correction Fig.~\ref{fig:cusp}, and $C_F t_F n_F$ is from $\text{Tr}\,(T^a T^b T^c T^b)$, and so the same for all particles, and does not depend on the particular particle whose cusp anomalous dimension is being computed.

The non-cusp part of the collinear anomalous dimension $B=2\Omega+\sigma$ does not obey Casimir scaling.  The one and two-loop terms for fermions are~\cite{jkps,jkps4}
\begin{eqnarray}
B^{(1)} &=& -6 C_F\nn
B^{(2)} &=& C_F^2\left(4 \pi^2-48 \zeta(3) -3\right)\nn
&&+ C_F C_A \left(52 \zeta(3)-\frac{11\pi^2}{3}-\frac{961}{27}\right)\nn
&&+C_F n_F T_F \left(\frac{4\pi^2}{3}+\frac{260}{27}\right)+C_F n_s \left(\frac{\pi^2}{6}+\frac{167}{54} \right)\nn
\label{37.11}
\end{eqnarray}
and the $C_F^2$ term in $B^{(2)}$ violates Casimir scaling already at two loops. In color operator notation, it is proportional to $(\mathbf{T}_i \cdot \mathbf{T}_i)^2$. The three-loop value $B^{(3)}$ is known for a gauge theory with only fermionic matter~\cite{moch:ns}.

The cusp anomalous dimension is the anomalous dimension of a Wilson loop with a point where the tangent vector is discontinuous (the cusp)~\cite{brandt}. Casimir scaling of Wilson line expectation values has been studied extensively in the literature, and there has been recent interest in this question~\cite{aybat1,aybat2,armoni,alday1,alday2,dms,gardi,becherneubert,alday3}. Casimir scaling of Wilson loops has also been studied on the lattice, to test whether the string tension is proportional to the color Casimir of the quark. Recent results seem to indicate there is deviation from Casimir scaling~\cite{teper} for the string tension. Casimir scaling of Wilson loops does not hold when non-perturbative effects are included. For example, large Wilson loops in the fundamental representation are confining and have an area law (in the absence of light fermions), whereas those in the adjoint representation are screened, and have a perimeter law. In the large $N$ limit, one can relate the two, since screening is $1/N^2$ suppressed~\cite{greensite,largen}.

The collinear anomalous dimension respects Casimir scaling at least till three loop order, by explicit computation~\cite{moch:ns}. Recently, Aybat et al.~\cite{aybat1,aybat2} have shown that the soft anomalous dimension \emph{matrix} at two-loop order also respects Casimir scaling,  i.e.\
\begin{eqnarray}
\bm{\gamma}_S^{(2)} &=& K_1 \bm{\gamma}_S^{(1)}
\end{eqnarray}
where $K_1$ is the same factor as for the collinear anomalous dimension, and is a number. Thus the entire two-loop anomalous dimension matrix can be obtained by multiplying the one-loop matrix by $(1+K_1 \alpha/(4\pi))$, as can the $D_L$ term in the two-loop matching.

For practical LHC computations, we only need the one-loop electroweak corrections, the one-loop QCD matching, and the two-loop QCD running. The one-loop terms have the nice $\mathbf{T}_i \cdot \mathbf{T}_j$ and sum-on-pairs form. The results of Aybat et al.~\cite{aybat1,aybat2} tell us that this form can also be used for the two-loop anomalous dimension. The two-loop non-cusp collinear anomalous dimension does not obey Casimir scaling, but it is a one-particle term and  can also be trivially included 
in the results.

An interesting question currently being investigated is whether Casimir scaling continues to hold at higher orders. Becher and Neubert~\cite{becherneubert} have shown that the soft anomalous dimension should continue to have the $\mathbf{T}_i \cdot \mathbf{T}_j \log (p_i \cdot p_j)$ form at three-loops, and have suggested that this form persists to all orders. If this form holds at higher orders, then the form derived in Sec.~\ref{sec:csfn} in terms of collinear and universal soft functions will continue to hold at higher order. For the electroweak problem, we need not only the anomalous dimension, but also the low-scale matching $\mathbf{D}$, and it would be interesting to investigate the form of $\mathbf{D}$ at higher orders. The Becher-Neubert~\cite{becherneubert} argument that there are no color structures at two and three-loops other than $\mathbf{T}_i \cdot \mathbf{T}_j$ implies that the low-scale matching should continue to have this form to three-loops.

\subsection{Sum on Pairs}\label{sec:sumonpairs}

The results in Secs.~\ref{sec:anomdim},\ref{sec:lowscale} have two  properties --- they are proportional to $\mathbf{T}_i \cdot \mathbf{T}_j$, and they are linear in $\log n_i \cdot n_j$, or equivalently in $\log p_i \cdot p_j$. Linearity in $\log p_i \cdot p_j$ implies that the amplitude can only depend linearly on the cross-ratios $\conf(ij)$; higher powers would induce $(\log p_i \cdot p_j)(\log p_k \cdot p_l)$ terms. The entire EFT one-loop result, including the low-scale matching, has linearity in both $\mathbf{T}_i \cdot \mathbf{T}_j$ and in $\log p_i \cdot p_j$. The color structure follows trivially from the color factors of a one-loop graph, and linearity in $\log p_i \cdot p_j$ can be seen by explicit computation. As shown in Ref.~\cite{CKM}, an amplitude of this form can also be written as a sum-on-pairs over two-particle Sudakov form factors weighted with $\mathbf{T}_i \cdot \mathbf{T}_j$,
\begin{eqnarray}
\log \amp &=& -\sum_{\vev{ij}} \mathbf{T}_i \cdot \mathbf{T}_j F(-p_i \cdot p_j).
\label{eq:pairs}
\end{eqnarray}
Thus all the information in the scattering amplitude is contained in the Sudakov form factor.

Becher and Neubert~\cite{becherneubert} give a proposed form for the all orders anomalous dimension, which is both linear in $\mathbf{T}_i \cdot \mathbf{T}_j$ and in $\log p_i \cdot p_j$, so that the sum-on-pairs formula Eq.~(\ref{eq:pairs}) would hold to all orders. Ref.~\cite{CKM} used the analytic regulator for SCET. On-shell soft graphs vanish with the analytic regulator, so the entire amplitude arises from collinear graphs. The analytic regulator has the big disadvantage that the $n$-collinear sector fields cannot be combined into the $W_n^\dagger \xi_n$ form of a single Wilson-line; instead one has separate $n$-collinear Wilson lines for each particle not in the $n$-direction, and the different collinear sectors have interlocking color structures~\cite{Delta}. This complication does not hold for amplitudes with a single color structure, such as the Sudakov form factor, or particular scattering processes such as $\bf{3} + \bf{\bar 3} \to \bf{6} +\bf{8}$ in QCD (quix production). There is also effectively only a single color structure in the large-$N$ limit, because color reordering of the operators is suppressed by $1/N$. The mixing is only suppressed for a suitable basis of operators, those written using products of color matrices in cyclic order.\footnote{In the double-line notation~\cite{thooft}, each color connected part of the operator should consist of quark lines  with alternating directions for the arrows, which are color-paired as one moves around the vertex in cyclic order, so that one has a planar vertex~\cite{coleman,largen}.} Thus in the $q \bar q \to gg$ example considered in Ref.~\cite{kidonakis}, one should use the basis $\bar q q\, \delta^{ab}$, $\bar q T^a T^b q$ and $\bar q T^b T^a q$ rather than $\bar q q\, \delta^{ab}$, $\bar q T^c q\, f^{abc}$ and $\bar q T^c q\, d^{abc}$ for the operator basis. One can then see that the anomalous dimension matrix in Ref.~\cite{kidonakis} becomes diagonal in the large-$N$ limit. In Ref.~\cite{CKM}, it was shown that for amplitudes with a single color structure, linearity in $\log p_i \cdot p_j$ holds to all orders. Thus a check of the Becher-Neubert proposal requires a four-loop computation away from the $N \to \infty$ limit.

Linearity in $\log p_i \cdot p_j$ implies that the amplitude and anomalous dimension have the form
\begin{eqnarray}
\log \bamp &=& \openone \sum_i J(\alpha(\mu),\lM,T_i^A) \log \frac{\bar  n_i \cdot p_i}{\mu} \nn
&&+\openone \sum_i   E(\alpha(\mu),\lM,T_i^A,m_i/\mu)\nn
&&- \sum_{\vev{ij}}   \mathbf{B}^{(ij)}(\alpha(\mu),\lM)  \log \frac{-n_i \cdot n_j-i0^+}{2}\nn
&&+\mathbf{G} (\alpha(\mu),\lM)
\label{eq186b}
\end{eqnarray}
where
\begin{eqnarray}
2 \sum_{{j \atop j \not = i}}  \mathbf{B}^{(ij)} +  J(\alpha(\mu),\lM,T_i^A) \openone =0
\end{eqnarray}
and do not depend on the cross-ratios. Imposing in addition linearity in $\mathbf{T}_i \cdot \mathbf{T}_j$ for the $\log n_i \cdot n_j$ terms and in $\mathbf{T}_i \cdot \mathbf{T}_i$ for the rest reduces the amplitude back to Eq.~(\ref{eq86b}). Only using linearity in $\mathbf{T}_i \cdot \mathbf{T}_j$ gives the general form Eq.~(\ref{eq:form}) with a restriction on the color structure of the terms, but no constraint on the dependence on cross-ratios.

\section{Collinear and Soft Functions}\label{sec:csfn}

The factorization form Eq.~(\ref{eq:form}) of the EFT scattering amplitude allows us to efficiently compute the radiative corrections to high energy scattering processes, such as those needed for the LHC. The factorization structure studied in Sec.~\ref{sec:fac} provides a more compact way of writing our previous EFT results, by dividing the EFT scattering amplitude into a sum of  one-particle collinear contributions, and a universal soft function. The procedure we follow is that used by  Aybat, Dixon and Sterman~\cite{aybat2} to separate the amplitude into jet and soft functions. We will use the name collinear function, rather than jet function, since the term jet function has already been used in the SCET literature for a different quantity. This decomposition will be used both for the anomalous dimension and for the low-scale matching $D$. 

The SCET computation automatically breaks the amplitude into a product of collinear and soft contributions. The splitting, however, is regulator dependent, and each piece depends, for example, on $\delta_i$ if one uses the $\Delta$-regulator. This is a generic feature of any factorization procedure which cannot be avoided, because the $\log Q^2$ term in the SCET anomalous dimension is generated precisely by this regulator 
dependence. It is nicer to reorganize the amplitude into regulator-independent terms which have the same factorization properties as the SCET collinear and soft contributions. The basic idea is to write the effective theory amplitude as the product of the square-roots of the Sudakov form factor in SCET for each external particle, and the rest. The square-root of the Sudakov form factor in the effective theory (i.e. the form factor without the high-scale matching) is called the collinear contribution, and the rest is the soft contribution. Since the total amplitude and the SCET Sudakov 
form factor are both regulator-independent, this provides a regulator independent split of the amplitude into collinear and soft parts, which, as we will see, respects the factorization structure of Eq.~(\ref{eq:form}). Our definition of the collinear function is identical to that of Aybat et al.~\cite{aybat2} for the jet function. Aybat et al.\ define the soft function in terms of a ratio of web functions regulated by taking the directions $n_i$ away from the null direction, and then taking the light-like limit. We have used a different regulator and defined the soft function in terms of SCET 
amplitudes, but the final result is the same.

The Sudakov form factor for particle $i$ is the amplitude for $p_1 \to p_2$, where both particles are of type $i$, and $\bar n_1 \cdot p_1 = \bar n_2 \cdot p_2=\bar n_i \cdot p_i$ in the Breit frame.\footnote{i.e.\ the Sudakov form factor is evaluated at the same energy that particle~$i$ has in the $r$-particle scattering process. An alternate choice is to evaluate all  the functions at some common momentum transfer, such as $\sqrt{\hat s}$.} The collinear anomalous dimension is defined as one-half the anomalous dimension for the amplitude, Eq.~(\ref{eq:run}), and can be written using Eq.~(\ref{eq:srun}) as
\begin{eqnarray}
\gamma_{Ci} &=& \Gamma(\alpha(\mu),T_i^A) \log \left(\frac{\bar n_i \cdot p_i}
{\mu}\right)+\Omega(\alpha(\mu),T_i^A,s_i)\nn
&&+\frac12 \sigma(\alpha(\mu),T_i^A)\, .
\label{eq72}
\end{eqnarray}
Half the log of the Sudakov low-scale matching gives the log of the collinear matching
\begin{eqnarray}
D_{Ci} &=& \frac12 D_{0,i} + \frac 12 D_{1,i} \log\frac {(\bar n_i \cdot p_i)^2}{\mu^2}\nn
&=& J(\alpha(\mu),\lM,T_i^A) \log \frac{\bar n_i \cdot p_i}{\mu} \nn
&&+E\left(\alpha(\mu),\lM,T_i^A,m_i/\mu\right)\nn
&& +
\frac12g(\alpha(\mu),\lM,T_i^A)
\label{eq73}
\end{eqnarray}
using Eq.~(\ref{eq:slowD}).

Now consider an arbitrary hard scattering process with $r$ external lines, where the particles have energies $2E_i \sim \bar n_i  \cdot p_i$. We will define the collinear anomalous dimension and collinear low-scale  matching by summing Eq.~(\ref{eq72}) and Eq.~(\ref{eq73}) evaluated at $\bar n_i \cdot p_i$, and summed over all external lines. They are proportional to the unit matrix in  color space. The total collinear anomalous dimension is the sum of one-particle terms
\begin{eqnarray}
\bm{\gamma}_C &=& \openone \sum_{i=1}^r \Biggl[ \Gamma(\alpha(\mu),T_i^A) 
\log \left(\frac{\bar n_i \cdot p_i}{\mu}\right)+\Omega(\alpha(\mu),T_i^A,s_i)\nn
&&+\frac12 \sigma(\alpha(\mu),T_i^A)\Biggr]\,,
\label{eq72C}
\end{eqnarray}
as is the collinear low-scale matching
\begin{eqnarray}
\mathbf{D}_{C} &=& \openone\sum_{i=1}^r \Biggl[ J(\alpha(\mu),\lM,T_i^A) \log \frac{\bar n_i \cdot p_i}{\mu}\nn
&&+ E
\left(\alpha(\mu),\lM,T_i^A,m_i/\mu\right)\nn
&&+\frac12 g(\alpha(\mu),\lM,T_i^A) \Biggr]\,.
\label{eq73C}
\end{eqnarray}

The soft anomalous dimension $\gamma_S$ is defined as the difference of the total anomalous dimension and collinear anomalous dimension. 
Using Eq.~(\ref{eq:run}), we find
\begin{eqnarray}
\bm{\gamma}_S(\mu) &=&\openone \sum_{i=1}^r
\Gamma \left(\alpha(\mu),T_i^A\right) Y(i) \nn
&&-\frac12 \openone \sum_{i=1}^r \sigma(\alpha(\mu),T_i^A)\nn
&&+ \bm{\Sigma}\left(\alpha(\mu),
\left\{T_i^A\right\},\left\{\conf(ij)\right\}\right)\,.
\label{eq:runsoft}
\end{eqnarray}
The log of the soft matching $\mathbf{D}_S$ is the difference of the logs of the total matching $\mathbf{D}$ and the collinear matching $D_C$,
\begin{eqnarray}
\mathbf{D}_S 
&=& \openone \sum_{i=1}^r
J \left(\alpha(\mu),\lM,T_i^A\right) Y(i)\nn
&&-\openone \sum_{i=1}^r  g\left(\alpha(\mu),\lM,\left\{T_i^A\right\},\left\{\conf(ij)
\right\}\right)\nn
&& + \mathbf{G}\left(\alpha(\mu),\lM,\left\{T_i^A\right\},\left\{\conf(ij)\right\}
\right)
\label{eq:lowsoft}
\end{eqnarray}
and the total anomalous dimension and log of the low-scale matching are
\begin{eqnarray}
\bm{\gamma} &=& \bm{\gamma}_C + \bm{\gamma}_S\,,\nn
\mathbf{D} &=& \mathbf{D}_C + \mathbf{D}_S\,.
\label{102}
\end{eqnarray}

The decomposition of the anomalous dimension and matching into soft and collinear functions discussed above is that used by Aybat et al.~\cite{aybat2}.  The soft/collinear split is related to the decomposition of the SCET amplitude into soft and  collinear graphs, but is not identical. The basic difference is that in an $r$-leg process, the soft contribution of the Sudakov form factors for each external particle has been added to the collinear function and subtracted from the soft functions (the $\sigma$ and $g$ terms in Eq.~(\ref{eq72C},\ref{eq73C},\ref{eq:runsoft}, \ref{eq:lowsoft})). This cancels the regulator dependence of the SCET collinear and soft functions, and provides a regulator-independent definition for the two functions.

The collinear functions are process-independent, and can be computed directly from the Sudakov form factor. The cusp  part $\Gamma$ of $\bm{\gamma}_C$  and $J$ of $\mathbf{D}_C$ are of order $\LL$ in log-counting, whereas the remaining terms of $\bm{\gamma}_C$, $\mathbf{D}_C$, as well as $\bm{\gamma}_S$, $\mathbf{D}_S$ are order unity. One can move the $\log (\bar n_i \cdot p_i)$ terms from the collinear factors into the soft factors by rewriting $n_i \cdot n_j \to (p_i \cdot p_j)/\mu^2$. It is more convenient to leave the order $\LL$ terms in the collinear functions, where they are explicitly proportional to the unit matrix. In addition, the $n_i \cdot n_j $ form for the soft function remains valid for heavy quarks, so that the same soft function can be used for SCET and HQET fields.

Assuming the Casimir scaling form Eq.~(\ref{eq79}) allows $\bm{\gamma}_S$ and $\mathbf{D}_S$ to be written in a much simpler form. Using the general form given in Sec.~\ref{sec:casimir} for the anomalous dimension and amplitude, the collinear and soft anomalous dimensions and matching reduce to Eqs.~(\ref{eq86a},\ref{eq86c},\ref{eq86b},\ref{eq86d}) given in Sec.~\ref{sec:casimir}. This is the form we need for the NLL standard model computations, which use one-loop matching and two-loop running, for which Casimir scaling is known to hold.

\section{Heavy Quarks}\label{sec:hq}

\begin{figure}
\begin{center}
\includegraphics[width=4cm]{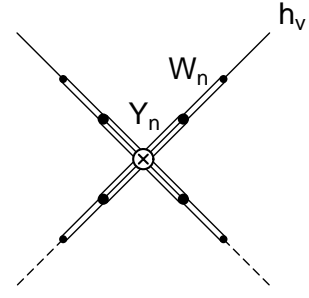} 
\end{center}
\caption{\label{fig:oph} Structure of the $r$-particle operator in the effective theory for external  light and heavy quarks. The dashed lines are collinear fields $\xi_n$, the solid lines are HQET fields $h_v$, the double lines are collinear Wilson lines $W_{n}^\dagger$, and the triple lines are ultrasoft Wilson lines $Y_n$. The gauge indices on the $r$ ultrasoft Wilson lines are combined into a color singlet at 
the vertex.}
\end{figure}

We can also compute the scattering amplitude for $r+s$ external legs, where $r$ legs are light, and treated using SCET fields $W_{n_i}^\dagger \xi_{n_i}$, $i=1,\ldots,r$ and $s \ge 1$ legs are heavy and treated using boosted HQET fields (bHQET) $W_{n_I}^\dagger h_{v_I}$, $I=1,\ldots, s$~\cite{top1,top2}. The SCET operators are given by Fig.~\ref{fig:oph}, where, for a heavy quark, the external dashed line is replaced by a HQET field.
%%-FIGURE--------------------------------------------------------------------------------------
%%
%------------------------------------------------------------------------------------------------------

The $n_i$ collinear graphs have the same form as before, Eq.~(\ref{eqcoll}). The on-shell HQET graphs are also a sum over the external heavy particles,
\begin{eqnarray}
\exp H_I\left(\alpha(\mu),\lM,T^A_I,\frac{\mu v_I^\mu}{\Delta_I^\prime} \right)\openone.
\label{eqh}
\end{eqnarray}
The HQET graph $H_I$ does not depend on the heavy particle mass, since it does not enter the HQET computation. The soft amplitude is given by graphs in which the gauge bosons couple to the soft Wilson lines $Y_{n_i}$ and $Y_{n_I}$ via eikonal interactions. The soft amplitude can be written (to all orders) as
\begin{eqnarray}
\exp\mathbf{S}\left(\alpha(\mu),\lM,\left\{T^A_i\right\},\left\{\mu n_i^\mu/\delta_i\right\},
\left\{\mu n_I^\mu/\delta_{I}\right\}\right).
\label{eqsofth}
\end{eqnarray}
The two amplitudes only depend on the combinations $v_I^\mu$, $n_i^\mu/\delta_i$ and $n_I^\mu/\delta_{I}$ as shown in Sec.~\ref{sec:regulator}.

The log of the scattering amplitude is
\begin{eqnarray}
&& \log \bamp\nn
&=& \mathbf{S}\left(\alpha(\mu),\lM,\left\{T^A_i\right\},\left\{
\frac {\mu^2 n_i \cdot n_j} {\delta_i \delta_j},\frac {\mu^2 n_i \cdot n_I} {\delta_i  \delta_{I}},
\frac{\mu^2 n_I \cdot n_J}{\delta_{I} \delta_{J}}\right\}\right)\nn
&& + \sum_{i=1}^r  I_i\left(\alpha(\mu),\lM,T^A_i,\Delta_i/\mu^2,m_i/\mu \right)\openone \nn
&& + \sum_{I=1}^s  H_I\left(\alpha(\mu),\lM,T^A_I,\Delta_I^\prime/\mu \right)\openone
\label{eq19h}
\end{eqnarray}
where matrix ordering is again irrelevant because the collinear and HQET terms are proportional to $\openone$, and we have written the $n$ and $v$ dependence in terms of Lorentz invariant dot products.

By a now familiar argument, the cancellation of the regulator in Eq.~(\ref{eq19h}) puts constraints on the form of the effective theory amplitude. Since the form Eq.~(\ref{eq19h}) is identical to Eq.~(\ref{eq19}) when there are no massive particles, the results of Sec.~\ref{sec:fac} hold. The only difference is that for heavy quarks, the collinear functions are replaced by HQET functions
\begin{eqnarray}
 H_I &=&  J \left(\alpha(\mu),\lM,T_I^A\right)\openone \mathcal{Z}(I) + E_h
\left(\alpha(\mu),\lM,T_I^A\right) \nn
\label{eq26bh}
\end{eqnarray}
analogous to Eq.~(\ref{eq26b}), where $\mathcal{Z}(I)=\log \Delta_I^\prime/\mu=\log (\delta_I/\mu)+\log(\bar n_I \cdot v_I)/2$. $E_h$ does not depend on $m_I$, since $m_I$ is not a parameter in bHQET. $E_h$ is not the same function as for a massless quark, but $J$ is. Eq.~(\ref{eq:form}) becomes
\begin{eqnarray}
\log \bamp &=& \sum_{i=1}^r
J \left(\alpha(\mu),\lM,T_i^A\right)\openone \left[\mathcal{Y}(i)+\log 
\left(\frac{\bar n_i \cdot p_i}{\mu}\right)\right]\nn
&&+ \sum_{I=1}^s J \left(\alpha(\mu),\lM,T_I^A\right)\openone \left[\mathcal{Y}
(I)+\log \left(\frac{\bar n_I \cdot v_I}{2}\right)\right]\nn
&&+ \sum_{i=1}^r  E\left(\alpha(\mu),\lM,T_i^A,m_i/\mu\right)\openone \nn
&&+ \sum_{I=1}^s  E_h\left(\alpha(\mu),\lM,T_I^A\right)\openone \nn
&& + \mathbf{G}\left(\alpha(\mu),\lM,\left\{T^A\right\},\left\{\conf(ij)\right\}\right)
\label{eq:formh}
\end{eqnarray}
where $\bar n_i \cdot p_i \to \bar n_I \cdot v_I/2=\gamma_I$ because the relation between $\Delta_I^\prime$ and $\delta_I$ is now Eq.~(\ref{5hh}). The soft function $\mathbf{G}$ depends on color factors or cross-ratios involving all the particles, and any three particles (heavy or light) can be chosen as the reference indices in defining $\mathcal{Y}$ and $\mathcal{C}$.

Equation~(\ref{eq:formh}) is the general form for the SCET amplitude and anomalous dimension, and can be used for processes involving any combination of heavy and light particles.The change on replacing a light quark by a heavy quark is $\bar n \cdot p \to \bar n \cdot v$ and $E \to E_h$ in the collinear function. Thus the results of Secs.~\ref{sec:casimir} and \ref{sec:csfn} can also be used for heavy quarks with this substitution.

\subsection{Matching from SCET to bHQET}\label{sec:bhqetmatch}

In processes where $Q \gg m \gg M$, the external particle is treated as an SCET field for $Q > \mu > m$, and as a bHQET field for $m > \mu > M$, and we need to compute the matching correction at the scale $\mu=\mu_m \sim m$ between the two theories. This result is need, for example, to compute QCD corrections to $t \bar t$ production.

The difference between graphs in the theory above and below $\mu_m$ is in the treatment of $n$-collinear interactions involving the heavy particle. In the theory above $\mu_m$, they are given by $W_n^\dagger \xi_n$, and in the theory below $\mu_m$, by $W_n^\dagger h_v$. In the amplitude Eq.~(\ref{eq19h}), the difference is that  $I_i$ is replaced by the corresponding $H_i$ for each external heavy particle. Thus for each particle making the transition from SCET to bHQET, the matching factor is
\begin{eqnarray}
V_i &=& \openone\biggl[I_i \left(\alpha(\mu),\epsilon,T^A_i,\Delta_i/\mu^2 \to 0,m_i/\mu \right)\nn
&&- H_i\left(\alpha(\mu),\epsilon,T^A_i,\Delta_i^\prime/\mu \to 0 \right)\biggr]\,.
\label{eqmatch}
\end{eqnarray}
The computation in the theory above $\mu_m$ is done on-shell with $p_i^2=m_i^2$, and since $m_i \gg M$ one can set $M=0$ to compute the matching. The $\Delta$-regulator is not needed, and dimensional regularization is sufficient to regulate the infrared divergences. The difference $I_i-H_i$ is infrared finite.

The matching $V_i$ only depends on the collinear functions for light and heavy particles. The soft function remains unchanged at the SCET $\to$ bHQET transition. The one-loop matching conditions are given in Table~\ref{tab:match},
\begin{table}
\begin{eqnarray*}
\renewcommand{\arraystretch}{2.0} 
\begin{array}{c|c|}
\text{Field} & D \\
\hline
\psi & \frac{\alpha}{4\pi} \mathbf{T}\cdot \mathbf{T}\left[ \frac12 \lm^2-\frac12\lm+\frac{\pi^2}{12}+2 \right] \\
\phi &  \frac{\alpha}{4\pi} \mathbf{T}\cdot \mathbf{T}\left[ \frac12 \lm^2-\lm+\frac{\pi^2}{12}+2  \right] \\
B_\perp &   \frac{\alpha}{4\pi} \mathbf{T}\cdot \mathbf{T}\left[\frac12 \lm^2-\lm+\frac{\pi^2}{12}+2  \right] \\
\end{array}
\end{eqnarray*}
\caption{\label{tab:match} The SCET $\to$ bHQET matching correction for fermions, scalars, and transverse gauge bosons. $\lm=\log m^2/\mu^2$, where $m$ is the particle mass.}
\end{table}
and were obtained previously in Ref.~\cite{CGKM2}. Two-loop results for $t$-quarks can be found in Ref.~\cite{jain}.

\subsection{The Heavy-Light Sudakov Form Factor}\label{sec:hlff}

The heavy-light Sudakov form factor is the scattering of an incoming HQET field  (particle 1) to an outgoing collinear field (particle 2) by an external current, similar to  the $b \to s \gamma$ decay process studied in Ref.~\cite{BFL}. The result is given  using Eq.~(\ref{eq:formh}) for the case $r=s=1$. The anomalous dimension is
\begin{eqnarray}
 \bm{\gamma}(\mu) &=& \Gamma \left(\alpha(\mu),T^A\right) \left[
\log \frac{(\bar n_1 \cdot p_1)(\bar n_2 \cdot v_2)}{\mu}\right]\nn
&&+   \Omega\left(\alpha(\mu),T^A,s_1\right)+ \Omega\left(\alpha(\mu),T^A
,s_2\right)\nn
&& + \sigma\left(\alpha(\mu),T^A\right)\,,
\label{eq:srunhl}
\end{eqnarray}
where $s_{1,2}$ also encode whether the particle is heavy, and we have dropped the subscript on $T^A$ since $T^A_1+T^A_2=0$. Note that all heavy particles are the same, i.e.\ it does not matter if they are fermions, scalar, or gauge bosons. The low-scale matching is
\begin{eqnarray}
D &=& D_0 + D_1\log \frac{(\bar n_1 \cdot p_1)(\bar n_2 \cdot v_2)}{\mu}\,,\nn
D_1 &=&  J \left(\alpha(\mu),\lM,T^A\right)\,,\nn
D_0 &=&  E_{h}\left(\alpha(\mu),\lM,T^A\right)  +E\left(\alpha(\mu),\lM,T^A,m_2/\mu\right) \nn
&& + g \left(\alpha(\mu),\lM,\left\{T^A\right\}\right)\,.
\label{eq:slowDhl}
\end{eqnarray}

\subsection{The Heavy-Heavy Sudakov Form Factor}\label{sec:hhff}

The Sudakov form factor for two heavy external lines is
\begin{eqnarray}
 \bm{\gamma}(\mu) &=& \Gamma \left(\alpha(\mu),T^A\right) 
\log (2v_1 \cdot v_2)\nn
&&+   \Omega\left(\alpha(\mu),T^A,s_1\right)+ \Omega\left(\alpha(\mu),T^A
,s_2\right)\nn
&& + \sigma\left(\alpha(\mu),T^A\right)\,,
\label{eq:srunhh}
\end{eqnarray}
and the low-scale matrix element is
\begin{eqnarray}
D &=& D_0 + D_1\log (2v_1 \cdot v_2)\,, \nn
D_1 &=&  J \left(\alpha(\mu),\lM,T^A\right)\,,\nn
D_0 &=&  E_{h}\left(\alpha(\mu),\lM,T^A\right) +E_{h}\left(\alpha(\mu),\lM,T^A\right) \nn
&& + g \left(\alpha(\mu),\lM,\left\{T^A\right\}\right)\,.
\label{eq:slowDhh}
\end{eqnarray}

The heavy-light anomalous dimension and matching are the average of the heavy-heavy and light-light values.

\subsection{Form Factor Relations}

From the results in Secs.~\ref{sec:hlff}, \ref{sec:hhff}, we see that
\begin{eqnarray}
\frac{F_{h_1 \to h_2}(Q^2) F_{l_1 \to l_2}(Q^2)}{F_{h_1 \to l_2}(Q^2) F_{l_1 \to h_2}(Q^2)} &=& 1\,,
\label{eq:ffreln1}
\end{eqnarray}
where $l_{1,2}$ and $h_{1,2}$ are light and heavy particles, and $l_{1,2},h_{1,2}$ have the same color. This relation is satisfied by the EFT form factors. It is also satisfied by the full form factors in the limit $Q \gg m_{1,2},M$,
and
\begin{eqnarray}
\frac{F_{m_1 \to m_2}(Q^2) F_{m_3 \to m_4}(Q^2)}{F_{m_1 \to m_4}(Q^2) F_{m_3 \to m_2}(Q^2)} &=& 1 + \mathcal{O}\left( \frac{m^2}{Q^2}\right),
\label{eq:ffreln}
\end{eqnarray}
because the high scale matching condition is independent of particle masses, and cancels out, as does the matching between SCET and HQET for each external heavy line. Eq.~(\ref{eq:ffreln}) holds regardless of whether $Q$, $m$ and $M$ are widely separated, $Q \gg m \gg M$, or $m$ and $M$ are comparable. So it also holds for the massless gauge theory $M\to 0$ with the infrared divergences cancelling in the ratio.

\section{The Equivalence Theorem}\label{sec:equivthm}

The radiative corrections at high energy for longitudinally polarized massive gauge bosons can be most easily computed using the Goldstone boson equivalence theorem~\cite{cornwall,vayonakis,leequiggthacker,chanowitz,gounaris,bagger,yaoyuan,bohmbook}, which relates the gauge boson amplitude to that of the corresponding (eaten) Goldstone boson,
\begin{eqnarray}
&& \braket{A_L^{a_1} \ldots A_L^{a_n} F |S|A_L^{b_1} \ldots A_L^{b_m} I } \nn
&=& \left(-i \mathcal{E}\right)^n
\left(i \mathcal{E}\right)^m
\braket{\varphi^{a_1} \ldots \varphi^{a_n} F|\cdot | \varphi^{b_1} \ldots 
\varphi^{b_m}I}\nn
&&+ \mathcal{O}\left(\left\{\frac{M}{E_i}\right\}\right)
\label{equiv}
\end{eqnarray}
up to corrections which vanish in the high energy limit. The l.h.s.\ is the $S$-matrix for an initial state to scatter to a final state, and the r.h.s.\ is the corresponding amplitude (made more precise later) with the gauge bosons replaced by the corresponding (unphysical) Goldstone boson. $\mathcal{E}$ is a radiative correction factor which is one at tree-level, and $E_i$ are the energies of the gauge bosons. The existence of $\mathcal{E}$ was first pointed out by Bagger and Schmidt~\cite{bagger}. The equivalence theorem only holds for the $S$-matrix for physical states. \emph{All longitudinal gauge bosons $A_L^a$ must be simultaneously replaced by the corresponding $\varphi^a$ for the theorem to be valid.} Thus $I$ and $F$ do not contain any longitudinal gauge bosons.

We briefly review the derivation of the equivalence theorem including the radiative correction $\mathcal{E}$, and then give the result for $\mathcal{E}$ in the \msbar-scheme.

For a longitudinal gauge boson in the $\mathbf{n}$ direction with energy $E$, the momentum is $p^\mu=(E,p\, \mathbf{n})$ with $E^2=p^2+M_{\text{phys}}^2$, and the longitudinal polarization is chosen to be $\epsilon_L^\mu=(p, E\, \mathbf{n})/M_{\text{phys}}$, $\epsilon_L \cdot p=0$, $\epsilon_L^2=-1$, where $M_{\text{phys}}$ is the physical mass of the gauge boson. In the high energy limit $\epsilon_L^\mu \to p^\mu/M_{\text{phys}} + \mathcal{O}(M/E)$.

\begin{figure}
\begin{center}
\includegraphics[width=4cm]{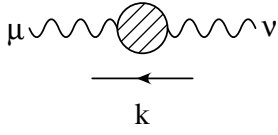}
\end{center}
\caption{Two-point function. \label{fig:prop}}
\end{figure}

The general form of the two-point function in the Goldstone boson/longitudinal gauge boson sector is
\begin{eqnarray}
i \left[ \begin{array}{cc} \Gamma_L \proj^L_{\mu \nu}+\Gamma_T \proj^T_{\mu \nu} 
& -i  k_\mu \Gamma^{A\varphi}  \\ i k_\nu\Gamma^{\varphi A} & \Gamma^{\varphi
\varphi}  \end{array}\right]
\label{eq91}
\end{eqnarray}
where the projection operators are
\begin{eqnarray}
\proj^L_{\mu \nu} &=& \frac{k_\mu k_\nu}{k^2}\,,\nn
\proj^T_{\mu \nu} &=& g_{\mu \nu}- \frac{k_\mu k_\nu}{k^2}\,.
\label{eq92}
\end{eqnarray}
The momentum $k$ flows into the blob through the line corresponding to the right  index of the matrix (see Fig.~\ref{fig:prop}).
%----FIGURE-----------------------------------------------------------------------------
%----END--FIGURE--------------------------------------------------------------------------
At tree-level,
\begin{eqnarray}
\Gamma_T &=& M^2-k^2\,,\nn
\Gamma_L &=& M^2-\frac{1}{\xi} k^2\,, \nn
\Gamma^{A\varphi} &=& \Gamma^{\varphi A} = M \left(1-\frac{\xi^\prime}{\xi}\right)\,,
\nn
\Gamma^{\varphi\varphi} &=&k^2-\frac{(\xi^\prime)^2}{\xi} M^2\,,
\label{eq:tree}
\end{eqnarray}
where $\xi$ and $\xi^\prime$ are the parameters in the gauge-fixing term, and $\xi=\xi^\prime$ in $R_\xi$ gauge at tree-level, and we have used the notation of Ref.~\cite{bohmbook}. The tree-level Goldstone boson mass is $M \xi^\prime/\sqrt{\xi}$.

The BRST Ward identity implies\footnote{See Ref.~\cite{bohmbook} for an excellent discussion of BRST Ward Identities and the Equivalence Theorem.}
\begin{eqnarray}
ik^\nu \vev{\underline{A_\nu} \cdots }   =  \frac{\Gamma_L +k^2/\xi } 
{\Gamma^{\varphi A} + M \xi^\prime/\xi }\vev{\underline{\varphi} \cdots }  
\label{eq68}
\end{eqnarray}
where $\underline{A_\nu}$ and $\underline{\varphi}$ are the truncated Green's function for emission of  a gauge boson and Goldstone boson with momentum $k$. This gives the equivalence theorem factor $C$
\begin{eqnarray}
C &=& \frac{1}{M} \frac{\Gamma_L +k^2/\xi } {\Gamma^{\varphi A} + M \xi^\prime/
\xi }\,.
\label{eqC}
\end{eqnarray}
Gauge bosons with polarization $k^\mu$ in truncated Green's functions are replaced by  $-i M C \varphi$.  Eq.~(\ref{eq:tree}) gives $C=1$ at tree-level.

The Ward identity Eq.~(\ref{eq68}) is exact. The scattering rates are given by the $S$-matrix, which is given by multiplying the  truncated Green's function evaluated at $k^2=M_{\text{phys}}^2$ by the wavefunction renormalization factor $\sqrt{\waver_A}$, and using the physical polarization $\epsilon_L^\mu$. $M/E$ corrections are introduced by replacing the polarization $\epsilon_L$ by the momentum, $\epsilon_L^\mu = k^\mu/M_{\text{phys}} + \delta \epsilon_L^\mu$, and then neglecting the $ \delta \epsilon_L^\mu$ term. The $k^\mu/M_{\text{phys}}$ term can be computed using the Ward identity:
\begin{eqnarray}
\braket{A_L \cdots | S | \cdots} &=& - i C \frac{M}{M_{\text{phys}}}  \waver_A ^{1/2} \left.
\vev{\underline{\varphi} \cdots }\right|_{k^2=M_{\text{phys}}^2}\,.\nn
\label{eq96}
\end{eqnarray}
The r.h.s.\ is evaluated at the physical gauge boson mass even in $R_\xi$ gauge where the Goldstone bosons have mass $M \xi^\prime/ \sqrt{\xi}$. We will write  Eq.~(\ref{eq96}) as
\begin{eqnarray}
\braket{A_L \cdots | S | \cdots} &=& - i \mathcal{E}\,
\waver_\varphi^{1/2}\, \left.\vev{\underline{\varphi} \cdots }\right|
_{k^2=M_{\text{phys}}^2}\nn
\mathcal{E} &=&  C\frac{M}{M_{\text{phys}}} \frac{\waver_A ^{1/2}}{\waver_\varphi^{1/2}}\,,
\label{eq97}
\end{eqnarray}
where $\waver_\varphi$ is defined as though it is the wavefunction renormalization factor 
for $\varphi$,
\begin{eqnarray}
\waver_\varphi^{-1} = \left.\frac{\rd \Gamma^{\varphi\varphi}}{\rd k^2}\right|_{k^2=k_0^2}\,,
\end{eqnarray}
where the derivative is taken at a fixed ($\mu$-independent) reference momentum. The quantity we need, the l.h.s.\ of Eq.~(\ref{eq97}) is independent of the choice of $\waver_\varphi$. $\waver_\varphi^{1/2} \left.\vev{\underline{\varphi} \cdots }\right|_{k^2=M_{\text{phys}}^2}$ is a fake $S$-matrix element for $\varphi$, and defines the quantity on the r.h.s.\ of Eq.~(\ref{equiv}).  It is computed like an $S$-matrix element, but it is gauge dependent because it is not the scattering amplitude for a physical state.\footnote{This is true at one-loop even if $\sqrt{\xi^\prime}/\xi=1$ so that the Goldstone boson and gauge boson masses are equal at tree-level.} The gauge dependence cancels that in $\mathcal{E}$, so that the left-hand side is gauge independent. We have checked this by computing both quantities in $R_\xi$ gauge. The l.h.s.\ of Eq.~(\ref{eq97}) is also $\mu$-independent. On the r.h.s.\ of Eq.~(\ref{eq97}), $\mathcal{E}$ and the $\varphi$ fake $S$-matrix element are both $\mu$-independent.

The results tabulated in Sec.~\ref{sec:results} give the $S$-matrix element for fermions and transverse gauge bosons. For longitudinal gauge bosons, we list the expression for the Goldstone boson $\varphi$, which is why we have included $\waver_\varphi^{1/2}$ so that it behaves as much as possible like an $S$-matrix element. The results in Table~\ref{tab:massless} are given in $R_{\xi=1}$ gauge, and $\waver_\varphi$ is evaluated at the reference momentum $k_0^2=M^2$, which is the tree-level gauge and Goldstone boson mass. 

The one-loop results we will need are:
\begin{eqnarray}
\frac{M_{\rm phys}}{M} 
&=& 1 + \frac{g^2}{16\pi^2} \biggl[ - \left(\frac{3z^2}{8}+\frac{59}{24}+\frac{9}{4 
z^2}-\frac13 n_F\right) \log \frac{M^2}{\mu ^2}\nn
&& -\frac{z^4}{24}+\frac{5 z^2}{8} +\frac{3}{4 z^2} -\frac{11 \pi\sqrt{3}  }{8}+
\frac{139}{18}-\frac{5}{9}n_F\nn
&& +\left(\frac{z^6}{48}-\frac{z^4}{8}\right) \log \left(z^2\right)  \nn
&& +\left(\frac{z^5}{12}-\frac{z^3}{3}+z\right)  \sqrt{4 - z^2 }  \tan^{-1} \sqrt{ \frac{2-
z}{2+z}  } \biggr] \nn
\end{eqnarray}
%-----------------------------------------------------------------------
where  $z = M_h/M$.  This expression, and the ones below can be analytically continued to $z>2$ using
\begin{eqnarray}
\sqrt{w}\  \tan^{-1} \sqrt{w} \leftrightarrow - \sqrt{-w} \ \tanh^{-1} \sqrt{-w }\nn
\frac{1}{\sqrt{w}}\  \tan^{-1} \sqrt{w} \leftrightarrow \frac{1}{ \sqrt{-w}}\  \tanh^{-1} \sqrt{-w }\,.
\label{continue}
\end{eqnarray}

The expression for $\mathcal{E}$ in 't~Hooft-Feynman gauge in the $\overline{\text{MS}}$ scheme is
\begin{eqnarray}
\label{E}
\mathcal{E}
&=& 1 + \frac{g^2}{16\pi^2} \biggl[ -\frac{z^2}{4}+\frac{47  \pi }{12\sqrt{3}}-\frac{73}
{12} + \frac13 n_F\nn
&&  +\left( \frac{z^4}{4} - \frac{7z^2}{8}+\frac{5}{8}\right) \log z \nn
&& +  \left(-\frac{z^4}{2}+\frac{11z^2}{4}-\frac{15}{4}\right) \frac{z}{\sqrt{4 - z^2 }}  
\tan^{-1} \sqrt{ \frac{2-z}{2+z}  } \biggr].\nn
\end{eqnarray}
Notice the expression is $\mu$ independent as discussed earlier.

\section{Results}\label{sec:results}

The results needed for a NLL analysis of hard scattering  processes in the $SU(2)$ theory are summarized here. The computations for external (massless and massive) fermion and scalar states were already discussed in detail in Refs.~\cite{CGKM1,CGKM2,CKM}. They were presented previously using the sum-on-pairs form discussed in Sec.~\ref{sec:sumonpairs}. They are tabulated here using the collinear+soft function form discussed in Sec.~\ref{sec:csfn}. The new computations are those for external gauge  bosons and Higgs bosons.   The standard model results can be easily derived using the results presented here; however, the expressions are considerably more involved than those for the $SU(2)$ theory because custodial $SU(2)$ is no longer a symmetry. Thus one has to include the effects of $\gamma-Z$ mixing, $M_W \not = M_Z$, and the $b-t$ mass difference. For this reason, the standard model results needed to compute \emph{all} standard model hard scattering processes are given in a companion paper~\cite{p2}.

The results for external transverse gauge bosons were computed by matching onto SCET operators where the gauge field is given by the SCET $B_\mu$ field,
%-----------------------------------------------------------------------
\begin{equation}
\label{Bop}
B^\mu_{n, p}  =  \frac{1}{g} [ W_n^\dagger iD_{n}^\mu W_n ] , \qquad
iD_n^\mu  =  i \partial_n + g A_{n,p},
\end{equation}
%-----------------------------------------------------------------------
The field $B_\mu$ does not produce longitudinal gauge bosons to leading order in the SCET power counting. The amplitude for longitudinal gauge bosons is computed using the Goldstone boson equivalence theorem discussed in Sec.~\ref{sec:equivthm}.

\subsection{Collinear functions for the $SU(2)$ theory}\label{sec:collsm}

\begin{figure}
\begin{center}
\includegraphics[width=3cm]{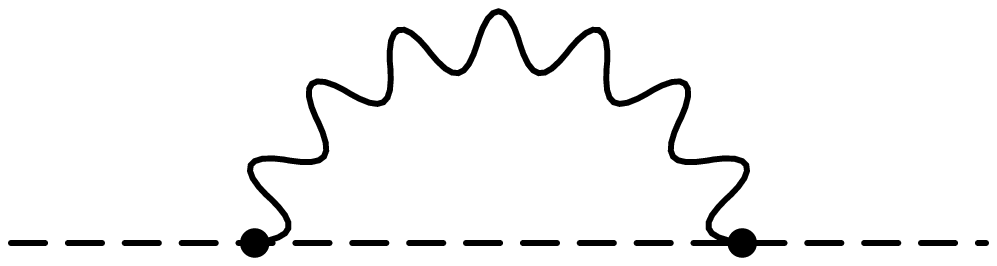} \\
\vspace{1cm}
\includegraphics[width=3cm]{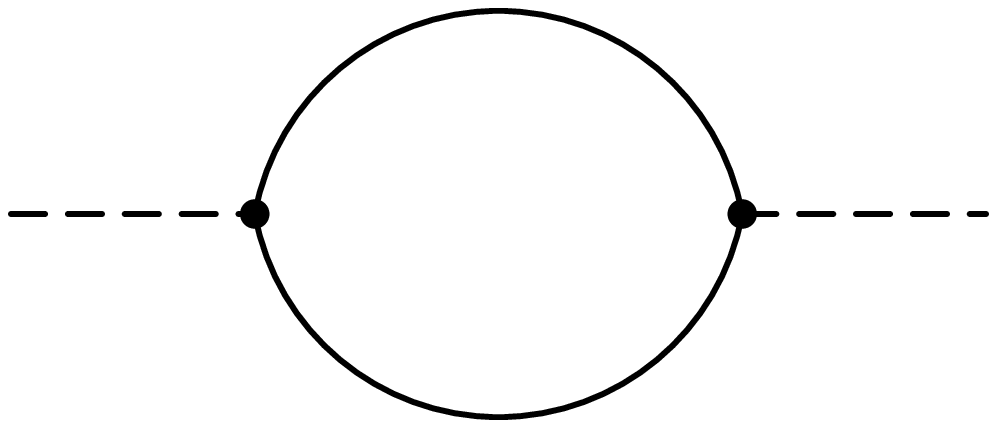} 
\end{center}
\caption{\label{fig:wave} The upper graph shows the topology of wavefunction corrections included in Table~\ref{tab:massless}.  Topologies such as those in the lower graph are not included.}
\end{figure}

\begin{table*}
\begin{eqnarray*}
\renewcommand{\arraystretch}{1.5} 
\begin{array}{c|c|c|}
\text{Field} & \gamma_C & D_C \\
\hline
\psi & \frac{\alpha }{4\pi}\mathbf{T}\cdot\mathbf{T}\left[ 4 \lp - 3\right] & \frac{\alpha }{4\pi}\mathbf{T}\cdot\mathbf{T}\left[2 \lM \lp - \frac12 \lM^2-\frac32 \lM - \frac{5\pi^2}{12} + 
\frac94\right]\\
\phi &  \frac{\alpha }{4\pi}\mathbf{T}\cdot\mathbf{T}\left[ 4\lp - 4 \right] &  \frac{\alpha }{4\pi}\mathbf{T}\cdot\mathbf{T}\left[ 2 \lM \lp - \frac12 \lM^2-2 \lM - \frac{5\pi^2}{12} + \frac74 \right] \\
h_v & \frac{\alpha }{4\pi}\mathbf{T}\cdot\mathbf{T}\left[ 4\log (2\gamma) -2 \right] & \frac{\alpha }{4\pi}\mathbf{T}\cdot\mathbf{T}\left[ 2\lM \log2\gamma-\lM \right] \\
\hline
\psi & \frac{\alpha }{4\pi}\mathbf{T}\cdot\mathbf{T}\left[ 4 \lp - 4\right] +\gamma_\psi & \frac{\alpha }{4\pi}\mathbf{T}\cdot\mathbf{T}\left[2 \lM \lp - \frac12 \lM^2-2 \lM - \frac{5\pi^2}{12} + 
2\right] +\frac12 \delta \waver_\psi\\
\phi &  \frac{\alpha }{4\pi}\mathbf{T}\cdot\mathbf{T}\left[ 4\lp - 2 \right] +\gamma_\phi &  \frac{\alpha }{4\pi}\mathbf{T}\cdot\mathbf{T}\left[ 2 \lM \lp - \frac12 \lM^2- \lM - \frac{5\pi^2}{12} + 1\right] +\frac12 \delta \waver_\phi\\
h_v & \frac{\alpha }{4\pi}\mathbf{T}\cdot\mathbf{T}\left[ 4\log (2\gamma) \right]+\gamma_h & \frac{\alpha }{4\pi}\mathbf{T}\cdot\mathbf{T}\left[ 2\lM \log2\gamma \right] +\frac12 \delta \waver_{h_v}\\
%\hline
W_\perp  &  \frac{\alpha }{4\pi}\mathbf{T}\cdot\mathbf{T}\left[4 \lp - 2\right]+\gamma_W &   \frac{\alpha }{4\pi}\mathbf{T}\cdot\mathbf{T}\left[ 2 \lM \lp - \frac12 \lM^2- \lM - \frac{5\pi^2}{12} + 
1 +
f_S(1,1)\right] +\frac12 \delta \waver_W\\
H &  \frac{\alpha }{4\pi}\mathbf{T}\cdot\mathbf{T}\left[4 \lp - 2\right] + \gamma_H &  \frac{\alpha }{4\pi}\mathbf{T}\cdot\mathbf{T}\left[2 \lM \lp - \frac12 \lM^2- \lM - \frac{5\pi^2}{12} + 1
+f_S(m_h^2/M^2,1)\right]+\frac12 \waver_H\\
\varphi^a &  \frac{\alpha }{4\pi}\mathbf{T}\cdot\mathbf{T}\left[4 \lp-2\right]+\gamma_\varphi &  \frac{\alpha }{4\pi}\mathbf{T}\cdot\mathbf{T}\left[2 \lM \lp - \frac12 \lM^2- \lM - \frac{5\pi^2}{12} 
+ 1
+\frac23 f_S(1,1)+\frac13 f_S(1,m_h^2/M^2)\right]+\frac12 \delta \waver_\varphi\\
\hline
\end{array}
\end{eqnarray*}
\caption{\label{tab:massless} The collinear anomalous dimension and low-scale matching.  $\lM=\log(M^2/\mu^2)$, $\lp=\log (\bar n \cdot p)/\mu$, and $\gamma=E/m$. The rows are $\psi$: fermion, $\phi$ non-Higgs scalar multiplet, $h_v$ HQET field, $B_\perp$: transverse gauge boson, $h$: Higgs, $\varphi^a$: Goldstone bosons (i.e.\ longitudinal gauge bosons using the equivalence theorem and multiplying by $\mathcal{E}$). The results are in $R_{\xi=1}$ gauge. $\gamma_{W,h,\varphi}$ and $\waver_{W,h,\varphi}$ are the wavefunction contributions.}
\end{table*}
The one-loop collinear functions are summarized in Table~\ref{tab:massless}.  The function $f_{S}$ is given in Appendix~B of Ref.~\cite{CKM}. The $\psi$ row is for fermions, $\phi$ for scalars, $W_\perp$ for external  transversely polarized gauge bosons, $H$ for the physical Higgs field, and $\varphi^a$ for the unphysical Goldstone bosons, which are used to compute longitudinally polarized gauge bosons using the equivalence theorem. The upper section of the table gives the results for fermions, scalars and bHQET fields, including wavefunction renormalization, assuming that the only wavefunction graphs are gauge boson corrections to the propagator with the topology shown in the upper graph in Fig.~\ref{fig:wave}. 
%%%----FIGURE--------------------------------------------------------------------------------------
%%
%------------------------------------------------------------------------------------------------------
The lower section of the table gives the results for fermions, scalars, bHQET fields, transverse gauge bosons, the Higgs, and unphysical Goldstone bosons, where the wavefunction contribution has been separated out explicitly. $\gamma_\phi$, etc.\ are the anomalous dimensions of the fields and $\delta\waver$ is the finite wavefunction factor needed in the $\overline{\text{MS}}$ scheme, from the residue of the pole in the two-point Green's function function, $(1+\delta \waver)/(p^2-m^2)$.

\subsubsection{Running from $\mu_h$ to $\mu_l \sim M_Z$}

The collinear anomalous dimensions are given using the $\gamma_C$ column in Table~\ref{tab:massless}. 

For the fermions $\psi$:
\begin{eqnarray}
\frac{\alpha}{4\pi}C_F\left(4 \log \frac{\bar n \cdot p}{\mu} - 3 \right)
\end{eqnarray}
with $C_F=3/4$.

The gauge field anomalous dimension at one-loop in $R_{\xi=1}$ gauge is
\begin{eqnarray}
\gamma_W &=& \frac{\alpha }{4\pi}\left(2 C_A - b_0\right)
\end{eqnarray}
where $b_0$ is the coefficient of the first term in the $\beta$-function,
\begin{eqnarray}
\mu \frac{\rd g}{\rd \mu} &=& -b_0 \frac{g^3}{16\pi^2}+\ldots
\end{eqnarray}
so that the collinear factor $\gamma_C$ for transverse gauge bosons is

$W_T$ (transverse $W^{1,2,3}$):
\begin{eqnarray}
\frac{\alpha }{4\pi}\left(4 C_A \lp - b_0\right)\,.
\label{131}
\end{eqnarray}
with $b_0=11/3 C_A-2/3 n_F$ and $C_A=2$. The final expression is gauge invariant. There is a factor of the coupling constant $g(\mu_h)$ in the high-scale matching for each gauge boson. The $-b_0$ term in Eq.~(\ref{131}) converts that to $g(\mu_l)$. The $b_0$ term is absent in the expression for $\zeta$ in Ref.~\cite{kuhnW}.

The unphysical Goldstone bosons and Higgs are renormalized together as elements of the scalar doublet $\phi$,\\
$\varphi, H$:
\begin{eqnarray}
\frac{\alpha}{4\pi} C_F \left(4 \log \frac{\bar n \cdot p}{\mu} -4\right)
\end{eqnarray}
with $C_F=3/4$.

\subsubsection{Matching at $\mu_l \sim M_Z$}\label{sec:matchZ}

The matching corrections at $\mu_l \sim M_Z$ have to be computed in the broken electroweak theory, using Table~\ref{tab:massless}. The collinear matching $D_C$ depends on the mass spectrum of the broken theory. The low-scale matching for the fermions is:
\begin{eqnarray}
\left[W^\dagger L_L\right] & \rightarrow & \left[ \begin{array}{c} 
\exp D_C^{(L)}\ \nu_L \\[5pt]
\exp D_C^{(L)}\ E_L \\
\end{array}\right]\,.
\end{eqnarray}
%%%----FIGURE--------------------------------------------------------------------------------------
\begin{figure}
\begin{center}
\includegraphics[width=4cm]{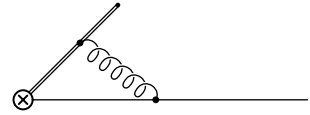} 
\end{center}
\caption{\label{fig:colmatching}Collinear matching graphs for $[W^\dagger \psi]$. The $\otimes$ is the $[W^\dagger \psi]$ operator, the solid line is $\psi$ and the double line is $W^\dagger$.}
\end{figure}
%%
%------------------------------------------------------------------------------------------------------
One can think of $[W^\dagger L]$ as a single collinear object which is matched in the effective theory onto the fields $\nu_L$ and $E_L$ (see Fig.~\ref{fig:colmatching}). Custodial $SU(2)$ invariance implies that $\nu_L$ and $E_L$ get the same collinear matching. In the standard model, custodial $SU(2)$ is broken, and the corresponding expression reads
\begin{eqnarray}
\left[W^\dagger_{\text{EW}} L_L\right] & \rightarrow & \left[ \begin{array}{c} 
\exp D_C^{(L \to \nu )}\ \left[W^\dagger_\gamma \nu_L\right]  \\[5pt]
\exp D_C^{(L \to E)}\ \left[W^\dagger_\gamma E_L\right]  \\
\end{array}\right]\,.
\label{134}
\end{eqnarray}
The collinear Wilson-line $W^\dagger_{\text{EW}} $ on the l.h.s.\ contains $W$ and $B$ fields, whereas on the r.h.s.\ $W^\dagger_\gamma $ contains only photon fields. The two $D_C$ functions in Eq.~(\ref{134}) are no longer equal. One also has different collinear functions $D_C$ for the left and right-handed quarks, top quarks, etc. For this reason, the complete standard model expressions are deferred to Ref.~\cite{p2}.

The collinear matching for the scalar doublet is:
\begin{eqnarray}
\left[W^\dagger \phi\right] & \rightarrow & \left[ \begin{array}{c} 
\exp D_C^{(\phi \to \varphi)}\ \varphi^+ \\[5pt]
\frac{1}{\sqrt 2} \exp D_C^{(\phi \to H)}\ H-\frac{i}{\sqrt 2} 
\exp D_C^{(\phi \to \varphi)}\ \varphi^3 \\
\end{array}\right]\nn
\end{eqnarray}
where $\varphi^+,\varphi^3$ have the same collinear matching by custodial-$SU(2)$ invariance. There is an interesting subtlety in the scalar doublet matching for the standard model, which is discussed in Ref.~\cite{p2}.
 
The matching for the transverse gauge bosons is:
\begin{eqnarray}
\left[W^\dagger W^a_\perp \right] & \to & \exp D_C^{(W \to W)} W^a_\perp\,.
\end{eqnarray}
Again, custodial $SU(2)$ ensures that all $W^a$ have the same collinear matching, and we do not have to worry about $W^3-B$ mixing.

The $D_C$ functions are:
\begin{eqnarray}
D_C^{(L)}&=&\frac{\alpha}{4\pi} C_F \left[ F_W -\frac12 \log \frac{M_W^2}{\mu^2}+\frac54 \right]\,,\nn
D_C^{(\phi \to \varphi)} &=& 
\frac{\alpha}{4\pi} \Biggl[\frac34 F_W +\frac12 f_S\left(1,1\right)+\frac14  f_S\left(1,\frac{M_H^2}{M_W^2}\right) \Biggr]\nn
&&+\frac12 \delta \waver_{\varphi}\,,\nn
D_C^{(\phi \to H)} &=& 
\frac{3\alpha}{4\pi}\left[F_W + f_S\left(\frac{M_H^2}{M_W^2},1\right)\right]+\frac12 \delta \waver_{H}\,,\nn
D_C^{(W \to W)} &=& \frac{\alpha_W}{4\pi}\left[2F_W +2 f_S\left(1,1\right)\right]+\frac12 \delta \waver_{W}\,,\nn
\label{87}
\end{eqnarray}
where
\begin{eqnarray}
F_W &=& 2\log \frac{M_W^2}{\mu^2} \log \frac{\bar n \cdot p}{\mu}- \frac12\log^2 \frac{M_W^2}{\mu^2}\nn
&&\qquad -  \log \frac{M_W^2}{\mu^2}-\frac{5\pi^2}{12}+1\,.
\end{eqnarray}

The wavefunction factors $\waver_{\varphi,H,W}$ are those for the standard model with $\sin^2\theta_W \to 0$, and can be found in Ref.~\cite{bohmbook,fleischer,hollik}. The $\lambda \left(\phi^\dagger \phi\right)^2$ coupling enters the calculation through the $m_H$ dependence of the wavefunction factors.

\subsection{Universal Soft Function}\label{sec:universal}

The universal soft functions is
\begin{eqnarray}
U_S(n_i,n_j) &=& \log\frac{-n_i \cdot n_j-i0^+}2
\label{eq:U}
\end{eqnarray}
in terms of which, the soft anomalous dimension and low-scale matching are
\begin{eqnarray}
\bm{\gamma}_S &=& \Gamma(\alpha(\mu))\left[- \sum_{\vev{ij}} \mathbf{T}_i \cdot 
\mathbf{T}_j \
U_S(n_i,n_j)\right]\nn
\mathbf{D}_S &=& J(\alpha(\mu),\lM)\left[- \sum_{\vev{ij}}  \mathbf{T}_i \cdot 
\mathbf{T}_j \ U_S(n_i,n_j)\right]\,.
\label{152}
\end{eqnarray}
For the Sudakov problem, there are only two particles, and $-n_1 \cdot n_2=2$, so that $U_S=0$, and the soft contribution vanishes. The Sudakov computation is thus entirely given by the sum of the collinear functions for the two external legs.

The soft function has a  universal form when written in the operator form Eq.~(\ref{152}). For numerical computations, it is more convenient to choose a basis of gauge invariant operators, and write the soft-anomalous dimension and matching as a matrix in the chosen basis. We compute the soft factor $\sum_{\vev{ij}}\mathbf{T}_i \cdot \mathbf{T}_j U_S(n_i,n_j)$ for some simple cases below for an $SU(N)$ gauge theory, to give the well-known results for the soft anomalous dimension for QCD~\cite{kidonakis}. We need certain standard soft matrices which will then be used to compute the soft-running in the broken gauge theory.

\subsubsection{Anomalous Dimension for $q\bar q \to q\bar q$}

\begin{figure}
\begin{center}
\includegraphics[width=4cm]{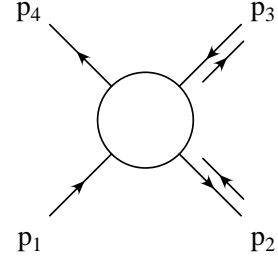}
\end{center}
\caption{\label{FgC1} Pair production $q(p_1) + \bar{q}(p_2) \to  q^\prime(p_4)+\bar{q}^\prime(p_3)$. Time runs vertically. This also defines the variable for $q + \bar q \to g +g $ if particles $3,4$ are replaced by gauge bosons. }
\end{figure}

As a sample computation, consider the soft function for $q \bar q \to q \bar q$ for fermions in the fundamental representation of an $SU(N)$ gauge theory. The kinematics for $q\bar q\to q^\prime  \bar q^\prime$ is illustrated schematically in Fig.~\ref{FgC1} where the incoming and outgoing particles have momenta $p_1$, $p_2$  and $p_3$, $p_4$, respectively, and we work in the limit $s,t,u \gg M^2 \gg m_i^2$. The external particles are all on-shell $(p_i^2$ = $m^2_i)$.  The Mandelstam variables are $s=(p_1+p_2)^2$, $t=(p_4-p_1)^2$ and $u=(p_3-p_1)^2$. 
%%%----FIGURE--------------------------------------------------------------------------------------
%%%------------------------------------------------------------------------------------------------------

An operator basis for gauge invariant operators is
\begin{eqnarray}
O_1 &=& \bar \psi_4 T^a \psi_3\, \bar \psi_2 T^a \psi_1 = T^a \otimes T^a \nn
O_2 &=&  \bar \psi_4  \psi_3 \, \bar \psi_2  \psi_1 = 1 \otimes 1
\label{153}
\end{eqnarray}
where only the group theory structure has been shown.

The group theory factors can be simplified using
%----------------------------------------------------------------
\begin{eqnarray}
T^a T^a &=& C_F\ 1 \nn
T^a T^b T^a &=& \left(C_F-\frac12 C_A\right) T^b \nn
T^a T^b \otimes T^a T^b &=& C_1\ 1 \otimes 1 +\frac14\left(C_d-C_A\right)T^a \otimes T^a\nn
T^a T^b \otimes T^b T^a &=& C_1\ 1 \otimes 1 +\frac14\left(C_d+C_A\right)T^a \otimes T^a\nn
\label{31}
\end{eqnarray}
%----------------------------------------------------------------
in the notation of Ref.~\cite{Manohar:2000hj}. The group invariants are listed in Table~\ref{tab:casimir}.
\begin{table}
\begin{eqnarray*}
\renewcommand{\arraystretch}{1.5} 
\begin{array}{c|c|c|c}
\text{Invariant} & SU(2) & SU(3) & SU(N) \\
\hline
C_A & 2 & 3 & N \\
C_F & \frac{3}{4} & \frac{4}{3} & \frac{N^2-1}{2N} \\
C_d & 0 & \frac{5}{3} &  \frac{N^2-4}{N}\\
C_1 & \frac{3}{16} & \frac{2}{9} &  \frac{N^2-1}{4N^2}\\
\end{array}
\end{eqnarray*}
\caption{\label{tab:casimir} Group theory invariants for $SU(2)$, $SU(3)$ and $SU(N)$.}
\end{table}

For scattering kinematics, $s>0$, $t<0$, and $u<0$, and the variables $T,U$ are defined by~\cite{kidonakis}
\begin{eqnarray}
T &=& \log \frac{-t}{s} + i \pi\,,\nn
U &=& \log \frac{-u}{s} + i \pi\,.
\end{eqnarray}

The object that enters the soft running and matching is ($U_{ij}\equiv U_S(n_i,n_j)$):
\begin{eqnarray}
&&- \sum_{\vev{ij}} \mathbf{T}_i \cdot 
\mathbf{T}_j\ U_S(n_i,n_j)\nn
&=&C_F\left[U_{12}+U_{34}\right] \openone+ \left[ \begin{array}{cc} 
\frac14 C_d r_1+\frac14C_A r_2  & r_1 \\
C_1r_1 
 & 0
 \end{array} \right]\nn
 r_1 &=&  \left[ U_{14}+U_{23}\right]-\left[ U_{13}+U_{24}\right]\nn
 r_2 &=&\left[ U_{14}+U_{23}\right]+\left[ U_{13}+U_{24}\right] -2\left[U_{12}+U_{34}\right]\nn
\end{eqnarray}
where
\begin{eqnarray}
U_{12} &=& U_{34} =-i\pi\nn
U_{13} &=& U_{24} =\log \frac{-u}{s}=U-i\pi\nn
U_{14} &=& U_{23} =\log \frac{-t}{s}=T-i\pi\nn
 r_1 &=&  2(T-U)\nn
 r_2 &=& 2(T+U)
\end{eqnarray}
so that
\begin{eqnarray}
&&- \sum_{\vev{ij}} \mathbf{T}_i \cdot 
\mathbf{T}_j\ U_S(n_i,n_j)\nn
&=&-2i\pi C_F \openone+ \left[ \begin{array}{cc} 
\frac12 C_d (T-U) +\frac12C_A (T+U)  & 2(T-U) \\
2C_1 (T-U)
 & 0
 \end{array} \right]\nn
 \label{157}
\end{eqnarray}
which agrees with VA and VC of Ref.~\cite{CKM}, and Ref.~\cite{kidonakis}

We will need the soft matrix Eq.~(\ref{157}) when the fields are in the fundamental representation of $SU(3)$,
\begin{eqnarray}
\softm^{(3)}&=&- \sum_{\vev{ij}} \mathbf{T}_i \cdot 
\mathbf{T}_j\ U_S(n_i,n_j)\nn
&=&-\frac{8}{3}i\pi  \openone+ \left[ \begin{array}{cc} 
\frac73 T +\frac23 U & 2(T-U) \\
\frac{4}{9} (T-U)
 & 0
 \end{array} \right]
 \label{m3}
\end{eqnarray}
and when they are in the fundamental representation of $SU(2)$,
\begin{eqnarray}
\softm^{(2)}&=&- \sum_{\vev{ij}} \mathbf{t}_i \cdot 
\mathbf{t}_j\ U_S(n_i,n_j)\nn
&=&-\frac32i\pi \openone+ \left[ \begin{array}{cc} 
 (T+U)  & 2(T-U) \\
\frac38 (T-U)
 & 0
 \end{array} \right]\,.
 \label{m2}
\end{eqnarray}

If two of the fermions (say $1,2$) are in the fundamental of $SU(N)$, and the other two are singlets, then the only allowed operator has the form $\mathbf{1} \otimes \mathbf{1}$, and the soft factor is
\begin{eqnarray}
- \sum_{\vev{ij}} \mathbf{T}_i \cdot 
\mathbf{T}_j\ U_S(n_i,n_j)
&=&-i\pi C_F 
\end{eqnarray}
with values
\begin{eqnarray}
\softm^{(3)\, \prime} = -\frac43 i \pi,\qquad
\softm^{(2)\, \prime} = -\frac34 i \pi,
\end{eqnarray}
for $SU(3)$ and $SU(2)$, respectively. If all four fields are singlets, the soft factor vanishes.

For a $U(1)$ gauge theory:
\begin{eqnarray}
\softm^{(1)}(q_1,q_2,q_3,q_4)&=&- \sum_{\vev{ij}} \mathbf{T}_i \cdot 
\mathbf{T}_j\ U_S(n_i,n_j)\nn
&=& 
q_1 q_2 U_{12}+q_3 q_4 U_{34}+q_1 q_4 U_{14}\nn
&&+q_2 q_3 U_{23}- q_1 q_3 U_{13}-q_2 q_4 U_{24}\nn
&=& 
 -i  \frac{\pi} 2\left(q_1^2+q_2^2+q_3^2+q_4^2 \right)\nn
 &&+ \left(q_1 q_4 +q_2 q_3\right) T -\left( q_1 q_3+q_2 q_4 \right) U \nn
\label{158soft}
\end{eqnarray}
which agrees with VIIB of Ref.~\cite{CKM} and VIIC. In most cases, we need this result for  $q_1=q_2=q_i$ and $q_3=q_4=q_f$,
\begin{eqnarray}
\softm^{(1)}(q_f,q_i)&=&- \sum_{\vev{ij}} q_i \cdot 
q_j\ U_S(n_i,n_j)\nn
&=& 
 -i  \pi \left(q_i^2+q_f^2 \right)+ 2 q_i q_f \left(T-U \right)\,.\nn
\label{m1}
\end{eqnarray}

\subsubsection{Anomalous Dimension for $qq \to qq$}

\begin{figure}
\begin{center}
\includegraphics[width=4cm]{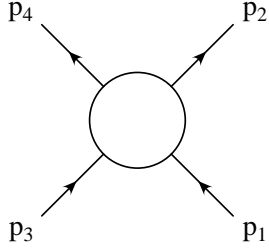}
\end{center}
\caption{\label{FgA0} Quark scattering $q(p_1)+q^\prime(p_3) \to q(p_2) + q^\prime(p_4)$. Time runs vertically. This also defines the variable for $q + g \to q +g $ if particles $3,4$ are replaced by gauge bosons.}
\end{figure}

The scattering kinematics is given in Fig.~\ref{FgA0}, and the kinematic variables are defined by
$s=(p_1+p_2)^2$, $t=(p_2-p_1)^2$ and $u=(p_4-p_1)^2$. 
For scattering, $ q q \to qq$, the soft matrix in the $t$-channel basis Eq.~(\ref{153}) becomes
%%%-----FIGURE--------------------------------------------------------------------------------------
%%%------------------------------------------------------------------------------------------------------
\begin{eqnarray}
&&- \sum_{\vev{ij}} \mathbf{T}_i \cdot 
\mathbf{T}_j\ U_S(n_i,n_j)\nn
&=& 2 C_F(T-i\pi) \openone+ \left[ \begin{array}{cc} 
\frac12 (C_d+C_A) U-C_A T  & 2U \\
2C_1 U
 & 0
 \end{array} \right]\nn
\end{eqnarray}
which agrees with VB of Ref.~\cite{CKM}.

\subsubsection{Anomalous Dimension for Heavy quark pair production}

The soft matrix is identical to that for $q \bar q \to q \bar q$, and
agrees with VC of Ref.~\cite{CKM}.

\subsubsection{Anomalous Dimension for $q \bar q \to gg$}

The scattering kinematics is given by Fig.~\ref{FgC1} with $s=(p_1+p_2)^2$, $t=(p_4-p_1)^2$ and $u=(p_3-p_1)^2$

The operator basis for $q \bar q \to gg $ is
\begin{eqnarray}
O_1 &=& 1 \otimes \delta^{AB} = \bar q_2 q_1 A_4^A A_3^A\nn
O_2 &=& T^C \otimes d^{ABC}=  d^{ABC} \bar q_2T^C q_1 A_4^A A_3^B\nn
O_3 &=& T^C \otimes i f^{ABC}=if^{ABC} \bar q_2 T^C q_1 A_4^A A_3^B\,.
\label{162}
\end{eqnarray}

The soft operator in this basis is
\begin{eqnarray}
&& -\sum_{\vev{ij}} \mathbf{T}_i \cdot 
\mathbf{T}_j\ U_S(n_i,n_j)\nn
&=&-i\pi \left(C_F+C_A\right) \openone\nn
&&+\left[ \begin{array}{ccc}
0 & 0 & U-T \\
0 & \frac12C_A(T+U) & \frac12 C_A (U-T)\\ 
2  (U-T)  & \frac12 C_d (U-T) & \frac12 C_A(T+U)\\ 
 \end{array} \right]\nn
\end{eqnarray}
which is the result of Ref.~\cite{kidonakis}.

For QCD, this gives the soft matrix
\begin{eqnarray}
\softm^{(3,g)}&=& -\frac{13}{3}i\pi  \openone\nn
&&+\left[ \begin{array}{ccc}
0 & 0 & U-T \\
0 & \frac32(T+U) & \frac32 (U-T)\\ 
2  (U-T)  & \frac56  (U-T) & \frac32(T+U)\\ 
 \end{array} \right]\,.\nn
\end{eqnarray}

For $SU(2)$, the $d$-symbol vanishes, so the operator basis reduces to
\begin{eqnarray}
O_1 &=& 1 \otimes \delta^{ab} = \bar q_2 q_1 W_4^a W_3^a\nn
O_2 &=& t^c \otimes i \epsilon^{abc}=i\epsilon^{abc} \bar q_2 t^c q_1 W_4^a W_3^b\,.
\label{162a}
\end{eqnarray}
and the soft matrix is given by using the $SU(2)$ values for the group constants, and dropping
the second row and column,
\begin{eqnarray}
\softm^{(2,g)}&=&-\frac{11}{4}i\pi \openone+\left[ \begin{array}{ccc}
0  & U-T \\
2  (U-T) & (T+U)\\ 
 \end{array} \right]\,.
 \label{158}
\end{eqnarray}

\subsubsection{Matching for Fermion Scattering}

The soft contribution to the low-scale matching can be computed using the above results. Consider, first the case of a four-quark operator, where all the fields are $SU(2)$ doublets, for example
\begin{eqnarray}
&& C_{1} \bar \psi^{(\mu)}_4 t^a \gamma^\alpha P_L \psi^{(\mu)}_3 \bar  \psi^{(e)}_2 t^a \gamma_\alpha P_L \psi^{(e)}_1\nn
&&+C_{2} \bar \psi^{(\mu)}_4  \gamma^\alpha P_L \psi^{(\mu)}_3  \bar \psi^{(e)}_2 \gamma_\alpha P_L \psi^{(e)}_1
\label{85}
\end{eqnarray}
where the index is $1$ for $t^a \otimes t^a$ and $2$ for $\mathbf{1} \otimes \mathbf{1}$ in $SU(2)$,

The group theory sum needed for the soft anomalous dimension matrix is
\begin{eqnarray}
\softm_2&=& -\sum_{\vev{ij}} \mathbf{t}_i \cdot 
\mathbf{t}_j\ U_S(n_i,n_j)= \softm^{(2)} 
 \label{165}
 \end{eqnarray}
 where $\softm^{(2)}$ is the standard form Eq.~(\ref{m2}).
The soft anomalous dimension is
\begin{eqnarray}
\gamma_S &=& \frac{\alpha}{ \pi}\softm_2\,.
\label{eq172}
\end{eqnarray}
At the low scale $\mu_l \sim m_Z$, the operators Eq.~(\ref{85}) match onto a linear combination of
\begin{eqnarray}
\widehat{\mathcal{O}}_{12} &=& [\bar{\nu_\mu}_{L4}  \gamma_\alpha {\nu_\mu}_{L3}] [\bar{\nu_e}_{L2}  \gamma^\alpha {\nu_e}_{L1}]  \nn
\widehat{\mathcal{O}}_{22} &=&  [\bar{\nu_\mu}_{L4}   \gamma_\alpha  {\nu_\mu}_{L3}][\bar{e}_{L2}  \gamma^\alpha e_{L1}] \nn
\widehat{\mathcal{O}}_{32} &=& [\bar{\mu}_{L4}^\prime   \gamma_\alpha \mu_{L3}] [\bar{{\nu_e}}_{L2} \gamma^\alpha {\nu_e}_{L1}]  \nn
\widehat{\mathcal{O}}_{42} &=& [\bar{\mu}_{L4}   \gamma_\alpha \mu_{L3}] [\bar{e}_{L2} \gamma^\alpha  e_{L1}^\prime] \nn
\widehat{\mathcal{O}}_{52} &=&   [\bar{\mu}_{L4}   \gamma_\alpha  {\nu_\mu}_{L3}][\bar{{\nu_e}}_{L2} \gamma^\alpha  e_{L1}]\nn
\widehat{\mathcal{O}}_{62} &=&  [\bar{ {\nu_\mu}}_{L4}   \gamma_\alpha \mu_{L3}][\bar{e}_{L2}\gamma^\alpha  {\nu_e}_{L1}] 
\label{81}
\end{eqnarray}
with coefficients $\widehat C_{ij}$. The matching matrix is
\begin{eqnarray}
\widehat C_{i} &=& R_{ij} C_{j}\,.
\label{87a}
\end{eqnarray}
At tree-level
$R$ is
\begin{eqnarray}
R^{(0)} &=& \left[\begin{array}{rc} \frac14 &1 \\[5pt]
-\frac14 & 1\\[5pt]
 -\frac14 & 1 \\[5pt]
 \frac14 & 1 \\[5pt]
\frac12 & 0  \\[5pt]
\frac12 & 0  
\end{array}\right]\,.
\label{88}
\end{eqnarray}
The one-loop soft matching due to $W$ exchange is given using Eq.~(\ref{eq86e})
\begin{eqnarray}
R_{S,W}^{(1)} &=& \frac{\alpha_W}{4\pi} 2\log \frac{M_W^2}{\mu^2}\left[-\sum_{\vev{ij}}\left( \mathbf{t}_i\cdot \mathbf{t}_j\right)U_S(n_i,n_j)\right]\,.\nn
 \label{167}
\end{eqnarray}
Using Eq.~(\ref{165}), the one-loop soft matching is
\begin{eqnarray}
R_{S,W}^{(1)} &=&\frac{\alpha_W}{4\pi} 2\log \frac{M_W^2}{\mu^2} \left[R^{(0)}\softm_2\right]\,
 \label{167a}
\end{eqnarray}
and the total soft matching is $R^{(0)}+R_{S,W}^{(1)}$. Again, one can see the great simplification due to custodial $SU(2)$ invariance. Since $W^{1,2,3}$ all have the same mass, the soft matching depends on the $SU(2)$ invariant combination  $\mathbf{t}_i\cdot \mathbf{t}_j$, so that the soft matching can be written in terms of the same matrix $\softm_2$ that enters the soft anomalous dimension. In the standard model, the sum has to be broken into the charged $W$ and $Z$ contributions, and takes a more complicated form given in Ref.~\cite{p2}.

\subsubsection{Matching for Electroweak Gauge Boson Pair Production}

\begin{figure}
\begin{center}
\includegraphics[width=6cm]{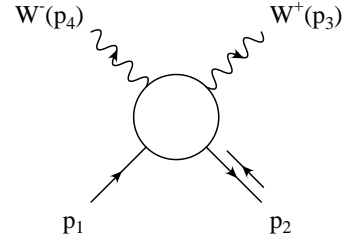}
\end{center}
\caption{\label{fig:W} Pair production $q(p_1) + \bar{q}(p_2) \to  W^+(p_3)+W^-(p_4)$. Time runs vertically. }
\end{figure}

The kinematics for the electroweak gauge boson pair-production process is shown in Fig.~\ref{fig:W}.
%%%----FIGURE--------------------------------------------------------------------------------------
%%%------------------------------------------------------------------------------------------------------
The operator basis is
\begin{eqnarray}
O_1 &=& \bar \psi_2 \psi_1 W^a_4W^a_3\,,\nn
O_2 &=& \bar \psi_2 t^c \psi_1  i \epsilon^{abc} W^a_4 W^b_3\,,
\label{185}
\end{eqnarray}
where only the gauge structure has been shown.  Note that $\epsilon^{abc}W^a_3 W^b_4\not=0$ since the two $W$ fields have momentum labels $p_3$ and $p_4$ which are different. Each gauge structure has 10 possible momentum structures, which are given in Ref.~\cite{sack}. The EFT running and matching is the same for all 10 amplitudes, so the momentum dependence can be ignored for our analysis.

In the basis Eq.~(\ref{185})
\begin{eqnarray}
\softm_2 &=& -\sum_{\vev{ij}} \mathbf{t}_i \cdot \mathbf{t}_j\ U_S(n_i,n_j) = \softm^{(2,g)}
\label{eq186}
\end{eqnarray}
where $\softm^{(2,g)}$ is given in Eq.~(\ref{158}). The soft anomalous dimension between the scales $Q$ and $\mu_l \sim M_Z$ is given by Eq.~(\ref{eq172}) with Eq.~(\ref{eq186}) for 
$\softm_2$.

At the low-scale $\mu_l \sim M_Z$, the operators Eq.~(\ref{185}) match onto
\begin{eqnarray}
\widehat O_1 &=& \bar \nu_{L2} \nu_{L1} W^+_4W^-_3 \nn
\widehat O_2 &=& \bar \nu_{L2} \nu_{L1} W^-_4 W^+_3\nn
\widehat O_3 &=& \bar \nu_{L2} \nu_{L1} W^3_4 W^3_3  \nn
\widehat O_4 &=& \bar e_{L2} e_{L1}W^+_4  W^-_3 \nn
\widehat O_5 &=& \bar e_{L2} e_{L1}  W^-_4 W^+_3 \nn
\widehat O_6 &=& \bar e_{L2} e_{L1}W^3_4  W^3_3 \nn
\widehat O_{7} &=& \bar \nu_{L2} e_{L1}W^+_4 W^3_3  \nn
\widehat O_{8} &=& \bar \nu_{L2} e_{L1} W^3_4  W^+_3 \nn
\widehat O_{9} &=& \bar e_{L2} \nu_{L1}W^3_4  W^-_3 \nn
\widehat O_{10} &=& \bar e_{L2} \nu_{L1}W^-_4  W^3_3 
\label{188}
\end{eqnarray}
where $W^\pm =(W^1\mp i W^2)/\sqrt{2}$.  The subscripts $3,4$ represent outgoing label momenta $p_3$ and $p_4$, and the gauge indices are to be treated as those on a quantum field, i.e.\ they represent the charge on the annihilation operator. These operators are used like terms in a Lagrangian. Thus $e \bar e \to W^+(k_1) W^-(k_2)$ is given by $\widehat C_4$ with $p_4=k_1$ and $p_3=k_2$, plus $\widehat C_5$ with $p_4=k_1$ and $p_3=k_2$.

The tree-level matching is
\begin{eqnarray}
 \widehat C_i &=&  \left(R^{(0)}\right)_{ij} C_j\,, \nn
 R^{(0)} &=&
\left[\begin{array}{cc} 
1 & \frac12\\
1 & -\frac12\\
1 & 0 \\
1 & -\frac12 \\
1 & \frac12 \\
1 & 0  \\
0 & -\frac{1}{\sqrt 2} \\
0 & \frac{1}{\sqrt 2} \\
0 & -\frac{1}{\sqrt 2}\\
0 & \frac{1}{\sqrt 2} \\
\end{array}\right]\,.
\end{eqnarray}
The one-loop soft matching due to $W$ exchange is given by Eq.~(\ref{167a}) using
the value of $\softm_2$ in Eq.~(\ref{eq186}).

\subsubsection{Matching for $q \bar q \to \varphi \varphi$}

For longitudinal $W$ production, we also need the results for external unphysical Goldstone boson $\varphi$ fields, which are contained in the Higgs multiplet $\phi$.
The operators are
\begin{eqnarray}
O_1 &=& \bar \psi t^a \psi\, \phi^\dagger_4 t^a \phi_3\,,\nn
O_2 &=& \bar \psi \psi\, \phi^\dagger_4 \phi_3\,.
\end{eqnarray}
The gauge current $i \left(\phi^\dagger T^a D_\mu \phi - D^\mu \phi^\dagger T^a \phi \right)$ produces operators of this form, weighted by a label  momentum factor $\mathcal{P}_4^\mu-\mathcal{P}_3^\mu$, which is included in the operator coefficients, and is antisymmetric in $3 \leftrightarrow 4$.

The group theory sums needed for the soft anomalous dimension matrix are
\begin{eqnarray}
\softm_2&=& -\sum_{\vev{ij}} \mathbf{t}_i \cdot 
\mathbf{t}_j\ U_S(n_i,n_j)= \softm^{(2)}\,.
\end{eqnarray}
At the low scale, the operators match onto
\begin{eqnarray}
\widehat O_1 &=& \bar \nu_{L2} \nu_{L1} \varphi^-_4 \varphi^+_3\nn
\widehat O_2 &=& \bar \nu_{L2} \nu_{L1} \varphi^3_4  \varphi^3_3 \nn
\widehat O_{3} &=& \bar \nu_{L2} \nu_{L1} H_4 \varphi^3_3\nn
\widehat O_{4} &=& \bar \nu_{L2} \nu_{L1} \varphi^3_4 H_3\nn
\widehat O_{5} &=& \bar \nu_{L2} \nu_{L1} H_4 H_3\nn
\widehat O_6 &=& \bar e_{L2} e_{L1} \varphi^-_4  \varphi^+_3\nn
\widehat O_7 &=& \bar e_{L2} e_{L1} \varphi^3_4 \varphi^3_3 \nn
\widehat O_{8} &=& \bar e_{L2} e_{L1} H_4 \varphi^3_3\nn
\widehat O_{9} &=& \bar e_{L2} e_{L1} \varphi^3_4 H_3\nn
\widehat O_{10} &=& \bar e_{L2} e_{L1} H_4 H_3\nn
\widehat O_{11} &=& \bar \nu_{L2} e_{L1} \varphi^3_4 \varphi^+_3 \nn
\widehat O_{12} &=& \bar \nu_{L2} e_{L1} H_4  \varphi^+_3\nn
\widehat O_{13} &=& \bar e_{L2} \nu_{L1} \varphi^-_4  \varphi^3_3\nn
\widehat O_{14} &=& \bar e_{L2} \nu_{L1}\varphi^-_4  H_3\,.
\label{188a}
\end{eqnarray}

The tree-level matching is
\begin{eqnarray}
R^{(0)} &=& \left[ \begin{array}{cc}  
 \frac 14 & 1 \\
 - \frac18 & \frac12\\
 \frac i 8 & -\frac i 2  \\
 -\frac i 8 & \frac i 2  \\
 - \frac 1 8 & \frac 12 \\
- \frac 14 & 1\\
 \frac 18 & \frac 12  \\
 - \frac i 8 & -\frac i 2  \\
 \frac i 8 & \frac i 2 \\
 \frac 1 8 & \frac 1 2 \\
 - \frac{1}{2 \sqrt 2} & 0\\
\frac{i}{2 \sqrt 2} & 0 \\
-\frac{1}{2 \sqrt 2} & 0\\
-\frac{i}{2 \sqrt 2} & 0\\
\end{array}\right]
\end{eqnarray}
and the one-loop matching is given by Eq.~(\ref{167a}) using the values of $R^{(0)}$ and $\softm_2$ for the current case.

\subsection{Final Result}

The final result for the scattering amplitude is then given by using Eq.~(\ref{eq:total}), with the running and matching given by the sum of the collinear and soft contributions, as in Eq.~(\ref{102}).

\subsection{Comparison with Previous Results}

The electroweak radiative corrections using SCET were computed previously in Refs.~\cite{CGKM1,CGKM2,CKM} for processes not involving external  gauge bosons. The results given here are much more compact, the final results are the collinear functions for each particle given in Sec.~\ref{sec:collsm} and the soft functions computed in Sec.~\ref{sec:universal}. The previous results were given in terms of kinematic variables such as $Q^2$ or $s$, $t$ and $u$, and summing over pairs of particles. The collinear and soft functions given here are more conveniently written in terms of $\bar n_i \cdot p_i$, and $n_i \cdot n_j$, which depends on the choice of $n$. The $n$ dependence cancels in the final result between the collinear and soft contributions. The kinematic conversions needed to compare the two forms of the result are given below.

In the Sudakov problem, $Q^2=-(p_1+p_2)^2$, where the two momenta are incoming. One can rewrite
\begin{eqnarray}
\log \frac{Q^2}{\mu^2} &=& \log \frac{\bar n_1 \cdot p_1}{\mu}
+\log \frac{\bar n_2 \cdot p_2}{\mu}
\end{eqnarray}
where $-\bar n_1 \cdot \bar n_2=2$, since $n_1=(1,\mathbf{n})$ and $n_2=-(1,-\mathbf{n})$.
Using this, the  results in Table~I of Ref.~\cite{CGKM1} and Table~I of Ref.~\cite{CGKM2} can be obtained by adding pairs of rows in Table~\ref{tab:massless}.

Table~III of Ref.~\cite{CGKM2} for the matching at heavy particle thresholds is equivalent to Table~\ref{tab:match}. If particle~1 is light and particle~2 is heavy, the relation
\begin{eqnarray}
\log \frac{Q^2}{\mu^2}-\frac12 \log \frac{m_2^2}{\mu^2}
&=&  \log \frac{\bar n_1 \cdot p_1}{\mu}
+  \log \frac{\bar n_2 \cdot p_2}{m_2} \nn
&=&  \log \frac{\bar n_1 \cdot p_1}{\mu}
+  \log (2\gamma_2) 
\end{eqnarray}
using $\bar n_2 \cdot v_2=2\gamma_2$ can be used to show the equivalence of the two forms. 

Table~IV of Ref.~\cite{CGKM2} for the heavy-heavy case is also obtained from Table~\ref{tab:match} using
\begin{eqnarray}
\log(2 w) = \log(2\gamma_1)+\log(2\gamma_2)
\end{eqnarray}
for $w=v_1\cdot v_2 \gg 1$.

\section{Sudakov Form Factor}\label{sec:sudakovff}

The SCET results are applied to the Sudakov form factor. This is an instructive example, and allows us to connect the SCET formulation with the results of Magnea and Sterman~\cite{ms} (and more recent work in Refs.~\cite{gluza,mitovmoch}) obtained using factorization methods. The SCET decomposition of the amplitude into collinear and soft is related to, but not exactly the same as, the factorization decomposition, and it is useful to study the difference. In addition, we can compare the Sudakov form factor in a theory with massive gauge bosons with the Magnea-Sterman analysis for QCD.

The spacelike Sudakov form factor $F(Q^2)$  is defined as the particle scattering amplitude  by an external current, with momentum transfer $Q^2=-q^2 >0$. It is convenient to compute the form factor in the Breit frame (see Fig.~\ref{fig:breit}), where the particle is back-scattered, and the momentum transfer $q$ has $q^0=0$. One can compute the Sudakov form factor in this frame using Eq.~(\ref{eq:form}) with $r=2$, where particle~1 and 2 have the same quantum numbers, and $n_1=-\bar n_2$, $n_2=-\bar n_1$. There is only one gauge invariant amplitude, so there is no matrix structure to the amplitude. We work in the large momentum transfer limit, so that $M^2/Q^2$ power corrections are dropped.

The form factors for fermion scattering by the vector current $j^\mu=\bar \psi \gamma^\mu \gamma$, the scalar current $j_S=\bar \psi \psi$ and the tensor current $j^{\mu\nu}_T=\bar \psi \sigma^{\mu \nu} \psi$ are defined as $F_V(Q^2,M,\mu)$, $F_S(Q^2,M,\mu)$ and $F_T(Q^2,M,\mu)$, where the operators are renormalized in the full theory at a scale $\mu$.\footnote{We also give results for gauge scattering by the field strength tensor $F^2=F_{\mu\nu}F^{\mu\nu}$ in Appendix~\ref{app:sudakov}. This provides a check on the collinear anomalous dimension for the gauge field.} They are normalized to one at tree-level. The anomalous dimensions of the operators in the full theory are
\begin{eqnarray}
\mu \frac{\rd}{\rd \mu} j^\mu &=& -\gamma_V(\alpha) j^\mu =0\,, \nn
\mu \frac{\rd}{\rd \mu} \bar \psi \psi &=&-\gamma_S(\alpha)\ \bar \psi \psi 
=6 {\alpha \over 4 \pi} C_F\ \bar \psi \psi \,,\nn
 \mu \frac{\rd}{\rd \mu} \bar \psi \sigma^{\mu\nu} \psi &=& -\gamma_T(\alpha)\ \bar 
\psi \sigma^{\mu\nu} \psi =  -2{\alpha \over 4 \pi} C_F\ \bar \psi \sigma^{\mu\nu} \psi \,,\nn
\label{eq2}
\end{eqnarray}
where we have given the one-loop values, and defined our sign convention for $\gamma$. The form factors must satisfy the equations
\begin{eqnarray}
\mu \frac{\rd}{\rd \mu} F_O(Q^2,M,\mu) &=& -\gamma_O(\alpha(\mu))\ 
F_O(Q^2,M,\mu)  \,.\nn
\label{eq3}
\end{eqnarray}

In SCET, the form factor is given by running the operator from $\mu$ to $\mu_h$ using the full theory anomalous dimension $\gamma_f$, matching onto SCET at $\mu_h$, running in SCET from $\mu_h$ to $\mul$, and then evaluating the matrix element at $\mul$. The result is more conveniently written for $\log F_O$, where $O=V,S,T$:
\begin{eqnarray}
&& \ln F_O(Q^2,M,\mu) =  \frac12 
\int^{\mu_h^2}_{\mu^2} \frac{\rd \tilde \mu^2}{\tilde \mu^2} \gamma_O\left(\alpha
\left(\tilde \mu\right)\right)\nn
&&+C_O\left(\alpha\left(\mu_h\right),\log \frac{Q^2}{\mu_h^2}\right)\nn
&&- \frac12 
\int^{\mu_h^2}_{\mul^2} \frac{\rd \tilde \mu^2}{\tilde \mu^2}\left[\Gamma\left(\alpha
\left(\tilde \mu\right)\right) \ln \frac{Q^2}{\tilde \mu^2}+B\left(\alpha\left(\tilde \mu
\right)\right) \right]\nn
&&+D_0\left(\alpha\left(\mul\right),\log \frac{M^2}{\mul^2}\right)+D_1\left(\alpha
\left(\mul\right),\log \frac{M^2}{\mul^2}\right) \ln \frac{Q^2}{\mul^2}\nn
\label{7}
\end{eqnarray}
where
\begin{eqnarray}
B &=& \Omega_1+\Omega_2+\sigma\,,
\end{eqnarray}
from Eq.~(\ref{eq:srun}). Explicit one-loop expressions for the form-factor and the different terms in Eq.~(\ref{7}) can be found in Appendix~\ref{app:sudakov}.

Note that there is a single-log term in the low-scale matrix element. The EFT quantities $\Gamma$, $B$, $D_{0,1}$ do not depend on the choice of fermion current, but the full theory anomalous dimension $\gamma_O$ and high-scale matching $C_O$ do. Eq.~(\ref{7}) is independent of $\mu_h$ and $\mu_l$, which implies the consistency conditions
\begin{eqnarray}
\Gamma\left(\alpha\left(\mu\right)\right) 
&=& -\mu \frac{\rd}{\rd \mu} D_1\left(\alpha\left(\mu\right),\log \frac{M^2}
{\mu^2}\right) \,,\nn
B\left(\alpha\left(\mu\right)\right) &=&  2 D_1\left(\alpha\left(\mu\right),\log \frac{M^2}
{\mu^2}\right) \nn
&&- \mu \frac{\rd}{\rd \mu} D_0\left(\alpha\left(\mu\right),\log \frac{M^2}
{\mu^2}\right)\,,\nn
\gamma_O(\alpha(\mu)) &=&
-\Gamma\left(\alpha\left(\mu\right)\right) \ln \frac{\mu^2}{Q^2}+B\left(\alpha\left(\mu
\right)\right)\nn
&&- \mu \frac{\rd}{\rd \mu}C(\alpha(\mu),Q/\mu)\,.
\label{eq8}
\end{eqnarray}
Eqs.~(\ref{7},\ref{eq8}) are valid to all orders in perturbation theory.

Differentiating $F_O(Q^2,M,\mu)$ w.r.t.\ $\log Q^2$ gives
\begin{eqnarray}
\frac{\partial \ln F_O(Q^2,M,\mu)}{\partial \log Q^2} &=&
\frac12 \Biggl[K_O\left(\alpha(\mu),\log\frac{M^2}{\mu^2}\right)\nn
&& + G_O\left(\alpha(\mu),\log\frac{Q^2}{\mu^2}\right) \Biggr],
\label{eq41}
\end{eqnarray}
where $K_O$ is the $\log M^2/\mu^2$ dependent part of the derivative, and $G_O$ is the rest. The form Eq.~(\ref{eq41}) is highly non-trivial, because the $\log M^2$ and $\log Q^2$ dependence has been separated into different terms, as shown below. Using Eq.~(\ref{7}), one gets
\begin{eqnarray}
K_O +G_O &=& 2 D_1\left(\alpha(\mul),\log \frac{M^2}{\mul^2}\right)\nn 
&&+2 \frac{\partial \ln C_O\left(\alpha(\mu_h),\log \frac{Q^2}{\mu_h^2}\right)}{\partial \ln 
Q^2}\nn
&&-\int_{\mul^2}^{\mu_h^2}
\frac{\rd \tilde \mu^2}{\tilde \mu^2} \Gamma\left(\alpha\left(\tilde \mu\right)\right).
\label{17}
\end{eqnarray}
The r.h.s.\ of Eq.~(\ref{17}) is independent of $\mu_h$ and $\mu_l$ due to the consistency conditions Eq.~(\ref{eq8}), and can be written in terms of $\alpha(\mu)$ and $\lM=\log M^2/\mu^2$ using the $\beta$-function to transform all coupling constants to the scale $\mu$. Equivalently, one can choose $\mu_h=\mu_l=\mu$ in Eq.~(\ref{17}),
\begin{eqnarray}
K_O +G_O &=& 2 D_1\left(\alpha(\mu),\log \frac{M^2}{\mu^2}\right)\nn 
&&+2 \frac{\partial \ln C_O\left(\alpha(\mu),\log \frac{Q^2}{\mu^2}\right)}{\partial \ln Q^2}\,,
\label{17a}
\end{eqnarray}
so that
\begin{eqnarray}
K  &=& 2 D_1\left(\alpha(\mu),\log \frac{M^2}{\mu^2}\right)\,,\nn 
G_O &=& 2 \frac{\partial \ln C_O\left(\alpha(\mu),\log \frac{Q^2}{\mu^2}\right)}{\partial \ln 
Q^2}\,.
\label{17b}
\end{eqnarray}
$D_1$ and hence $K$ are independent of the operator $O$, so we can drop the subscript on $K$. Eq.~(\ref{17b}) justifies the form used in Eq.~(\ref{eq41}). Eq.~(\ref{17b}) follows from the low scale matching which depends on $M$, $D_0+D_1 \log Q^2/\mu_l^2$ being linear in $\log Q^2$, as proved in Refs.~\cite{dis,Bauer:2003pi}. The derivative of the low-scale matching w.r.t.\ $\log Q^2$ is then independent of $Q^2$, and gives the expression for $K$.

The low scale single-log matching $D_1$ is independent of the operator $\mathcal{O}$, and is universal. $D_1$ is the universal function $f_{i,2}^{(1)}$ of Ravindran, Smith and van Neerven~\cite{ravindran,moch:ff} for a massive gauge theory. $D_1$ is related to the cusp anomalous dimension by Eq.~(\ref{eq8}), and so should have the maximally non-Abelian color structure of the cusp anomalous dimension. This is true at two-loops, where $D_1^{(2)}$ for an $SU(2)$ gauge theory with degenerate Higgs and gauge bosons is~\cite{CGKM2} from the results of Ref.~\cite{jkps4}
\begin{eqnarray}
D_1^{(2)}(\mu=M) &=& C_F n_F T_F \left(\frac{4\pi^2}{9}+\frac{112}{27}\right)+\frac{13}{2}\sqrt{3}\text{Cl}_2(\pi/3)\nn
&& -5 \zeta(3) +\frac{15 \sqrt{3} \pi}{4}-
\frac{391}{18}\,.
\label{37.12}
\end{eqnarray}
$D_1$ depends on the full mass spectrum of the theory.

Differentiating the form-factor evolution equation
\begin{eqnarray}
\mu \frac{\rd}{\rd \mu}\log F_O &=& -\gamma_O(\alpha(\mu))
\label{18}
\end{eqnarray}
w.r.t. $\log Q^2$ gives~\cite{ms}
\begin{eqnarray}
\mu \frac{\rd}{\rd \mu} \left[K+G_O\right]&=& 0\,.
\label{19}
\end{eqnarray}
This is clearly satisfied by Eq.~(\ref{17}) since $\mu$ does not appear on the r.h.s. However, $K$ and $G_O$ separately depend on $\mu$ when the r.h.s. is written in terms of $\alpha(\mu)$, as in Eq.~(\ref{17b}). Eq.~(\ref{19}) gives
\begin{eqnarray}
\mu \frac{\rd}{\rd \mu} K &=& - \gamma_K\left(\alpha\left(\mu\right)\right)\,,\nn
\mu \frac{\rd}{\rd \mu} G_O &=&  \gamma_K\left(\alpha\left(\mu\right)\right)\,,
\label{20}
\end{eqnarray}
since $\alpha(\mu)$ is the only common variable to both $K$ and $G_O$. $\gamma_K$ does not depend on $O$ or $Q^2$, since $K$ does not, and it does not depend on $M^2$ since $G_O$ does not. Using $K$ in Eq.~(\ref{17b}) and Eq.~(\ref{7}) gives~\cite{ms}
\begin{eqnarray}
\gamma_K \left(\alpha\left(\mu\right)\right) &=& 2\, \Gamma \left(\alpha\left(\mu
\right)\right)\,,
\label{20a}
\end{eqnarray}
which shows that the single-log term in the SCET anomalous dimension is related to the cusp anomalous dimension. Eq.~(\ref{20}) also follows from Eq.~(\ref{eq8}) using Eq.~(\ref{17b}).

Integrating Eq.~(\ref{eq41}) gives
\begin{eqnarray}
\ln F_O(Q^2,M,\mu) &=& \ln F_O(Q_0^2,M,\mu)\nn
&&\hspace{-1cm}+\frac{1}{2} \int_{Q_0^2}^{Q^2}
\frac{\rd \xi^2}{\xi^2}\Biggl[ K \left(\alpha(\mu),\log\frac{M^2}{\mu^2}\right)\nn
&& + G_O\left(\alpha(\mu),\log\frac{\xi^2}{\mu^2}\right)\Biggr]\,,
\end{eqnarray}
and integrating Eq.~(\ref{20}) gives
\begin{eqnarray}
G_O\left(\alpha(\xi),0\right) &=& G_O\left(\alpha(\mu),\log\frac{\xi^2}{\mu^2}\right)+
\frac{1}{2} \int_{\mu^2}^{\xi^2}
\frac{\rd \tilde \mu^2}{\tilde \mu^2}\gamma_K(\alpha(\tilde \mu)),\nn
\end{eqnarray} so that
\begin{eqnarray}
\ln F_O(Q^2,M,\mu) &=& \ln F_O(Q_0^2,M,\mu)\nn
&&\hspace{-1cm}+\frac{1}{2} \int_{Q_0^2}^{Q^2}
\frac{\rd \xi^2}{\xi^2}\Biggl[ K \left(\alpha(\mu),\log\frac{M^2}{\mu^2}\right)\nn
&&\hspace{-1cm} + G_O\left(\alpha(\xi),0\right)-\frac{1}{2} \int_{\mu^2}^{\xi^2}
\frac{\rd \tilde \mu^2}{\tilde \mu^2}\gamma_K(\alpha(\tilde \mu))\Biggr]\,\nn
\label{eq62}
\end{eqnarray}
which is the expression for the Sudakov form factor in a massive gauge theory written as in Ref.~\cite{ms}. This equation is equivalent to the SCET form Eq.~(\ref{7}); a detailed comparison can be found in Sec.~IV of Ref.~\cite{CGKM2}.

\subsection{Massless QCD}\label{sec:masslessqcd}

The entire analysis above can be repeated for a massless gauge theory such as QCD. In Eq.~(\ref{7}), the high scale running, high scale matching and running between $\mu_h$ and $\mu_l$ using the SCET anomalous dimension are unchanged by the infrared structure of the theory, and are {\sl identical} in the broken and unbroken gauge theory. The only difference is the low scale matrix element $D$. In QCD, the low-scale matrix element $D$ is infrared divergent, and has the form
\begin{eqnarray}
&&D_0\left(\alpha\left(\mul\right),\epsilon\right)+D_1\left(\alpha\left(\mul\right),
\epsilon\right) \ln \frac{Q^2}{\mul^2}
\label{7d}
\end{eqnarray}
where the $1/\epsilon$ terms are infrared divergent, and are not cancelled by renormalization counterterms. So all the results of the previous section hold with the replacement of a functional dependence on $\log M^2/\mu^2$ by one on $1/\epsilon$.\footnote{This does not mean that the functions are identical. Using $M$ as an IR regulator instead of $\epsilon$ can change the subleading divergences and finite parts. Compare Eq.~(\ref{eqA6}) and (\ref{eqA8})} Explicit one-loop expressions for the massless form factor are given in Appendix~\ref{app:sudakov}.

The form-factor computed for a massless gauge theory in pure dimensional regularization is IR divergent, and has the functional form
\begin{eqnarray}
F_O\left(\alpha_s(\mu),\epsilon, \frac{Q^2}{\mu^2}\right)\,.
\end{eqnarray}
The explicit $\mu$ dependence in $F_V(Q^2,\mu^2)$ (i.e.\ for the vector form factor) is cancelled by the implicit $\mu$ dependence in $\alpha$, so that $F_V$ is $\mu$-independent. The EFT computation splits the form factor into a high scale part which depends on $Q$, but is IR finite, and  low-scale matrix elements $D_{0,1}$ which are IR divergent, but $Q$ independent. The total low-scale matrix element has the linear-in-log form Eq.~(\ref{7d}).

The Magnea-Sterman expression for the Sudakov form factor in massless QCD is given by Eq.~(\ref{eq62}) with the replacement $K \to K(\alpha(\mu,\epsilon))$. Magnea and Sterman studied only the vector form-factor, for which $F_V(Q_0^2=0)=1$, since it is the forward matrix element of the quark number current. They dropped $\ln F(Q_0^2)$ in Eq.~(\ref{eq62}) and set $Q_0 \to 0$ in the limits of integration. We have used SCET in our computation, and been able to compute the form factor for any current. We cannot take the limit $Q_0 \to 0$, since by construction, the SCET results are only valid for large $Q$, and power corrections are neglected. In the massive gauge theory, this is because we have explicitly dropped $M^2/Q^2$ power corrections. In massless QCD in perturbation theory, it might seem that there is no other scale in  the problem, so we could use SCET for $Q \to 0$. However, this apparent scale independence is illusory; perturbation theory knows about the scale $\lqcd$ through the coupling constant, and we have neglected non-perturbative $\lqcd^2/Q^2$ corrections from higher dimension operators.

\section{Conclusion}\label{sec:concl}

We have extended our previous results on high-energy scattering to include external longitudinally and transversely polarized Higgs bosons. This allows us to compute any high energy scattering process in the standard model.

The factorization structure of the effective theory allows us to write the results in terms of single-particle collinear functions which only depend on the energy of each particle and a universal soft function that only depends on the particle directions. This holds for both the anomalous dimension and the finite matrix element in the broken gauge theory.

The electroweak corrections are substantial and increase rapidly with energy. They are largest for processes involving transversely polarized electroweak gauge bosons.  The one-loop electroweak corrections are more important than the two-loop QCD corrections, in the cases we have studied. The EFT method allows us to compute the hard cross-section with a theoretical  uncertainty of better than 1\%, except for transverse $W$ production, where the uncertainty is $\alt 10$\%.

The EFT result can be used to give the renormalization group improved cross-section if the high-scale matching is known. The high-scale matching is two powers of $\LL$ smaller than the other corrections in the log-counting. Numerically, the largest high-scale matching we have found (including QCD) is about 10\% (for $W_T$ production), but this is also the case where the other EFT corrections are large, and the \emph{purely} electroweak corrections are several times the matching even at 2~TeV. As the energy increases, the matching corrections decrease slowly, whereas the EFT corrections increase very rapidly. One can compute the cross-section in cases where the high-scale matching corrections are not known by simply using the tree-level values. This introduces an uncertainty $\alt 10\%$ in the rate, but now we can compute the radiative corrections to \emph{any} standard model process with an arbitrary number of external particles. The collinear corrections are given by simply multiplying the contributions from each external particle. The only tedious computation is the soft group theory sum over pairs of particles, but this is easily automated.

The EFT results neglect $M^2/Q^2$ power corrections. The largest power corrections are $M_Z^2/Q^2$ power corrections to the electroweak contributions. The $\lqcd^2/Q^2$ corrections to to the QCD contribution are much smaller. As discussed in Sec.~\ref{sec:power}, the unknown power corrections are only in the one-loop terms; the tree-level power corrections can be included using the exact tree-level amplitude and phase space. Thus the EFT results can be used close to threshold. For the Sudakov form factor, the power corrections are 2\% exactly at threshold.

The collinear functions for all standard model particles, and the soft functions for some important $2 \to 2$ scattering processes, are given in Ref.~\cite{p2}. The results in this paper sum the NLL series and the largest NNLL terms. The missing terms for a complete NNLL computation are the scalar pieces in the three-loop cusp and two-loop non-cusp anomalous dimension, which can be computed in an unbroken gauge theory, and the two-loop $\log Q^2/\mu^2$ term $D_L$ in the low-scale matching, which depends on the mass spectrum of the theory. The two-loop matching for $W_T$ production is required to reduce the $\mu_h$ uncertainty in the cross-section.

The partonic cross-sections computed here have to be convoluted at a factorization scale $\mu_f$ with parton distribution functions and possibly jet and soft functions to get hadronic cross-sections. This is a non-trivial task, but has been well-studied and is independent of the computations given in this paper.

\begin{appendix}

\section{The Sudakov Form Factor}\label{app:sudakov}

We summarize the one-loop expressions for $O=V,S,T$, and the high scale matching for $O=F^2$ here.

\subsection{Massive Gauge Theory}\label{app:ffmassive}

The form factors for $O=V,S,T$ are
\begin{eqnarray}
F_V &=& 1 + \frac{\alpha\, C_F}{4 \pi}
\Bigl[ -\lQM^2 + 3 \lQM - \frac72-\frac{2\pi^2}{3}\Bigr]\,,\nn
F_S &=& 1 + \frac{\alpha\, C_F}{4 \pi}
\Bigl[ -\lQM^2 -3\lM +  \frac52-\frac{2\pi^2}{3}\Bigr]\,,\nn
F_T &=& 1 + \frac{\alpha\, C_F}{4 \pi}
\Bigl[ -\lQM^2 + 3 \lQM +\lQ - \frac72-\frac{2\pi^2}{3}\Bigr]\,.\nn
\end{eqnarray}

The full theory anomalous dimensions are:
\begin{eqnarray}
\gamma_V\left(\alpha\left(\mu\right)\right) &=& 0\,, \nn
\gamma_{S}\left(\alpha\left(\mu\right)\right) &=& -6 {\alpha(\mu) \over 4 \pi} C_F\,, \nn
\gamma_{T} \left(\alpha\left(\mu\right)\right) &=& 2 {\alpha(\mu) \over 4 \pi} C_F \,.
\label{9a}
\end{eqnarray}

The high-scale matching is:
\begin{eqnarray}
C_V(\mu) &=& {\alpha\left(\mu\right) \over 4 \pi} C_F \Biggl[-\lQ^2+3\lQ+{\pi^2 \over 
6}-8 \Biggr]\,,\nn
C_S(\mu) &=& {\alpha\left(\mu\right) \over 4 \pi} C_F \Biggl[-\lQ^2+{\pi^2 \over 6}-2 
\Biggr]\,,\nn
C_T(\mu) &=& {\alpha\left(\mu\right) \over 4 \pi} C_F \Biggl[-\lQ^2+4\lQ+{\pi^2 \over 
6}-8 \Biggr]\,.
\label{9b}
\end{eqnarray}

The SCET anomalous dimension is:
\begin{eqnarray}
\Gamma\left(\alpha\left(\mu\right)\right) &=& 4 {\alpha(\mu) \over 4 \pi} C_F = - A\left(\alpha\left(\mu\right)\right)  \,,\nn
B\left(\alpha\left(\mu\right)\right) &=& -6 {\alpha(\mu) \over 4 \pi} C_F \,.
\label{9c}
\end{eqnarray}

The low-scale matrix element is:
\begin{eqnarray}
D_0\left(\alpha\left(\mul\right),\lMl\right)+D_1\left(\alpha\left(\mul\right),\lMl\right) \ln 
\frac{Q^2}{\mul^2}\,,\nn
\end{eqnarray}
with
\begin{eqnarray}
D_0 &=& \frac{\alpha(\mu_l)\, C_F}{4 \pi}
\Bigl[-\log^2 \frac{M^2}{\mul^2}-3\lMl+\frac92-\frac{5\pi^2}{6}\Bigr]\,,\nn
D_1 &=& \frac{\alpha(\mu_l)\, C_F}{4 \pi}
\Bigl[2\lMl\Bigr]\,.
\label{eqA6}
\end{eqnarray}

The effective theory form factors are all equal to
\begin{eqnarray}
&&F_{\text{EFT}}(Q^2,M,\mu)=\frac{\alpha(\mu)\, C_F}{4 \pi}
\Bigl[-\log^2 \frac{M^2}{\mu^2}\nn
&&+2\log \frac{M^2}{\mu^2}\log \frac{Q^2}{\mu^2}
-3\log \frac{M^2}{\mu^2}+\frac92-\frac{5\pi^2}{6}\Bigr]
\label{eqA6a}
\end{eqnarray}
and differ from the full theory form-factors by the high-scale matching 
Eq.~(\ref{9b}).

\subsection{Massless Gauge Theory}\label{app:ffmassless}

The form factors are
\begin{eqnarray}
F_V(Q^2,\mu) &=&1 + {\alpha(\mu) \over 4 \pi} C_F \Biggl[
-{2 \over \epsilon^2} + \frac{2}{\epsilon} \lQ - \frac{3}{\epsilon} \nn
&& -\lQ^2 +3 \lQ +{\pi^2 \over 6} -8\Biggr] \,,\nn
F_S(Q^2,\mu) &=&1 + {\alpha \over 4 \pi} C_F \Biggl[
-{2 \over \epsilon^2} + \frac{2}{\epsilon} \lQ - \frac{3}{\epsilon} \nn
&&  -\lQ^2  +{\pi^2 \over 6} -2\Biggr]\,, \nn
 F_T(Q^2,\mu) &=&1 + {\alpha \over 4 \pi} C_F \Biggl[
-{2 \over \epsilon^2} + \frac{2}{\epsilon} \lQ - \frac{3}{\epsilon} \nn
&&  -\lQ^2 +4 \lQ  +{\pi^2 \over 6} -8\Biggr] \ . 
\label{eq5}
\end{eqnarray}
Note that the $1/\epsilon$ and $\lQ$ terms do not have the same coefficient for 
$F_S$ and $F_T$ since there is also a $1/\epsilon$ UV divergence that is cancelled 
by the renormalization counterterm, and leads to the anomalous dimensions in 
Eq.~(\ref{eq2}).

The full theory anomalous dimension, high-scale matching, and SCET anomalous 
dimension have the same values as for the massive case, Eqs.~(\ref{9a},\ref{9b},
\ref{9c}).

The low-scale matrix element is:
\begin{eqnarray}
D_0\left(\alpha\left(\mul\right),\epsilon \right)+D_1\left(\alpha\left(\mul\right),\epsilon
\right) \ln \frac{Q^2}{\mul^2}\,,
\end{eqnarray}

with
\begin{eqnarray}
D_0 &=& {\alpha \over 4 \pi} C_F \Biggl[ -{2 \over \epsilon^2}- \frac{3}{\epsilon} 
\Biggr]\,,\nn
D_1 &=& {\alpha \over 4 \pi} C_F \Biggl[   \frac{2}{\epsilon}  \Biggr]\,.
\label{eqA8}
\end{eqnarray}

$D_{0,1}$ can be computed directly using the EFT. The EFT graphs are scaleless, 
and vanish, so $D_{0,1}$ are given purely by the counterterm contributions which 
cancel the UV divergences in the effective theory, so that $D_{0,1}$ contain purely 
IR singularities in $1/\epsilon$.

The effective theory form-factors are all equal to
\begin{eqnarray}
&& F(Q^2,\epsilon,\mu)={\alpha(\mu) \over 4 \pi} C_F \Biggl[ -{2 \over \epsilon^2}
+ \frac{2}{\epsilon} \log \frac{Q^2}{\mu^2}- \frac{3}{\epsilon}
\Biggr]\nn
\label{eqA8c}
\end{eqnarray}
and differ from the full theory form-factors by the high-scale matching 
Eq.~(\ref{9b}).

\subsection{$F^2$}
 
For gauge boson scattering by the field-strength tensor, the anomalous dimension in the full theory is
\begin{eqnarray}
\mu \frac{\rd}{\rd \mu} F^2 &=&  -\gamma_{F^2}(\alpha) F^2\,,\nn
\gamma_{F^2} &=& \frac{\partial \beta(g)}{\partial g}-\frac{\beta(g)}{g} = -\frac{\alpha}{4\pi} 2b_0\,,
\label{eq2g}
\end{eqnarray}
and the high-scale matching is:
\begin{eqnarray}
C_{F^2}(\mu) &=& {\alpha\left(\mu\right) \over 4 \pi} C_A \Biggl[-\lQ^2+{\pi^2 \over 
6}-8 \Biggr]\,.
\label{9bg}
\end{eqnarray}
The last result gives a consistency check on the $-b_0$ term in the collinear anomalous dimension for gauge fields, using Eq.~(\ref{eq8}).

\section{Color Identities}\label{app:color}

The operator is a color singlet, so
\begin{eqnarray}
\sum_i \mathbf{T}_i^A &=& 0\,.
\end{eqnarray}
Then
\begin{eqnarray}
0 &=& \left(\sum_i \mathbf{T}_i\right)\cdot \left(\sum_j f_j \mathbf{T}_j\right)\nn
&=& \sum_i f_i \mathbf{T}_i \cdot \mathbf{T}_i + \sum_{\vev{ij}} \left(f_i+f_j\right) \,
\mathbf{T}_i \cdot \mathbf{T}_j \,,
\end{eqnarray}
so that
\begin{eqnarray}
\sum_{\vev{ij}}( f_i+f_j) \,\mathbf{T}_i \cdot \mathbf{T}_j &=& -
\sum_i  f_i\, \mathbf{T}_i \cdot \mathbf{T}_i\,.
\label{sumiden}
\end{eqnarray}
Similarly,
\begin{eqnarray}
0 &=& \mathbf{T}_i \cdot  \left(\sum_j \mathbf{T}_j\right)\nn
&=&   \mathbf{T}_i \cdot \mathbf{T}_i + \sum_{{j \atop j\not=i}}  \,\mathbf{T}_i \cdot 
\mathbf{T}_j \,,
\end{eqnarray}
so that
\begin{eqnarray}
\sum_{{j \atop j\not=i}}  \,\mathbf{T}_i \cdot \mathbf{T}_j &=& - \mathbf{T}_i \cdot 
\mathbf{T}_i\qquad (\text{no sum on}\ i)\,.
\label{sumiden2}
\end{eqnarray}

\section{Symmetric Form for the Amplitude}\label{app:sudform}

We follow the argument below Eq.~(\ref{eq25}), but starting with the initial form Eq.~(\ref{eq79}). In Eq.~(\ref{eq79}), treat $I_i$ as a function of $\mathcal{Z}(i)=\log(\Delta_i/\mu^2)=\log(\bar n_i \cdot p_i)/\mu+\log \delta_i/\mu$, and $S(n_i \cdot n_j)$ as a function of $\mathcal{X}_{ij}=\log (n_i \cdot n_j)-\log (\delta_i/\mu)-\log(\delta_j/\mu)$. Then $\delta_i$ independence of the total amplitude gives
\begin{eqnarray}
\mathbf{T}_i \cdot \mathbf{T}_i\frac{\partial I_i(\mathcal{Z}_i)}{\partial \mathcal{Z}_i}
&=& \sum_{j\not=i}\mathbf{T}_i \cdot \mathbf{T}_j\frac{\partial S(\mathcal{X}_{ij})}
{\partial \mathcal{X}_{ij}}\,.
\label{eq81}
\end{eqnarray}
Differentiating w.r.t. $\delta_j$, $j\not=i$ gives
\begin{eqnarray}
0 &=& \frac{\partial^2 S_{ij}(\mathcal{X}_{ij})}{\partial \mathcal{X}_{ij}^2}\,,
\end{eqnarray}
so that $S$ is linear in $\mathcal{X}_{ij}$,
\begin{eqnarray}
S(\mathcal{X}_{ij})=J(\alpha(\mu),M,\mu) \mathcal{X}_{ij}- g^\prime(\alpha(\mu),M,\mu)\,.
\end{eqnarray}
Eq.~(\ref{eq81}) then implies
\begin{eqnarray}
\mathbf{T}_i \cdot \mathbf{T}_i\frac{\partial I_i(\mathcal{Z}_i)}{\partial \mathcal{Z}_i}
&=& \sum_{j\not=i}\mathbf{T}_i \cdot \mathbf{T}_j J(\alpha(\mu),M,\mu) \nn
&=& -  \mathbf{T}_i \cdot \mathbf{T}_i\ J(\alpha(\mu),M,\mu)\,,
\end{eqnarray}
using the relations in Sec.~\ref{app:color} so that
\begin{eqnarray}
I_i(\alpha(\mu),M,\Delta_i,m_i,\mu) &=&  -J(\alpha(\mu),M,\mu) \mathcal{Z}(i)\nn
&&+
 E(\alpha(\mu),M,m_i,\mu)\nn
\end{eqnarray}
and the total amplitude is
\begin{eqnarray}
\log \bamp &=&J(\alpha(\mu),M,\mu)  \biggl[ \sum_i \mathbf{T}_i \cdot 
\mathbf{T}_i \, \log \frac{\bar  n_i \cdot p_i}{\mu} \nn
&&- \sum_{\vev{ij}}  \mathbf{T}_i \cdot \mathbf{T}_j \log (n_i \cdot n_j)\biggr]\nn
&&\hspace{-1cm}+\sum_i  \mathbf{T}_i \cdot \mathbf{T}_i\left[ E(\alpha(\mu),M,m_i,\mu)
+\frac12 g(\alpha(\mu),M,\mu)\right]\nn
\label{eq86old}
\end{eqnarray}
using Eq.~(\ref{sumiden}). It is convenient to replace $n_i \cdot n_j \to (- n_i \cdot n_j)/2$, and absorb the $\log(-2)$ term in $g$  so that
\begin{eqnarray}
\log \bamp &=&J(\alpha(\mu),M,\mu)  \biggl[ \sum_i \mathbf{T}_i \cdot 
\mathbf{T}_i \, \log \frac{\bar  n_i \cdot p_i}{\mu} \nn
&&- \sum_{\vev{ij}}  \mathbf{T}_i \cdot \mathbf{T}_j \log \frac{-n_i \cdot n_j-i0^+}{2}\biggr]\nn
&&+\sum_i  \mathbf{T}_i \cdot \mathbf{T}_i\left[ E(\alpha(\mu),M,m_i,\mu)
+\frac12 g(\alpha(\mu),M,\mu)\right]\nn
\label{eq86}
\end{eqnarray}
where the $i0^+$ is given by analytically continuing $n_i \cdot n_j$ for spacelike momentum transfer to timelike  momentum transfer.

This is the form of the general amplitude, Eq.~(\ref{eq:form}), and as promised, is written in a manifestly symmetric form in the particle labels. The only term not proportional to the unit matrix is
\begin{eqnarray}
- J(\alpha(\mu),M,\mu) \sum_{\vev{ij}}  \mathbf{T}_i \cdot \mathbf{T}_j \log \frac{-n_i \cdot n_j-i0^+}{2}\,.
\label{eq86z}
\end{eqnarray}
For the other terms, one can use $\mathbf{T}_i \cdot \mathbf{T}_i =C_{Fi} \openone$ to write them in terms of the color Casimir.

\end{appendix}

\bibliography{biblio}

\end{document}